\documentclass[useAMS,usenatbib]{mn2e}
\usepackage{natbib}
\usepackage{deluxetable}
\usepackage{threeparttable}
\usepackage{longtable}
\usepackage{environ}
\usepackage{graphicx}
\usepackage{textcomp}
\usepackage[fleqn]{amsmath}
\usepackage{lscape}
\usepackage{pdflscape}
\usepackage{dcolumn}
\usepackage{rccol}
\usepackage{threeparttablex}
\usepackage{amssymb}
\usepackage{comment} 

\newcommand{\omatter}{\ensuremath{\Omega_{\mathrm{M}}}}

\newcommand{\deltam}{\ensuremath{\Delta m_{15}}}

\title[Swift SNe~Ia]{Swift UVOT Grism Observations of Nearby Type~Ia Supernovae -- I. Observations and Data Reduction}
\author[Pan et al.]{
Y.-C.~Pan$^{1}$\thanks{E-mail:ypan6@ucsc.edu},
R.~J.~Foley$^{1}$,
A.~V.~Filippenko$^{2,3}$,
N.~P.~M.~Kuin$^{4}$
\\
  $^{1}$Department of Astronomy and Astrophysics, University of California, Santa Cruz, CA 95064, USA\\
  $^{2}$Department of Astronomy, University of California, Berkeley, CA 94720-3411, USA\\
  $^{3}$Miller Senior Fellow, Miller Institute for Basic Research in Science, University of California, Berkeley, CA 94720, USA\\
  $^{4}$Mullard Space Science Laboratory/University College London, Holmbury St. Mary, Dorking, Surrey, RH5 6NT, UK\\
}

\begin{document}

\maketitle

\label{firstpage}

\begin{abstract}
  Ultraviolet (UV) observations of Type~Ia supernovae (SNe~Ia) are
  useful tools for understanding progenitor systems and explosion
  physics.  In particular, UV spectra of SNe~Ia, which probe the
  outermost layers, are strongly affected by the progenitor
  metallicity.  In this work, we present 120 {\it Neil Gehrels Swift Observatory}
  UV spectra of 39 nearby SNe~Ia.  This sample is the largest UV
  ($\lambda < 2900$\,\AA) spectroscopic sample of SNe~Ia to date,
  doubling the number of UV spectra and tripling the number of SNe
  with UV spectra.  The sample spans nearly the
  full range of SN~Ia light-curve shapes ($\deltam(B)\approx 0.6$--1.8\,mag).  
  The fast turnaround of {\it Swift} allows us to obtain
  UV spectra at very early times, with 13 out of 39 SNe having their
  first spectra observed $\gtrsim1$~week before peak brightness and
  the earliest epoch being 16.5\,days before peak brightness.  The
  slitless design of the {\it Swift} UV grism complicates the data
  reduction, which requires separating SN light from underlying
  host-galaxy light and occasional overlapping stellar light. We
  present a new data-reduction procedure to mitigate these issues,
  producing spectra that are significantly improved over those of standard
  methods.  For a subset of the spectra we have nearly simultaneous
  {\it Hubble Space Telescope} UV spectra; the {\it Swift} spectra are
  consistent with these comparison data.
\end{abstract}


\section{Introduction}
\label{sec:introduction}

\begin{figure*}
	\centering
		\includegraphics[scale=0.49]{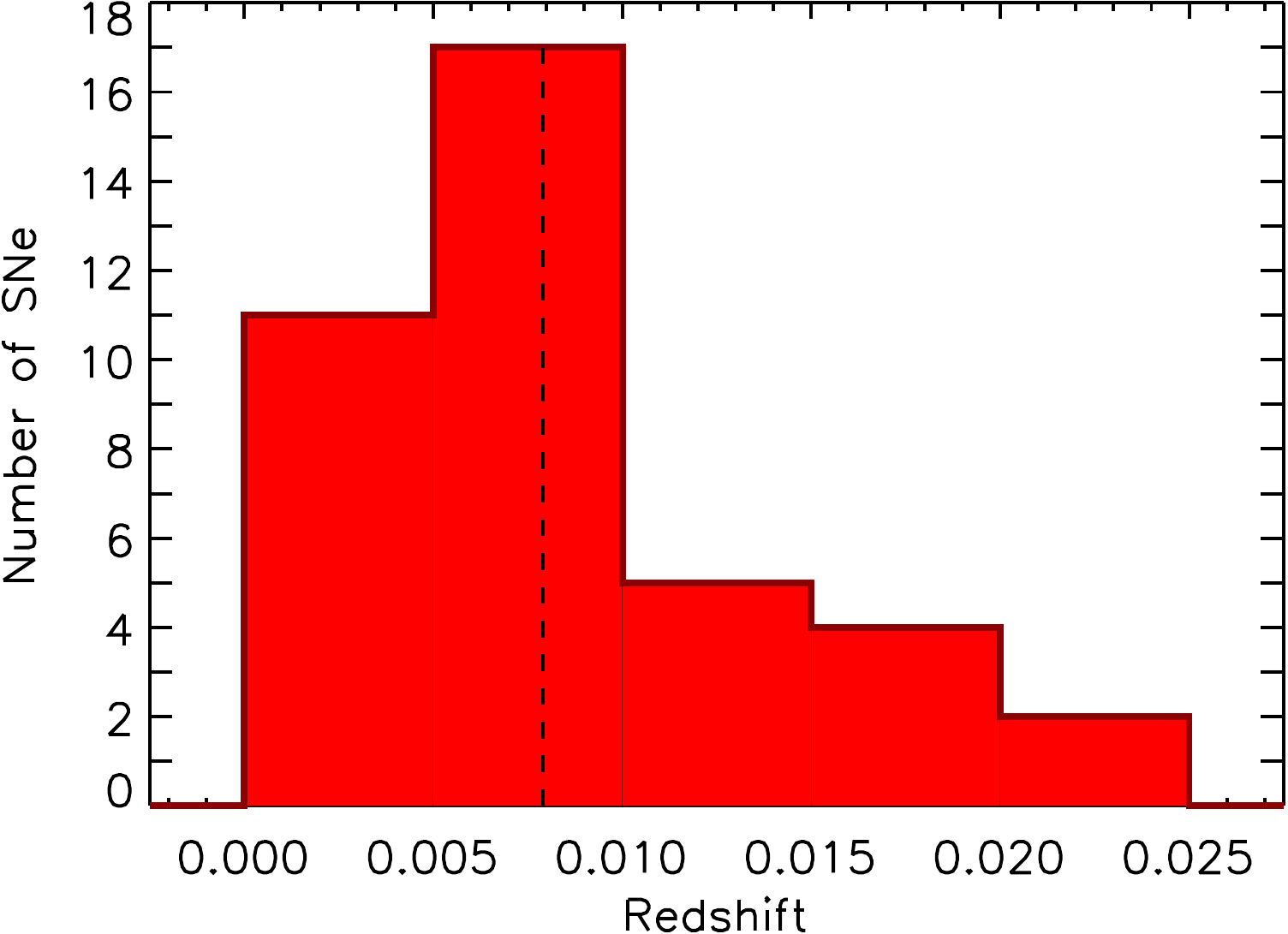}
		\hspace{0.25cm}
		\includegraphics[scale=0.49]{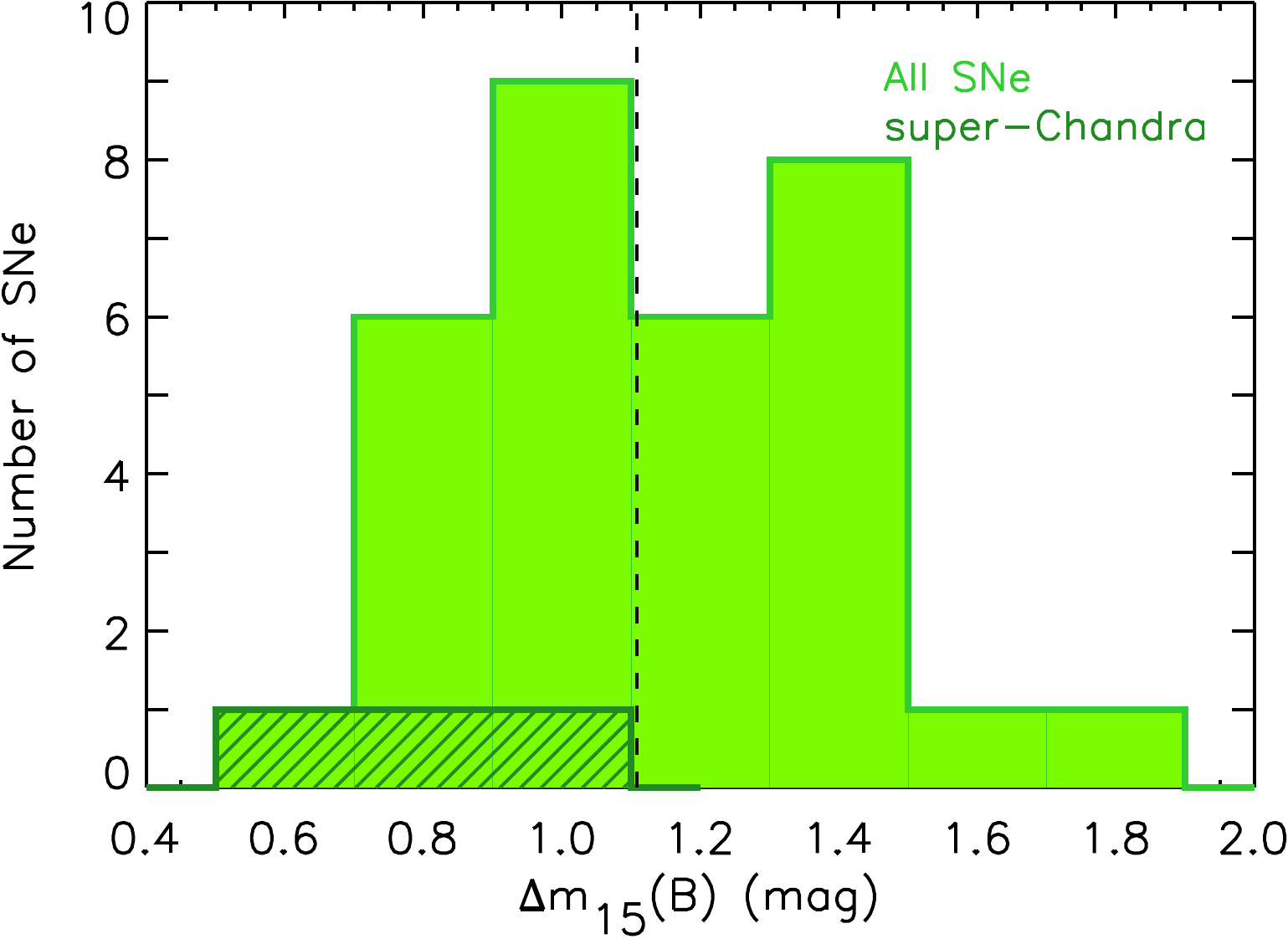}\\
		\vspace{0.25cm}
		\includegraphics[scale=0.49]{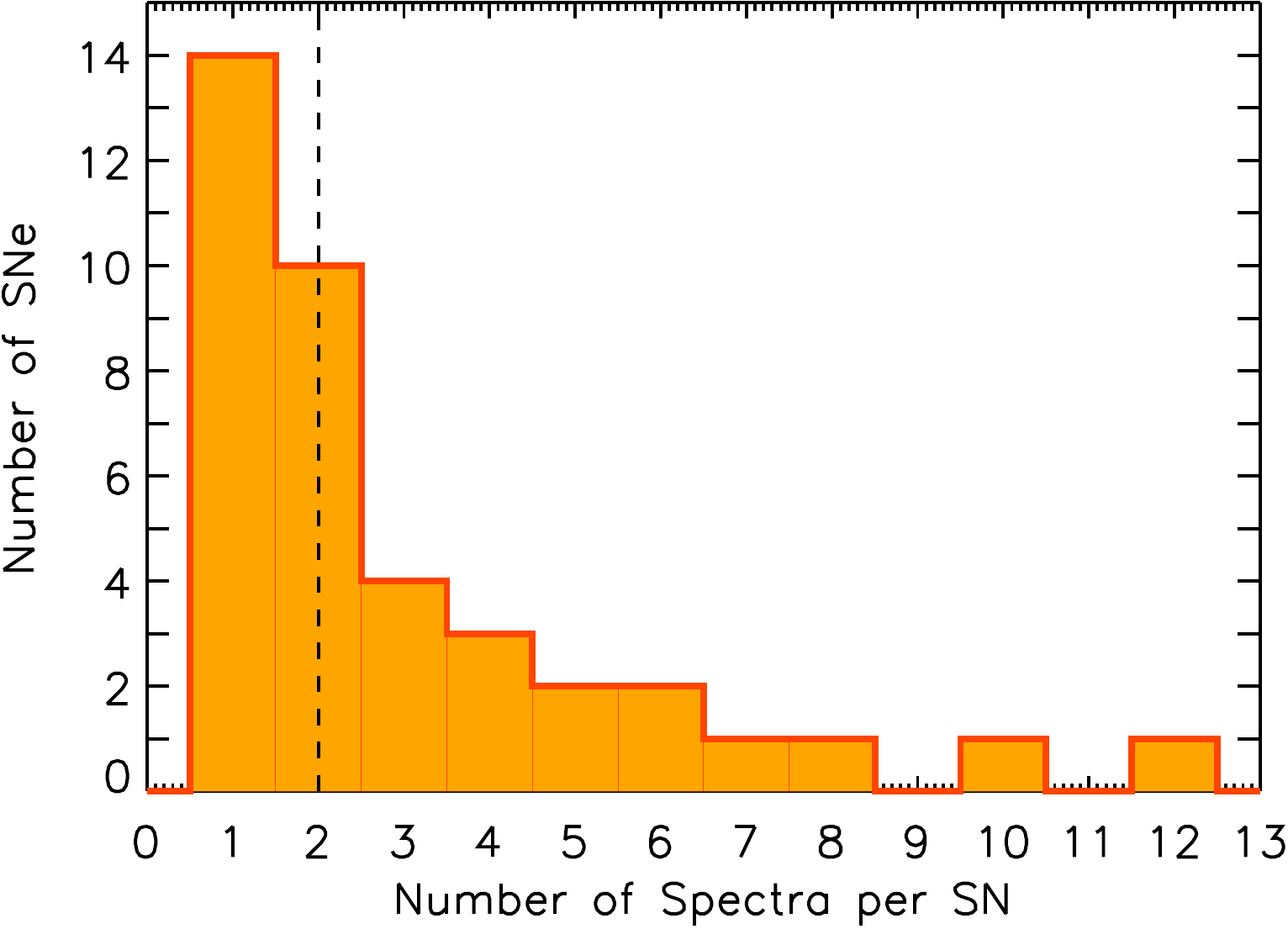}
		\hspace{0.25cm}
		\includegraphics[scale=0.45]{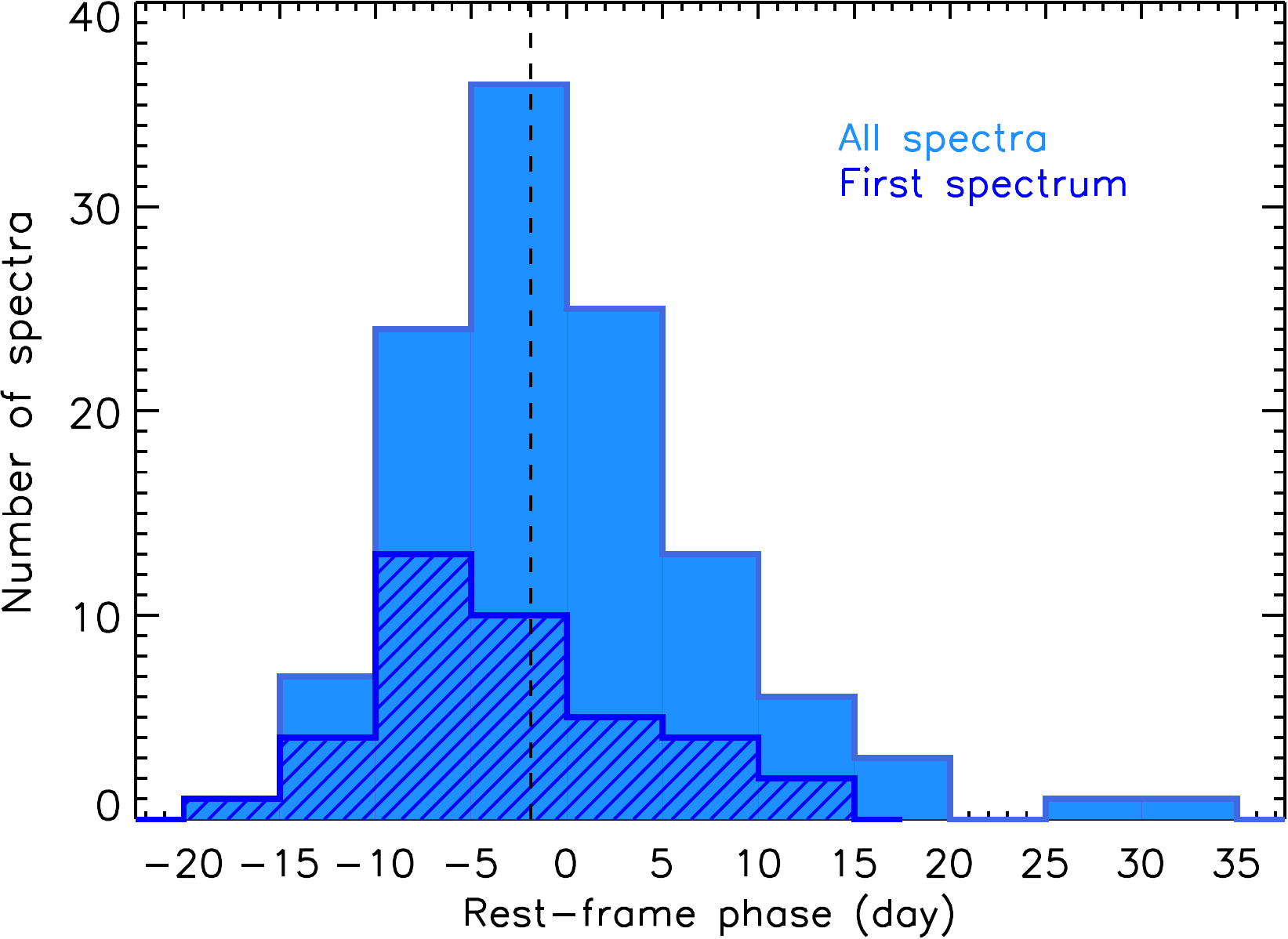}
                \caption{Redshift (upper left), \deltam($B$)
                  (upper right), number of spectra per SN
                  (lower left), and rest-frame phase (lower right)
                  distributions of the {\it Swift} UV SN~Ia sample
                  presented in this work.  \deltam($B$) represents the
                  $B$-band decline 15~days after peak $B$-band
                  brightness.  In the upper-right panel, we separate
                  our sample into those classified as normal SNe~Ia
                  (solid histogram) and ``super-Chandrasekhar'' SNe~Ia
                  (dashed histogram).  In the lower-right panel, we
                  also overplot the phase distribution of the first
                  spectrum (dashed histogram). The vertical dashed
                  line in each panel shows the median of the
                  distribution.  }
        \label{swift-dist}
\end{figure*}

Type Ia supernovae (SNe~Ia) are luminous distance indicators that were
used to first discover the accelerating expansion of the Universe
\citep{1998AJ....116.1009R, 1999ApJ...517..565P}.  SNe~Ia provide a
direct route to probe the nature of the dark energy that drives the
accelerated expansion.  While SNe~Ia are not perfect standard candles,
they can be standardised via the tight relation between SN~Ia
light-curve width and luminosity \citep[width-luminosity relation, WLR;][]{1993ApJ...413L.105P}
and between SN~Ia optical colour and luminosity
\citep{1996ApJ...473...88R} --- more luminous SNe~Ia are bluer and
have broader, slower evolving light curves.  After making these
corrections, we are able to use these {\it standardisable} candles for
cosmological inferences.

Observational evidence indicates that a SN~Ia is the result of the
thermonuclear explosion of an accreting carbon-oxygen white dwarf (WD)
star in a close binary system \citep[e.g.,][]{2000ARA&A..38..191H,
  2013FrPhy...8..116H, 2014ARA&A..52..107M}.  Theoretically, the
amount of $^{56}$Ni synthesised during the thermonuclear explosion
affects the optical opacity and changes the ``width'' of a SN light
curve, with slower declining SNe~Ia having more opacity and higher
luminosities \citep{1996ApJ...472L..81H}.  Making empirical
corrections based on light-curve width and colour, SNe~Ia become
exquisite distance indicators with a distance scatter below
8\% \citep[e.g.,][]{2008ApJ...681..482C,2009ApJ...699L.139W,2015Sci...347.1459K}.  
However, that remaining distance scatter is {\it intrinsic scatter}, beyond any
measurement error, and must be related to physics unaccounted for in
the standardisation.

Current SN cosmology analyses assume that SNe~Ia across all redshifts
have the same peak luminosity after standardising by the WLR. However,
new observations showed that even after standardisation, luminosity
still correlates with the large-scale host-galaxy environment
\citep[e.g.,][]{2010ApJ...715..743K,2010MNRAS.406..782S, 2010ApJ...722..566L,2014MNRAS.438.1391P}.  
SNe~Ia in galaxies of higher metallicities have (on average) higher corrected
luminosities than those with lower metallicities.  SNe~Ia with
identical progenitors except for metallicity are predicted to produce
different amounts of $^{56}$Ni. Higher progenitor metallicity will
result in a larger fraction of stable iron-group elements (IGEs) and 
less $^{56}$Ni in the SN explosion, and therefore fainter peak luminosity
\citep{2003ApJ...590L..83T}.

Theoretical studies indicate that the higher progenitor metallicity
will increase the IGEs in the outer layers of the SN, which will cause
greater UV line blanketing \citep{2000ApJ...530..966L}. Consequently,
the progenitor metallicity will not change the optical spectral energy
distribution (SED) of a SN~Ia
significantly, but will dramatically change its UV SED. This effect is
also seen in recent observations of the ``twin'' SN~2011by and SN~2011fe
\citep{2013ApJ...769L...1F}. The two SNe have nearly identical optical
light-curve widths and spectra but very different UV spectra. Thus,
the WLR is insufficient to calibrate the luminosity, resulting
in increased Hubble-diagram scatter. The UV SED is essential to detect this
metallicity effect.

The earliest UV observations of SNe date back to early 1980s,
with a handful of SNe~Ia observed by the {\it International Ultraviolet Explorer (IUE)} 
satellite \citep*[e.g.,][]{1995ESASP1189.....C}. 
The launch of the {\it Hubble Space Telescope} ({\it HST}) marked a milestone in 
obtaining high-quality UV spectra to study the progenitor composition 
and explosion mechanisms of SNe~Ia \citep[e.g.,][]{2012MNRAS.426.2359M,2015MNRAS.452.4307P, 
2016MNRAS.461.1308F}. However, the current {\it HST} sample is too small 
\citep[e.g., 9 SNe in][]{2016MNRAS.461.1308F} to cover all of the parameter space, such as
light-curve width, ejecta velocity, and progenitor metallicity. Thus,
we try to increase the sample of SNe~Ia with UV spectra obtained with
the {\it Neil Gehrels Swift Observatory} \citep{2004ApJ...611.1005G}.

A good amount of $\it Swift$ UV spectroscopy has been published by recent studies
\citep[e.g.,][]{2009ApJ...700.1456B,2012ApJ...744...38F,2014ApJ...787...29B,2016PASP..128c4501S,2017arXiv170703823G}.
Although {\it HST} has superior UV capabilities compared to {\it Swift}, 
the fast turnaround of {\it Swift} and its efficiency to obtain the data are unmatched.
The earliest UV spectrum of a SN~Ia published before this work was obtained from {\it Swift}
\citep[SN~2009ig;][]{2012ApJ...744...38F}; it was observed $\sim13$~days before the peak
brightness. Extremely early observations with {\it Swift} are complementary to the existing
{\it HST} UV sample.

However, the slitless design of the {\it Swift} observations makes the spectrum more likely to be
contaminated by nearby background sources. This not only complicates the data reduction, but also makes
the interpretation of {\it Swift} datasets difficult. Traditional methods using {\it Swift}/Ultraviolet--Optical 
Telescope \citep[UVOT;][]{2004SPIE.5165..262R} grism data-reduction software 
\citep[UVOTPY;][]{2014ascl.soft10004K} have been widely used, but they
become less reliable when reducing spectra that are seriously contaminated by nearby background sources.
A more effective decontamination technique has been developed by \citet{2016PASP..128c4501S}.
However, their method requires a template observation of the galaxy with the same spacecraft roll angle
at late times (usually $>1$~yr after SN explosion), which weakens the advantage of 
{\it Swift} (i.e., its fast turnaround). A relatively fast and correct reduction
of early-time data is useful for assessing and scheduling the follow-up observations of
young transients.

In this work, we present a SN sample
that is more than three times larger than that of the {\it HST} UV sample.  We
improve the {\it Swift} UVOT grism data-reduction procedure
to better extract SN spectra.  This allows us to produce more accurate
{\it Swift} spectra, which we will exploit in future analyses.

A plan of the paper follows. In Section~\ref{sec:obs} we introduce the
selection and observations of our SN~Ia sample.
Section~\ref{sec:reduction} discusses the data-reduction techniques.
We present the spectra in Section~\ref{sec:result} and summarise our
results in Section~\ref{sec:summary}. Throughout this paper, we assume
$\mathrm{H_{0}} = 70$\,km\,s$^{-1}$\,Mpc$^{-1}$ and a flat universe
with $\omatter = 0.3$. 

\section{Observations}
\label{sec:obs}
\subsection{SN sample selection}
\label{sec:sample}

\begin{figure}
	\centering
        \includegraphics[scale=0.40]{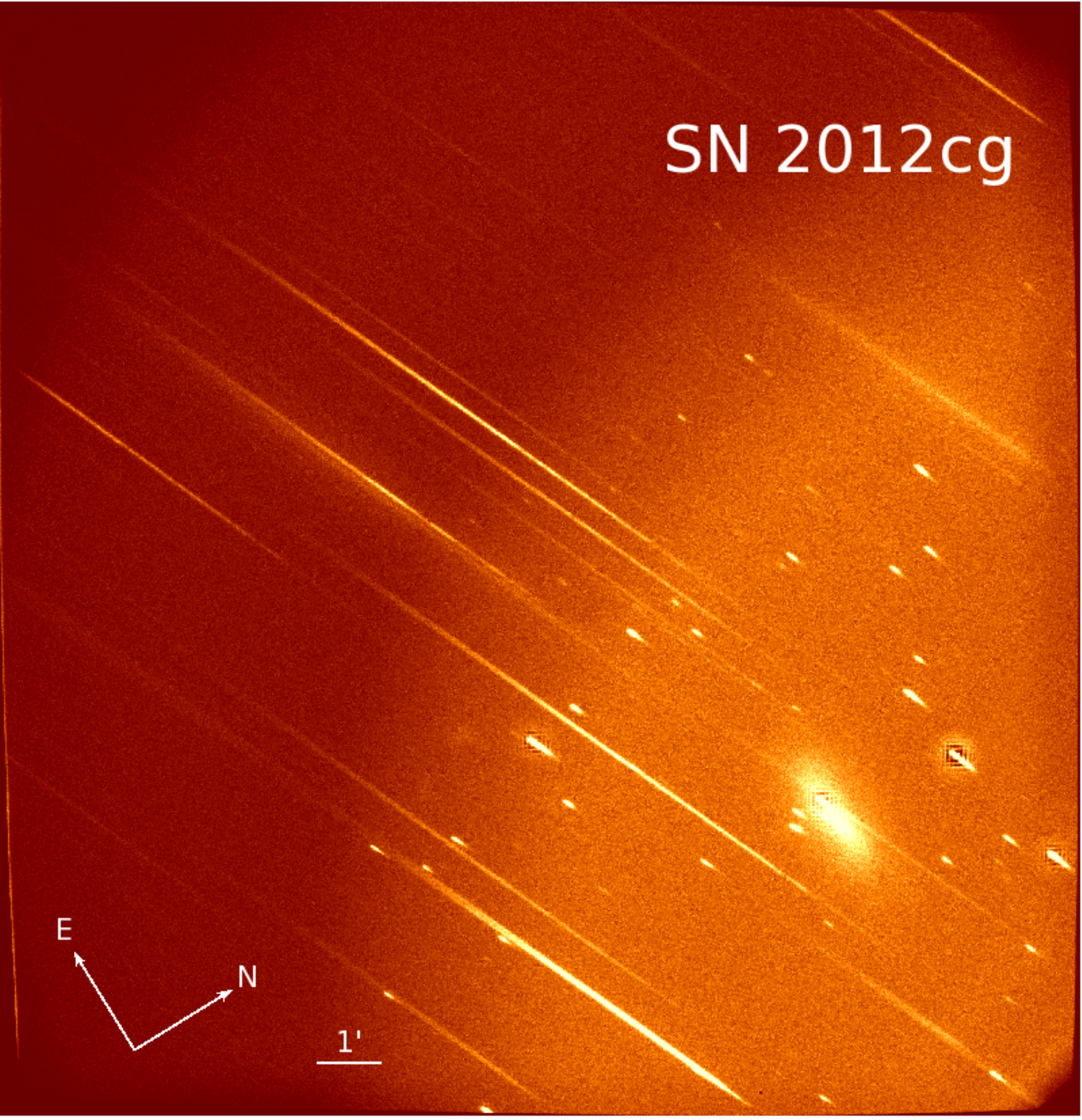}\\
        \vspace{0.2cm}
        \includegraphics[scale=0.40]{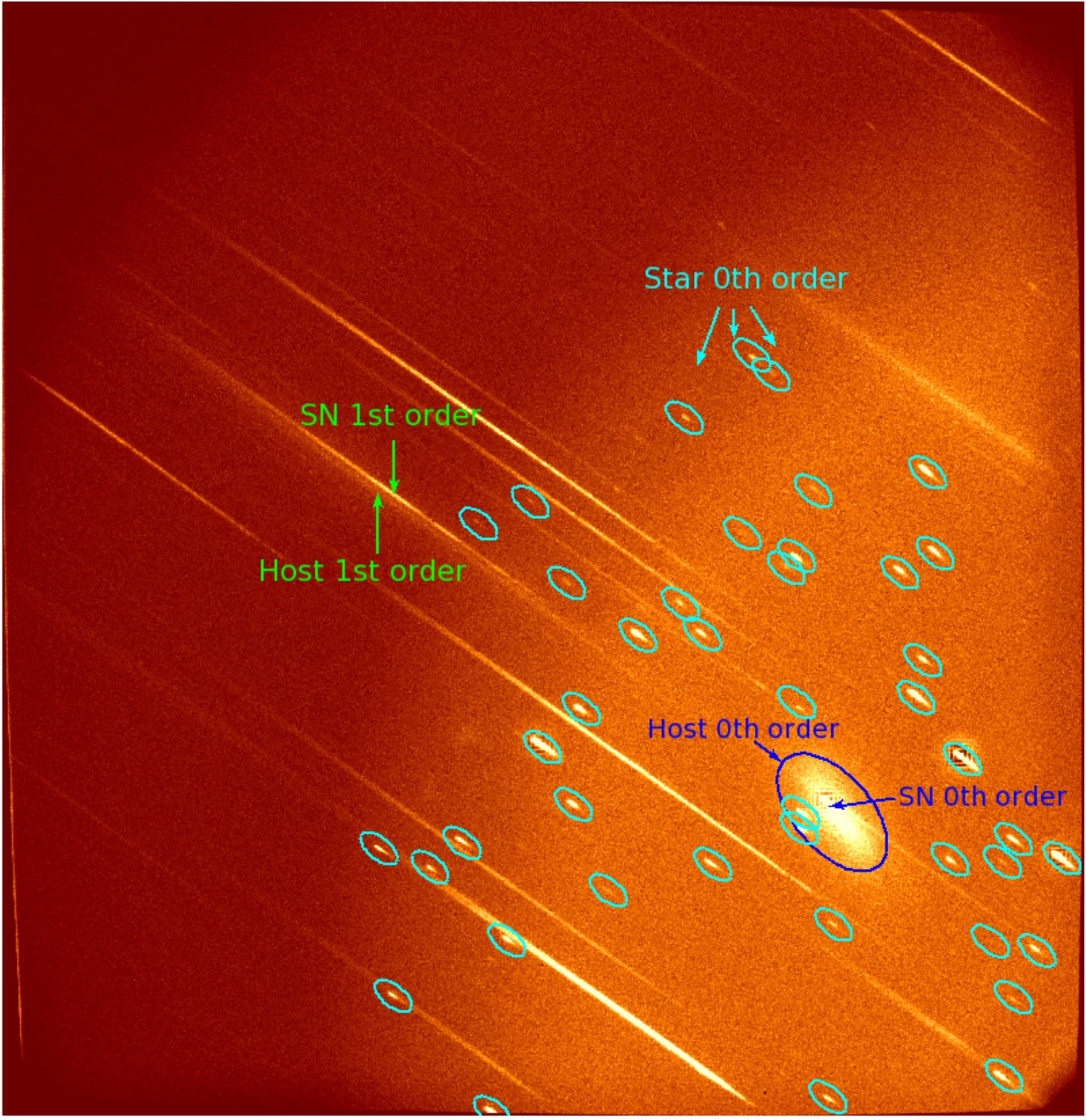}
        \caption{{\it Upper panel}: {\it Swift} data image of
          SN~2012cg. {\it Lower panel}: Same as upper panel, but with
          annotations. The first-order spectra from the SN and its
          host galaxy are marked with green arrows. The zeroth-order
          light from the SN and its host galaxy are marked with blue
          arrows.  All zeroth-order images from nearby bright stars
          are marked in cyan. The image is 17\arcmin$\times$17\arcmin.  }
        \label{swift-data1}
\end{figure}

\begin{figure*}
	\centering
	    \includegraphics[scale=0.8]{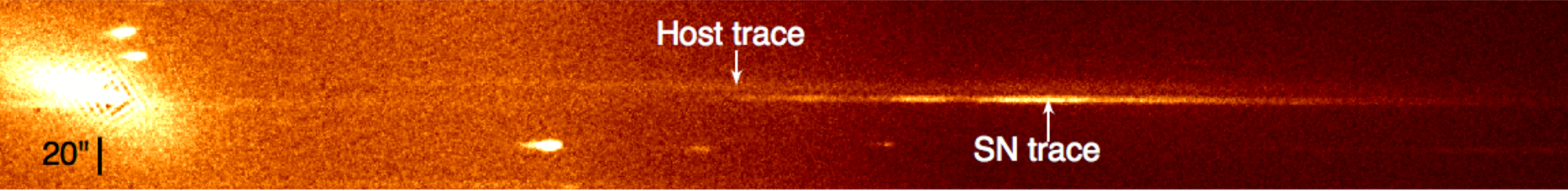}\\
        \vspace{0.25cm}
        \includegraphics[scale=0.8]{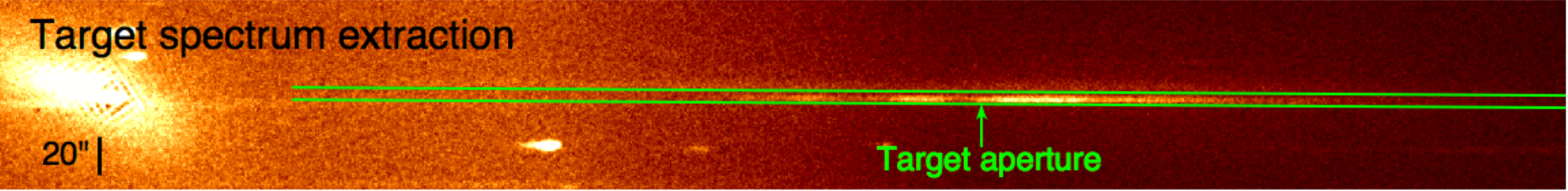}\\
        \vspace{0.25cm}
        \includegraphics[scale=0.8]{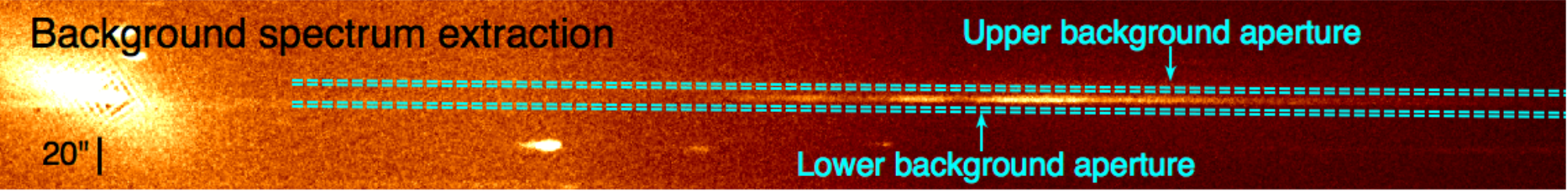}
        \caption{Rotated and expanded version of the SN~2012cg image
          shown in Figure~\ref{swift-data1}. The top, middle, and
          bottom panels mark the SN and host-galaxy spectra, the SN
          extraction aperture, and the background regions,
          respectively.  }
        \label{swift-data2}
\end{figure*}

\begin{figure}
	\centering
	    \includegraphics[scale=0.52]{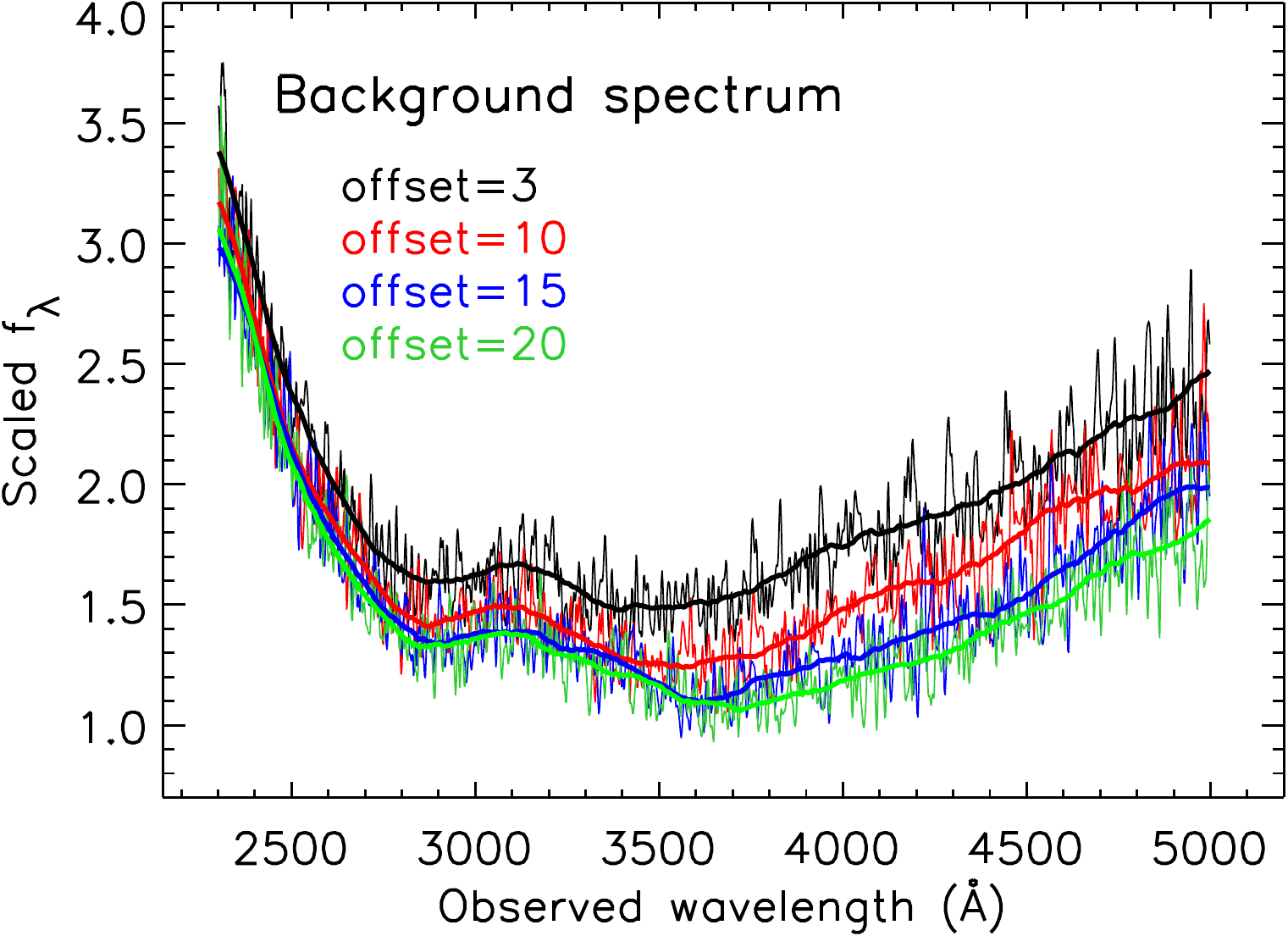}\\
	    \vspace{0.25cm}
	    \includegraphics[scale=0.52]{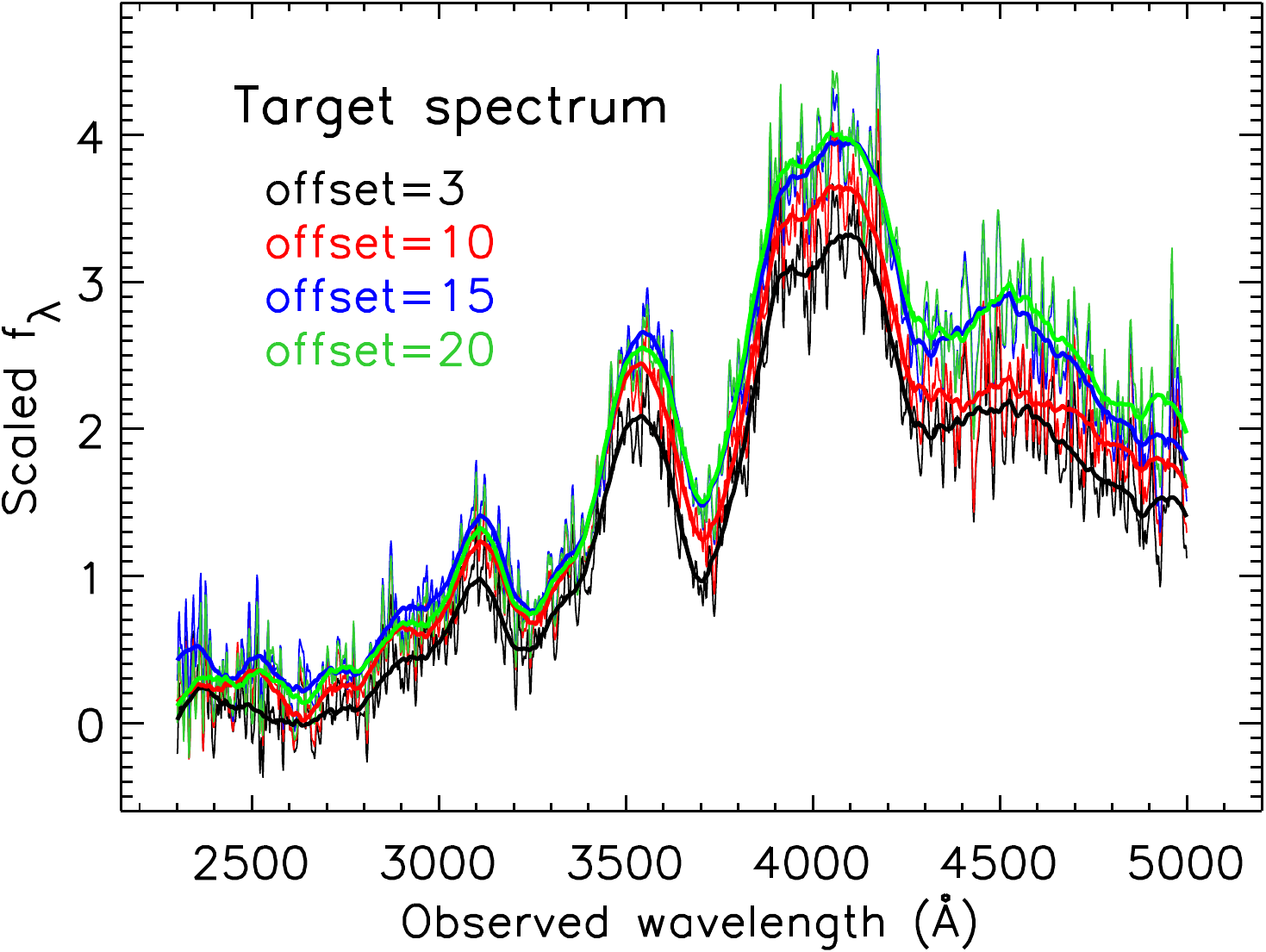}
            \caption{{\it Upper panel}: Comparison of interpolated
              background spectra by varying the offset (in pixels) of
              the background channel relative to the SN aperture as
              shown in Figure~\ref{swift-data2}. Here the background
              spectra with offsets of 3 (the value ultimately chosen
              for this spectrum), 10, 15, 20~pixels are shown as
              black, red, blue, and green curves, respectively. {\it
                Bottom panel}: The resulting SN spectra for the
             background offsets shown in the upper panel plotted in
              their corresponding colours. Savitzky-Golay smoothed
              spectra are overplotted, as thick lines, in both panels.
            }
        \label{bg-compare}
\end{figure}

Most of the SNe in our sample were observed as part of our dedicated
Guest Investigator programs (GI--04047, GI--5080130, PI Filippenko;
GI--6090689, GI--8110089, GI--1013136, GI--1215205, PI Foley).
However, the full sample contains all SNe~Ia observed with the UVOT/UV
grism by {\it Swift} (excluding spectra that are not useful for further analysis; see below
for more details).

The UVOT has a relatively small aperture (30~cm), and so only
relatively nearby SNe~Ia are sufficiently bright (distance modulus $\mu \lesssim 35$\,mag,
corresponding to V$\rm_{max} \lesssim15$\,mag) to produce high-quality
spectra in reasonable exposure times.  We desired spectra of the SNe
before or near peak brightness, and therefore most SNe in the sample
were discovered 1--2 weeks before peak brightness.

The sample presented here contains 120 {\it Swift}/UVOT Grism
spectra of 39 SNe~Ia, with 20 SNe observed through our {\it Swift}
programs and another 19 SNe selected from the {\it Swift} data archive
that have UV spectra. The complete list of SNe in this work
can be found in Tables~\ref{obs-log} and \ref{obs-log2}. Besides
these 120 spectra, there are 17 observations from which we cannot extract
any useful data or that were highly contaminated by background sources.
We summarise them in Table~\ref{bad-log}.

In Figure~\ref{swift-dist}, we present the redshift, \deltam($B$) (the
$B$-band decline 15~days after peak brightness), number of spectra
per SN, and rest-frame phase distributions of our sample.  The median
redshift of the sample is 0.0079, with the closest object having $z =
0.0006$ ($D = 3.3$\,Mpc; SN~2014J) and the most distant SN having $z =
0.0214$ ($D = 93$\,Mpc; SN~2009dc). The \deltam($B$) of our sample
ranges from 0.6 to 1.8\,mag (median of 1.1\,mag), with three objects
classified as super-Chandra SNe~Ia having \deltam($B$) = 0.72
(SN~2009dc), 0.59 (SN~2011aa), and 1.08\,mag (SN~2012dn).  
We lack sufficient data to measure \deltam($B$) of 7 SNe.

A large fraction of the SNe in our sample (24 out of 39) have multiple
epochs of UV spectra, with a median of 2 spectra per SN. This makes
our sample particularly useful in studying UV spectral evolution.  The
SN phase is relative to the epoch of peak optical brightness. For the
6 SNe where we do not have light curves, the epoch of peak brightness
is estimated by fitting the optical spectra with \textsc{snid}
\citep{2007ApJ...666.1024B}. The median phase (in the rest frame) of
the first observation of a SN and all spectra is $-4.5$ and $-1.9$\,days, respectively.
Among the 39 SNe, 13 have spectra $\gtrsim$1~week before peak brightness, 
29 have pre-peak spectra, and 34 have spectra before 5~days after peak brightness. The earliest
spectrum in our sample (of SN~2017cbv) was observed $16.5$\,days
before peak brightness. It was observed by {\it Swift} 
$<2$~days after explosion.  This shows the fast turnaround of {\it
  Swift} and its ability to obtain a spectrum at extremely early
times.

\subsection{Swift observations}
\label{sec:swift-obs}
\begin{figure}
	\centering
	    \includegraphics[scale=0.52]{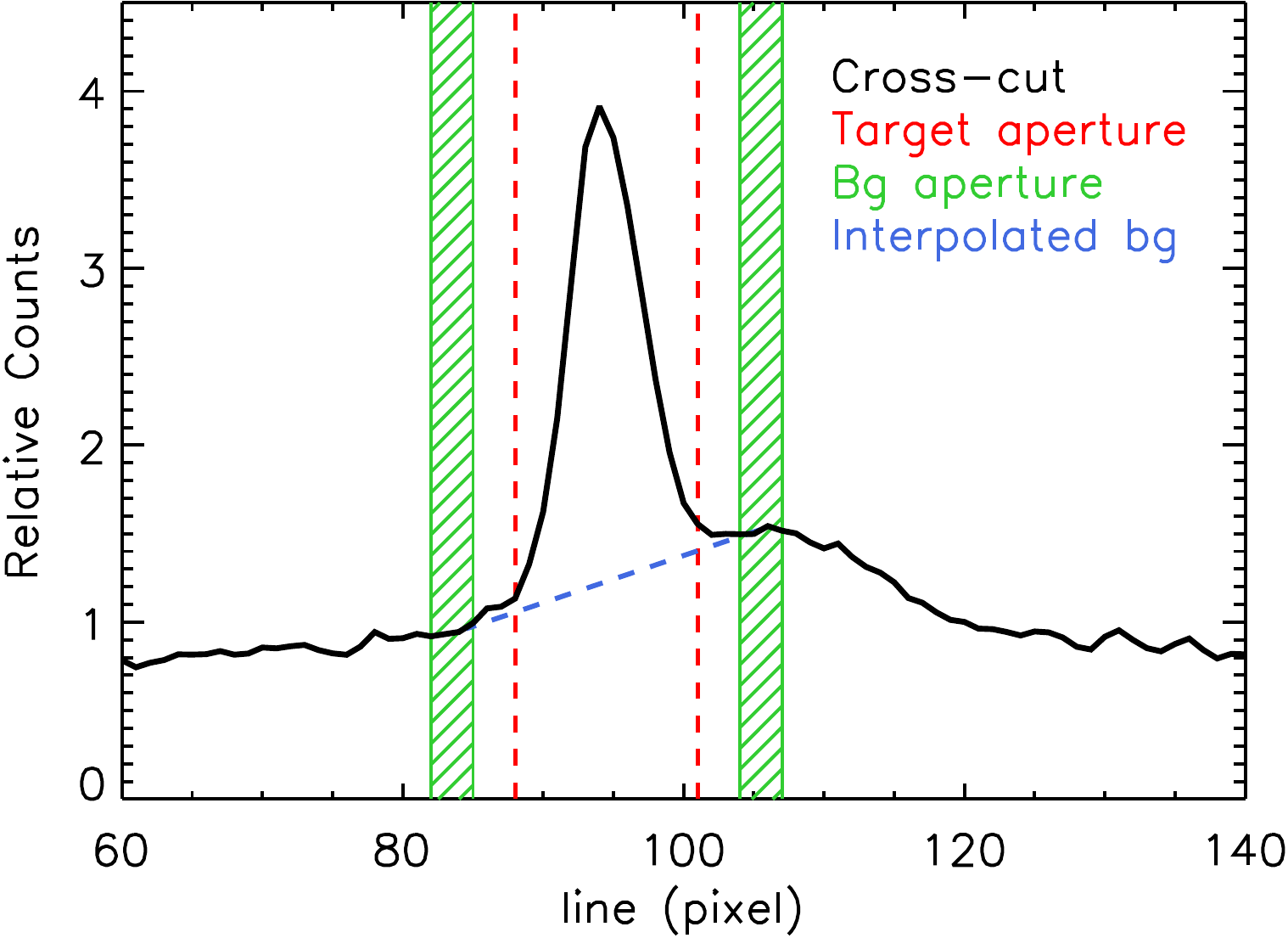}\\
            \caption{Spatial profile of the SN and background
              perpendicular to the SN trace (a ``cross cut'') for the
              data presented in Figure~\ref{swift-data2} (black
              curve).  The red dashed lines define the aperture used
              to extract the SN spectrum. The regions used to define
              the background are marked by hashed green regions. The
              interpolated background used for final background
              subtraction is shown as a blue dashed line. In this
              example, the size of the target aperture is set to
              13~pixels (1.8$\sigma$). The size of each background
              region is 3~pixels. The upper and lower background
              regions are offset by 3~pixels from the edge of the
              target aperture.}
        \label{cross-cut}
\end{figure}

The spectroscopic observations were performed by the {\it Swift} UVOT
\citep{2004SPIE.5165..262R}.  The UVOT provides UV and
optical spectroscopy with either slitless UV-grism or $V$-grism.
In this work, we focus on the UV-grism observations,
with a wavelength coverage of $\sim1700$--5000\,\AA\ and the UV response
optimised in the 2000--3400\,\AA\ region.

Owing to the slitless design of the UVOT for spectroscopy, the data image
contains both zeroth-order and higher-order emission (see
Figure~\ref{swift-data1} for an example). We asked our targets to be
observed under the ``clocked mode'' when possible. Observing in
clocked mode reduces the contamination from zeroth-order images of
field stars (e.g., those marked in the lower panel of
Figure~\ref{swift-data1}) where the first-order light falls on the detector
- and comes close to a slit spectrograph in the occulted region.

Given the brightness of our targets, we requested an average exposure
time $T_{\rm exp} = 15$\,ks in our own programs to obtain a good UV
spectrum near peak brightness; however, such long exposures were rarely
obtained because of various observing constraints.  The exposure times
ranged from 0.6 to 18\,ks with a median of 8.5\,ks.

\begin{figure}
	\centering
	\includegraphics[scale=0.54]{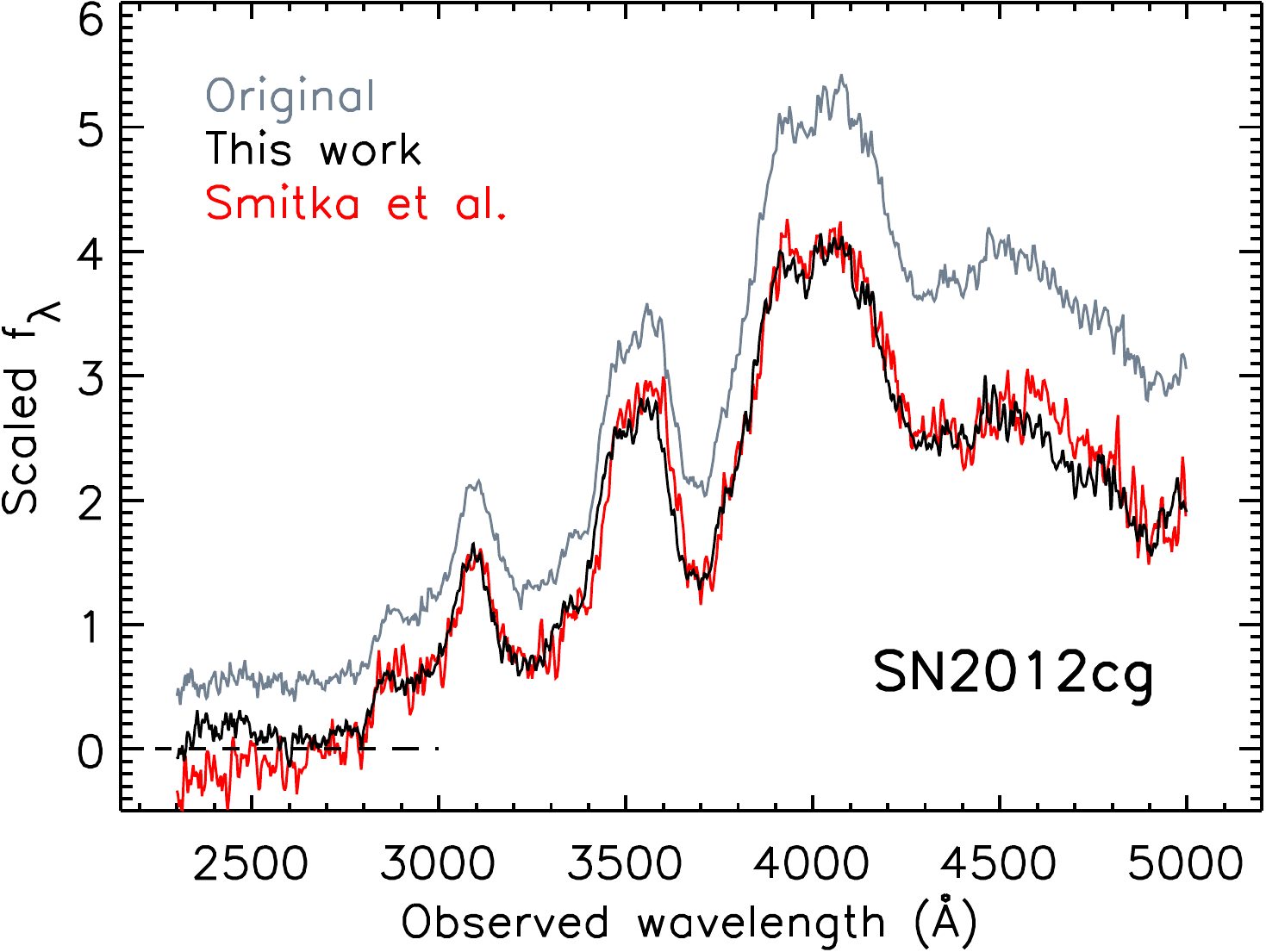}
        \caption{Comparison of our fully reduced spectrum of SN~2012cg
          (in black) to reduced spectra produced by the original 
          data-reduction pipeline (in grey) and by the
          \citet{2016PASP..128c4501S} decontamination technique (in
          red). The dashed line marks the level of zero flux.  }
        \label{swift-data3}
\end{figure}

\begin{figure*}
	\centering
	    \includegraphics[scale=0.51]{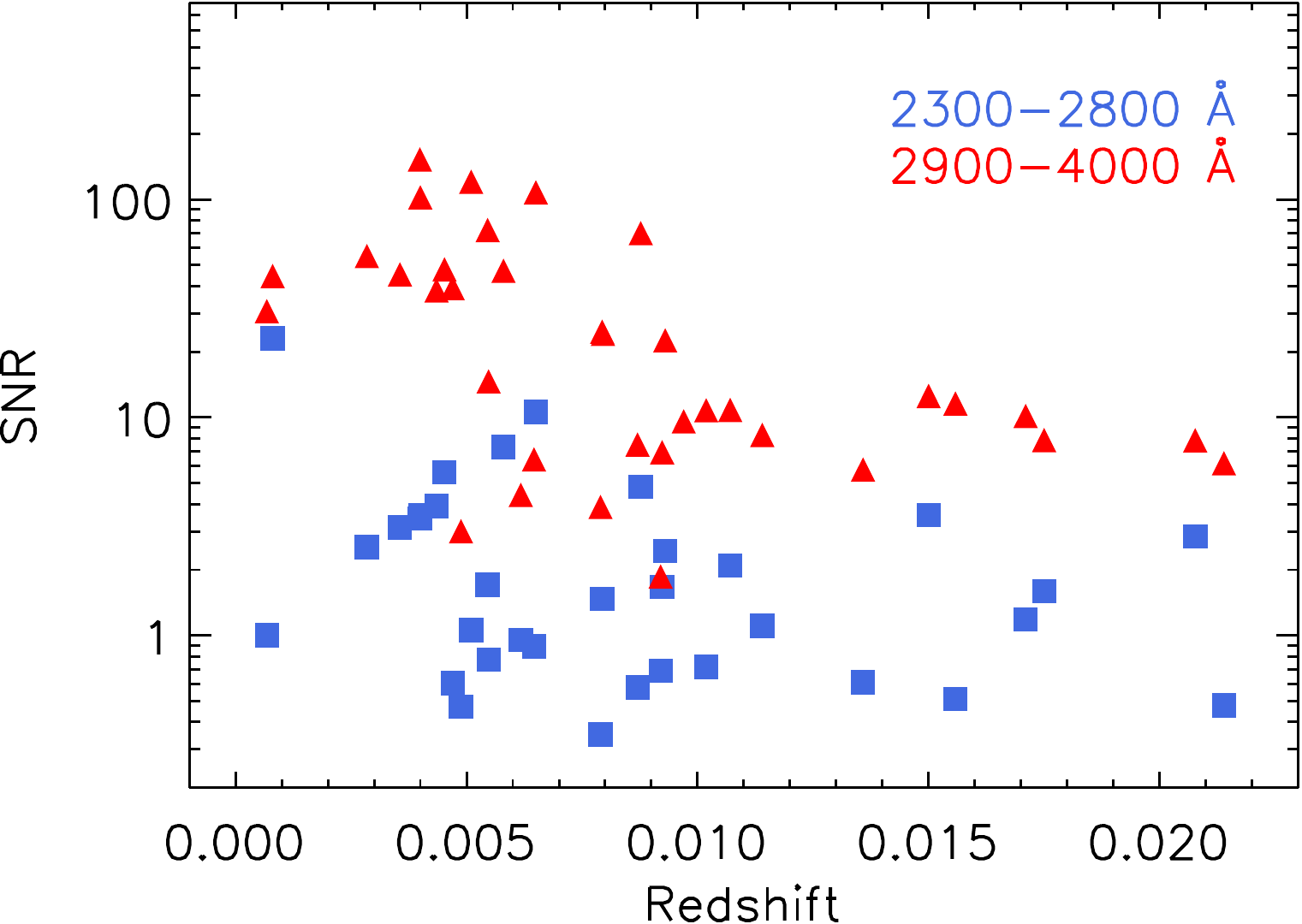}
	    \hspace{0.25cm}
	    \includegraphics[scale=0.51]{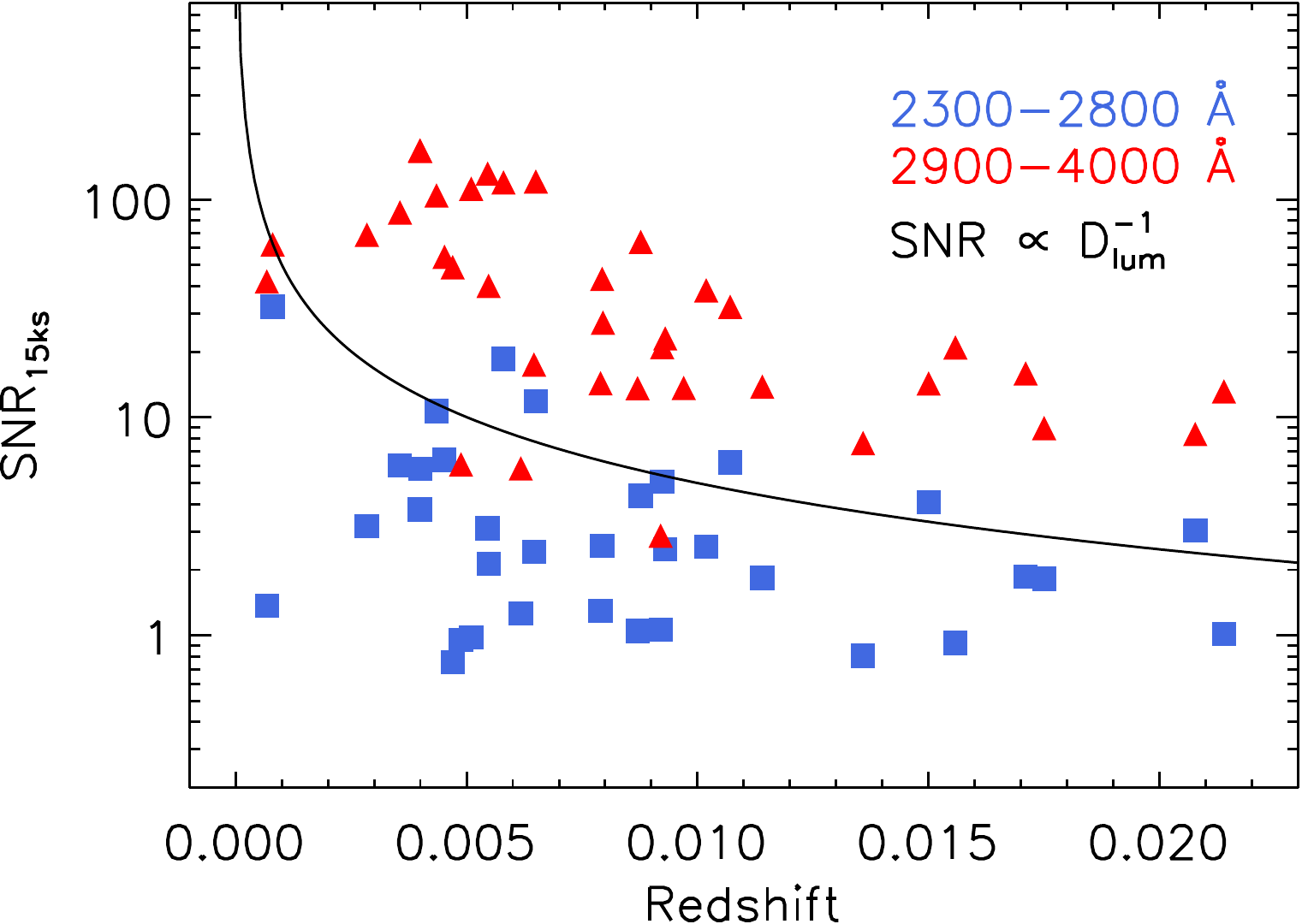}\\
	    \vspace{0.25cm}
	    \includegraphics[scale=0.51]{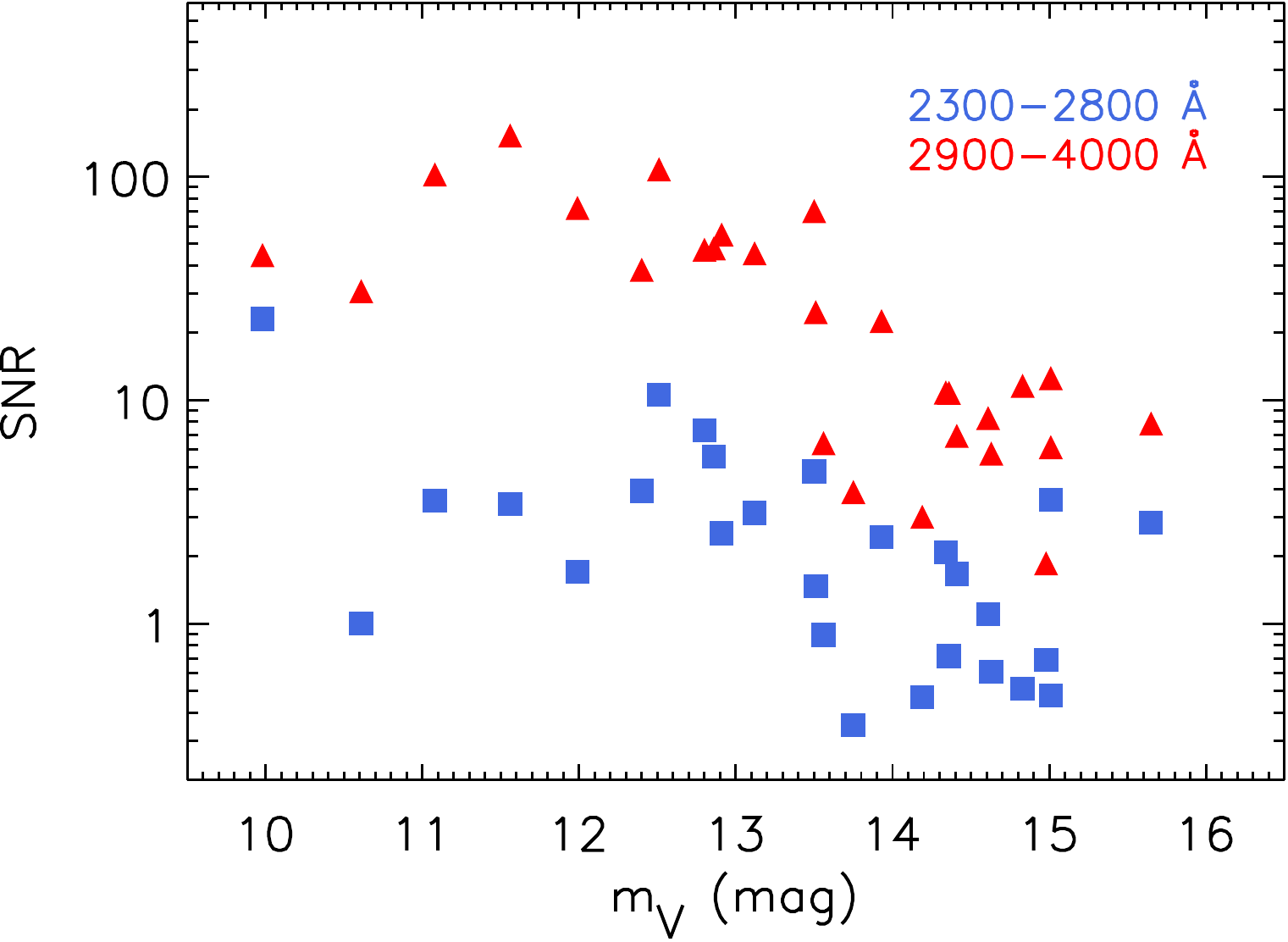}
	    \hspace{0.25cm}
	    \includegraphics[scale=0.51]{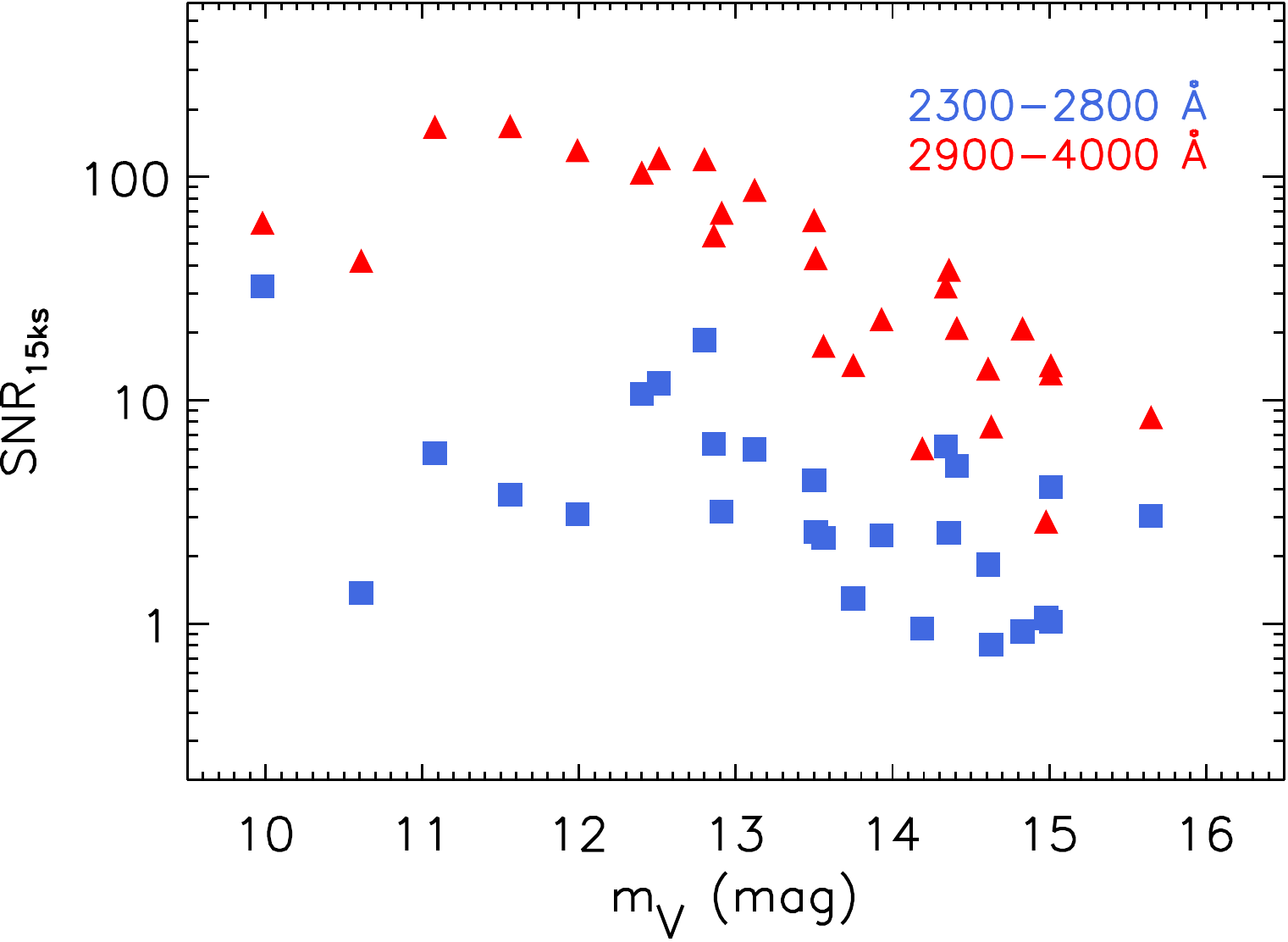}\\
	    \vspace{0.25cm}
		\includegraphics[scale=0.51]{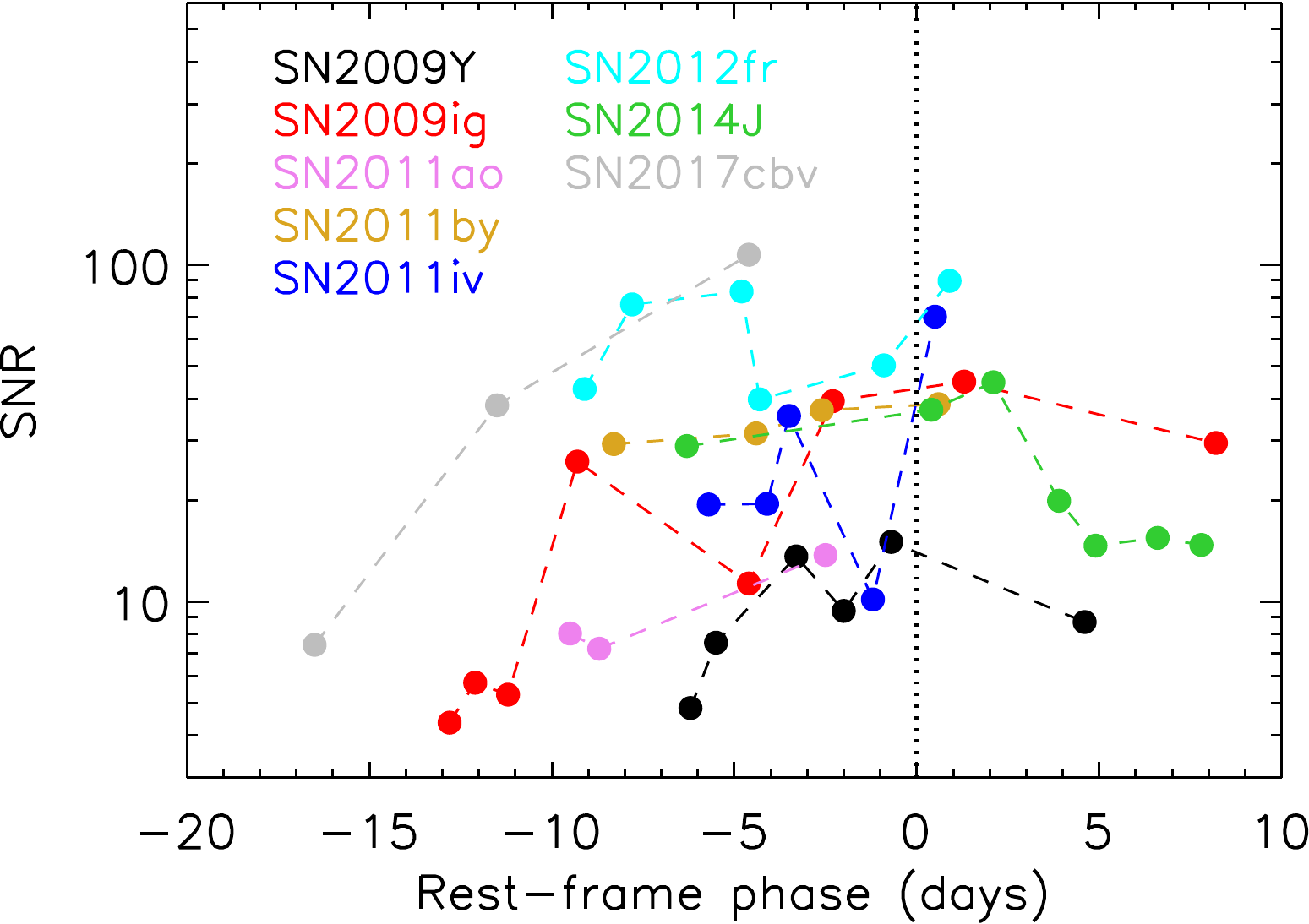}
		\hspace{0.25cm}
		\includegraphics[scale=0.51]{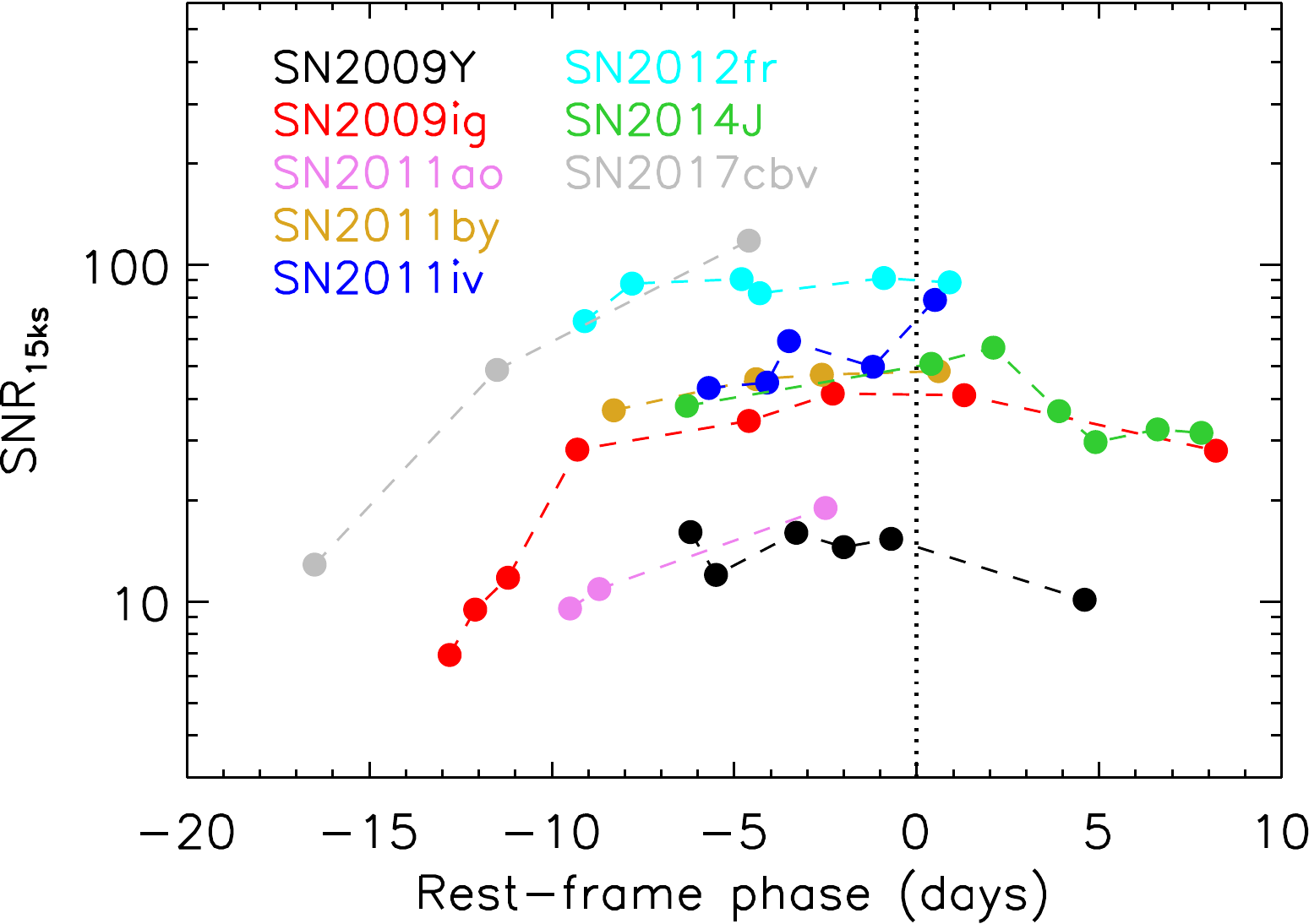}
                \caption{{\it Upper-left panel}: Signal-to-noise ratio
                  (SNR) of the {\it Swift} sample as the function of
                  redshift. Here we select the spectrum closest to the
                  peak brightness (within 5 days) of each SN for comparison. The SNRs
                  are determined for 2300--2800\,\AA\ (mid-UV) and
                  2900--4000\,\AA\ (near-UV) and displayed as blue squares and
                  red triangles, respectively. {\it Upper-right panel}:
                  Same as the upper-left panel, but scaling the SNR to
                  that expected with a $T_{\rm exp}=15$\,ks exposure
                  (see Section~\ref{sec:quality} for details). The
                  inverse-square law, flux $\propto D^{-2}$ (or SNR $\propto D^{-1}$),
                  is overplotted (the vertical position is arbitrary). 
                  Here the variable $D$ represents the
                  luminosity distance of the SN.
                  {\it Middle-left panel}: The SNR of the {\it Swift} 
                  sample as a function of $V$-band peak brightness.
                  As in the upper panels, the SNR is determined
                  from the near-peak spectrum of each SN.
                  {\it Middle-right panel}: Same as the upper-right panel, 
                  but with $V$-band peak brightness.
                  {\it Lower-left panel}: The
                  SNR of the {\it Swift} sample as a function of
                  rest-frame phase. We select the SNe which have
                  multi-epoch observations to show the temporal
                  variation of SNR. Here the SNR is determined for the
                  entire spectral range (2300--5000\,\AA). The
                  vertical dotted line marks the epoch of peak
                  brightness. {\it Lower-right panel}:
                  Same as the lower-left panel, but scaling the SNR to
                  that expected with a $T_{\rm exp}=15$\,ks exposure.}
        \label{snr}
\end{figure*}

\section{Data Reduction}
\label{sec:reduction}

Compared to typical slit spectroscopy, the slitless design of the {\it
  Swift} UVOT grism makes the target spectrum more likely to be
contaminated by nearby background sources. Although ``clocked mode''
is adopted for most of our observations, higher-order spectra (e.g.,
from either the host galaxy or field stars) could still overlay or fall
close to our targets, affecting the background subtraction.

The other major complication for slitless spectra is that the entire
galaxy along the dispersion direction, instead of the region directly
coincident with the SN, will contaminate the spectrum.  These
regions are spatially offset from the SN and other regions of the
galaxy, and therefore light from nearby wavelengths can contaminate
the SN spectrum (effectively ``smearing'' the galaxy spectrum in the
wavelength direction).

\citet{2016PASP..128c4501S} attempted to address these problems by
observing a template image of the galaxy long after the SN faded in
nearly the same configuration (e.g., clocking, pointing, and roll
angle) as the original data image. They measure the background at the
same location in the template image and then subtracted that from the
target spectrum. This decontamination technique is effective in
reducing any contamination in the SN spectrum.  However, the template
image can only be obtained after the SN has faded away (usually
$\gtrsim 1$ yr after peak brightness). Given the fast turnaround of
{\it Swift}, it would be ideal to be able to perform an effective data
reduction on a shorter timescale. Moreover, the exposure time for the
template spectrum must be long (i.e., comparable to that of the data image), 
so as to not adversely affect the SNR of the final host-subtracted SN spectrum.

We developed our own pipeline to reduce UVOT grism data, building upon
the {\it Swift} UVOTPY software \citep{2014ascl.soft10004K,2015MNRAS.449.2514K}.  
We follow the standard
procedures in UVOTPY, such as target extraction, background
subtraction, wavelength calibration, and flux calibration. However,
our pipeline measures the background in a different way. Instead of
creating a smoothed background image by averaging the regions above
and below the target spectrum, we extract background spectra above and
below the target with customisable offsets and sizes and then subtract
the interpolated (at the position of the target) background spectrum.

We illustrate the target and background extractions in
Figure~\ref{swift-data2}. The top panel shows the
first-order spectrum of SN~2012cg from Figure~\ref{swift-data1}. The
contamination from the host galaxy spectrum is clearly seen above the
SN trace. The aperture used to extract the SN spectrum is marked in
the panel.  For bright objects, we generally adopt an aperture size
similar to that of the UVOTPY default (i.e., 2.5\,$\sigma$). Here the
aperture size is controlled by $\sigma$, the standard deviation of a 
Gaussian distribution used to fit the count rate of the cross-dispersion 
profile of the spectrum \citep[for details, see][]{2015MNRAS.449.2514K}. Smaller
apertures are recommended for fainter sources or those with nearby
contamination.  For SN~2012cg, as displayed in
Figure~\ref{swift-data2}, we use an aperture size of 1.8\,$\sigma$.
When using smaller apertures, we rescale the flux to match that of the
default aperture size which was used for flux calibration and
coincidence-loss correction.

We demonstrate the background extraction in the bottom panel of
Figure~\ref{bg-compare}. The two apertures used to extract the
background spectra above and below the target are shown in the panel.
They are offset from the target along the dispersion axis to trace
the curvature of the spectrum.
The background spectrum can be sensitive to the offset of the
apertures from the target. Because of the contamination from the
underlying host-galaxy spectrum, the flux of the background spectrum
can vary by a factor of $\sim2$ depending on different offsets,
altering the final target spectrum significantly; this effect is shown
in Figure~\ref{bg-compare}. To accurately estimate the local
background, we generally extract the background spectrum as close to
the SN as possible, selecting an aperture size that does not overlap
with the SN aperture. This is achieved by inspecting the ``cross cut''
of the spectrum (i.e., the spatial flux distribution perpendicular to
the trace; see Figure~\ref{cross-cut}). For SN~2012cg, we offset the
background apertures from the edge of the target aperture by 3~pixels
to not only remove the host-galaxy light, but also not to include
the SN light. Larger offsets result in incorrect background subtraction 
and increase the contamination from the host galaxy.

After the target extraction and background subtraction, we wavelength-
and flux-calibrate each spectrum following the procedures in UVOTPY.
Given the large uncertainty in wavelength calibration \citep[the
accuracy is $\sim 9$\,\AA\ for the UV grism clocked mode;][]{2015MNRAS.449.2514K}, 
we shift the zeropoint of the wavelength solution for each individual
exposure by cross-correlating the spectrum to that of another spectrum
(either an {\it HST} or a ground-based spectrum) at a similar phase.

Each epoch of spectroscopy usually consists of several short
exposures, with each observed under slightly different conditions that
slightly change the flux and wavelength of the spectrum.  Therefore,
we repeat the same reduction procedure with each individual exposure
before combining all spectra into a single spectrum. If no comparison
spectrum is available, we simply shift the wavelength of all the
exposures to match the mean value. The shift of our {\it Swift} spectra ranges 
from $\sim-30$ to $+30$\,\AA, with an average of $-6$\,\AA\ relative to the {\it HST} spectra.

For bright zeroth-order contamination (e.g., field stars) falling
close to the target spectrum, our pipeline can identify those sources
(through the background spectrum) more easily than the old method, but it 
is generally inferior to
reductions using a template spectrum \citep[i.e.,][]{2016PASP..128c4501S}.  
We cross-check bright sources ($B < 18$\,mag) from the USNO-B1.0 catalog \citep{2003AJ....125..984M} 
and remove the affected pixels when producing the final spectrum.

\section{Results}
\label{sec:result}
\subsection{Comparison of data-reduction methods}
\label{sec:comp}

Figure~\ref{swift-data3} shows a comparison of spectra reduced using
the method described in Section~\ref{sec:reduction} to the same
spectra reduced using the default UVOTPY software and the
\citet{2016PASP..128c4501S} decontamination method.  The spectrum
produced through our method agrees well with that from
\citet{2016PASP..128c4501S}. However, the spectrum reduced by the
default pipeline is clearly offset in flux from the other two spectra.
This is likely caused by incorrect host-galaxy subtraction (see
Figures~\ref{swift-data2} and \ref{bg-compare}). The default UVOTPY
method underestimates the underlying background, resulting in a
spectrum with higher flux at all wavelengths (i.e., with additional
light from the host galaxy). Our new method removes the host
contamination and performs the background subtraction more correctly.

\begin{figure*}
	\centering
		\includegraphics[scale=0.53]{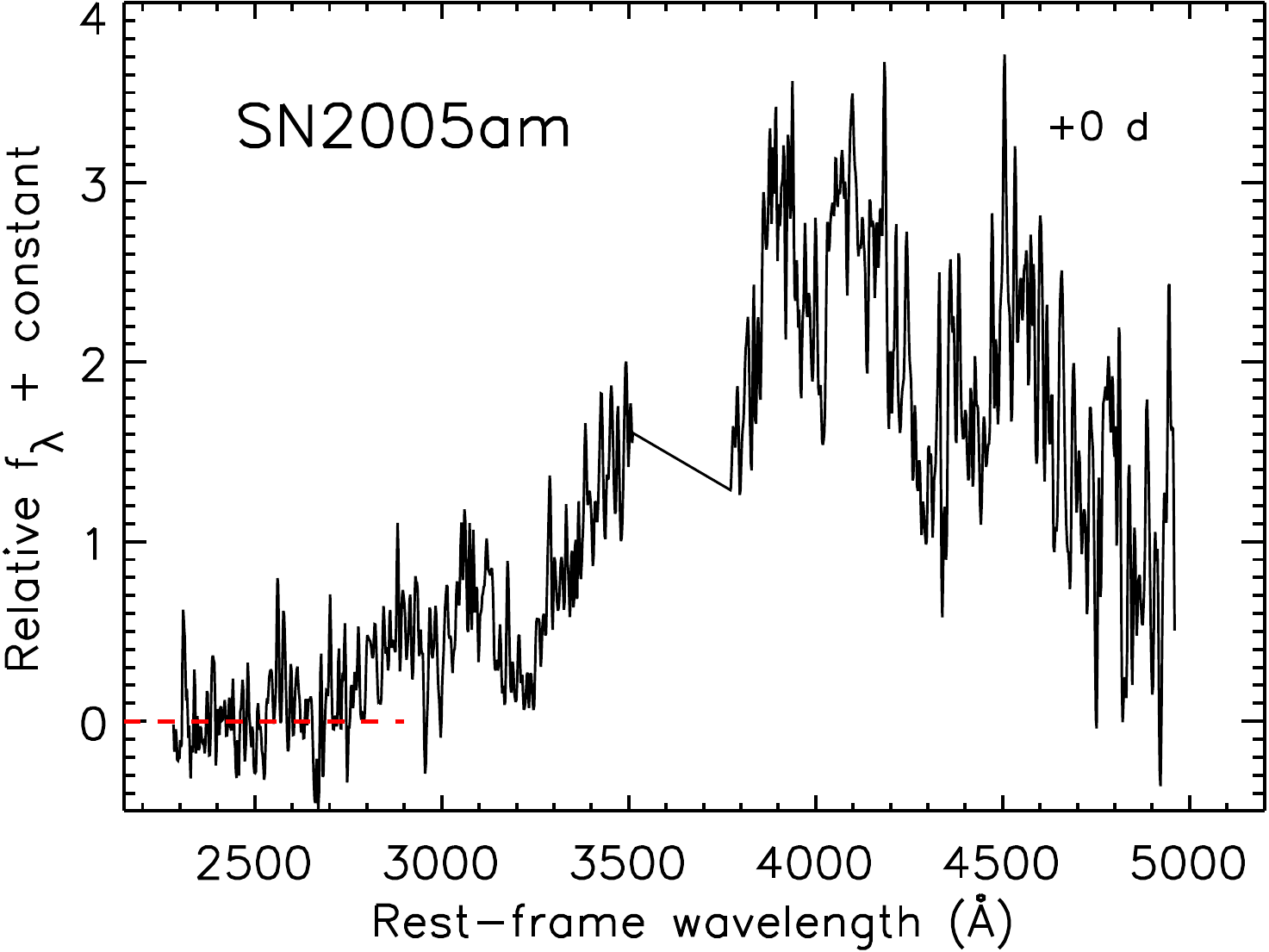}
		\hspace{0.25cm}
		\includegraphics[scale=0.53]{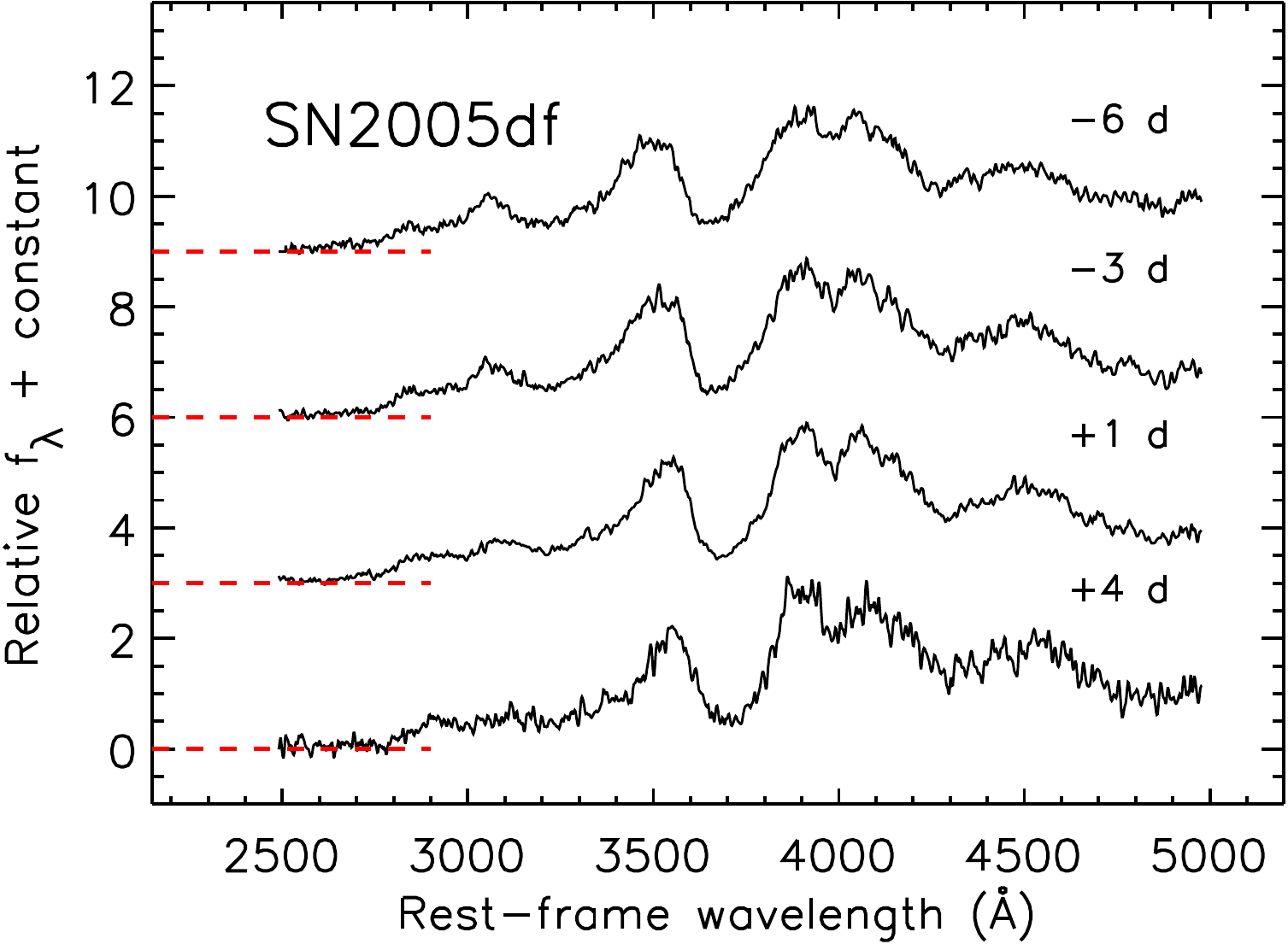}\\
		\vspace{0.25cm}
		\includegraphics[scale=0.53]{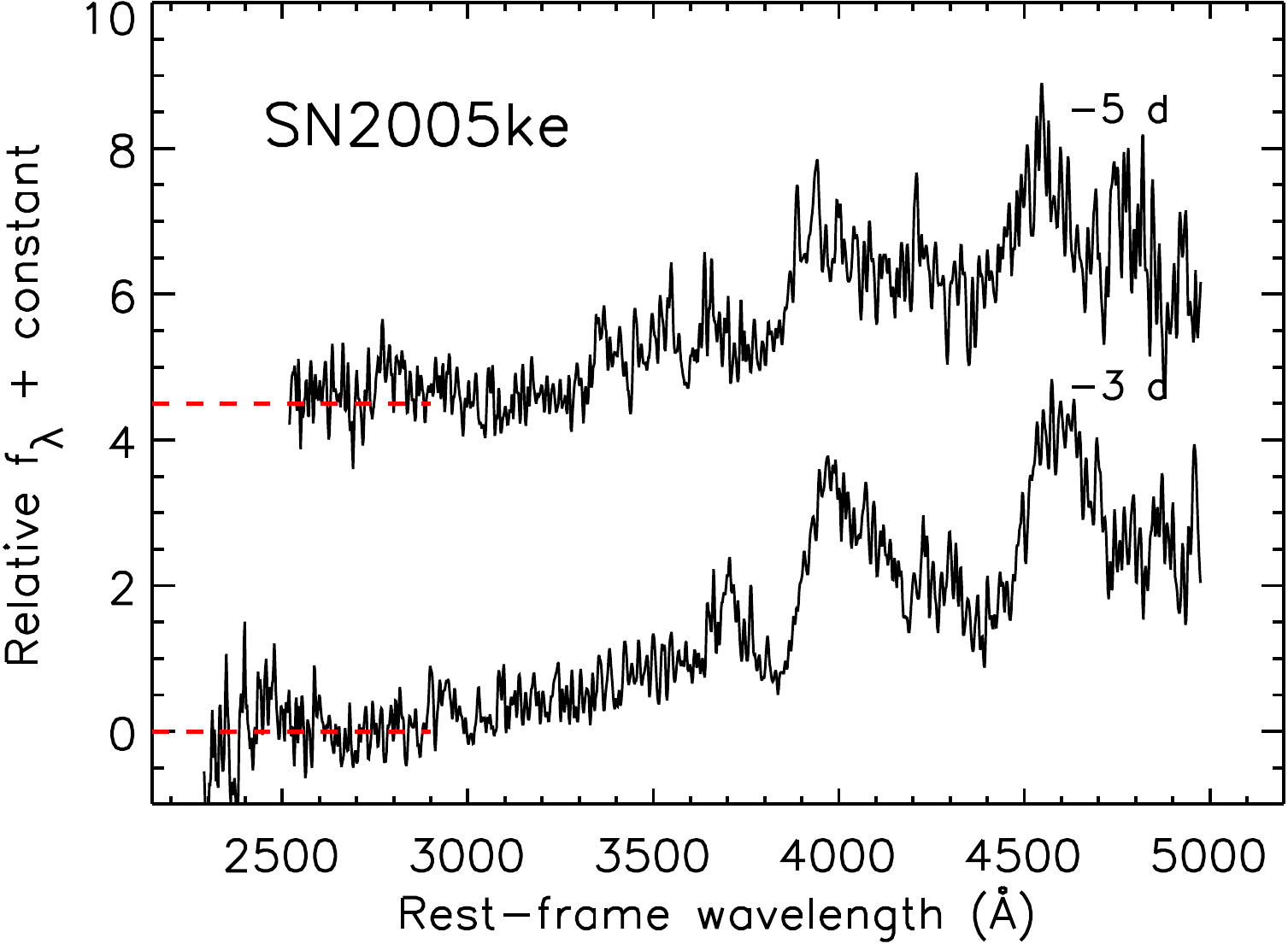}
		\hspace{0.25cm}
		\includegraphics[scale=0.53]{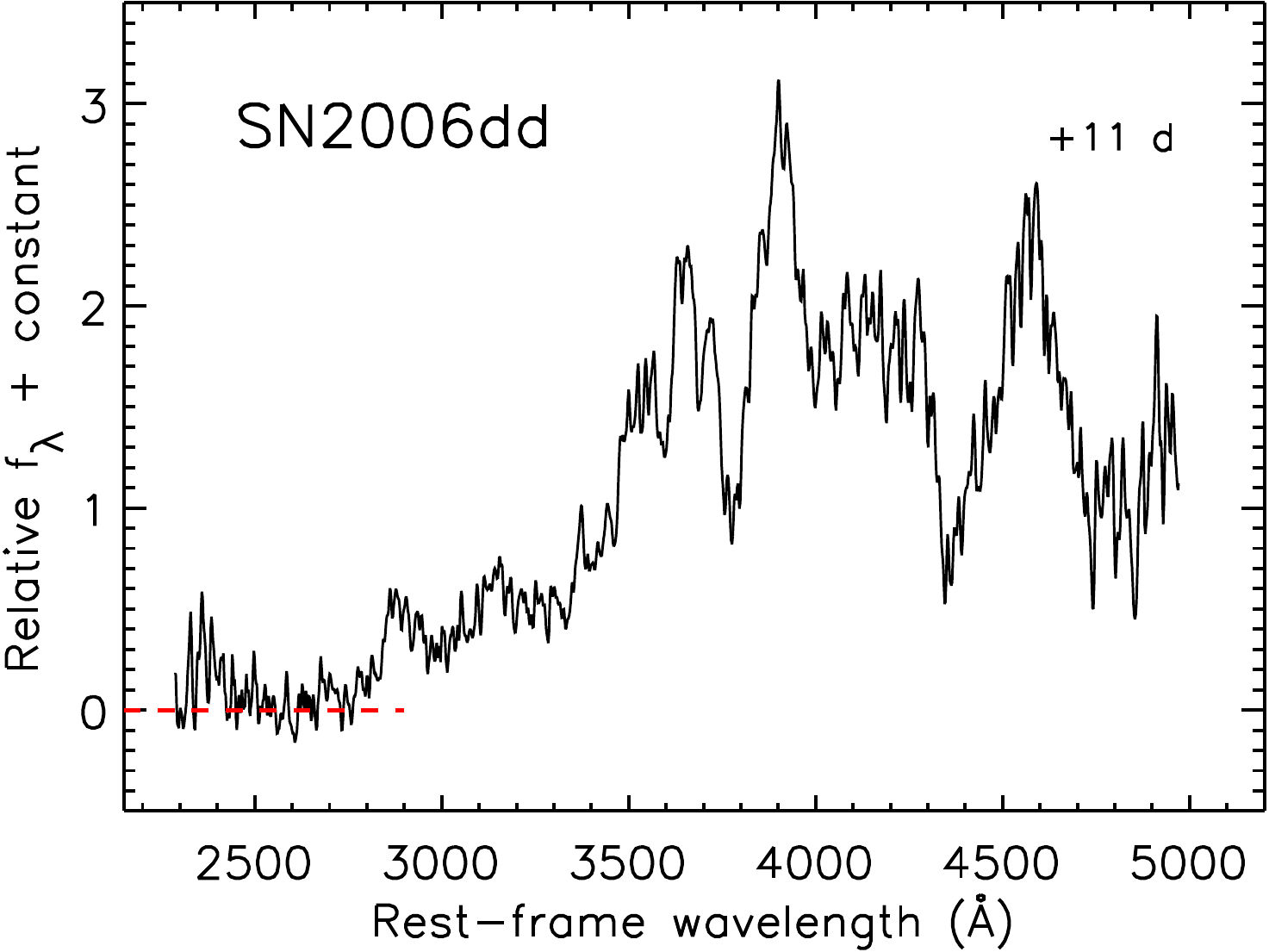}\\
		\vspace{0.25cm}
		\includegraphics[scale=0.8]{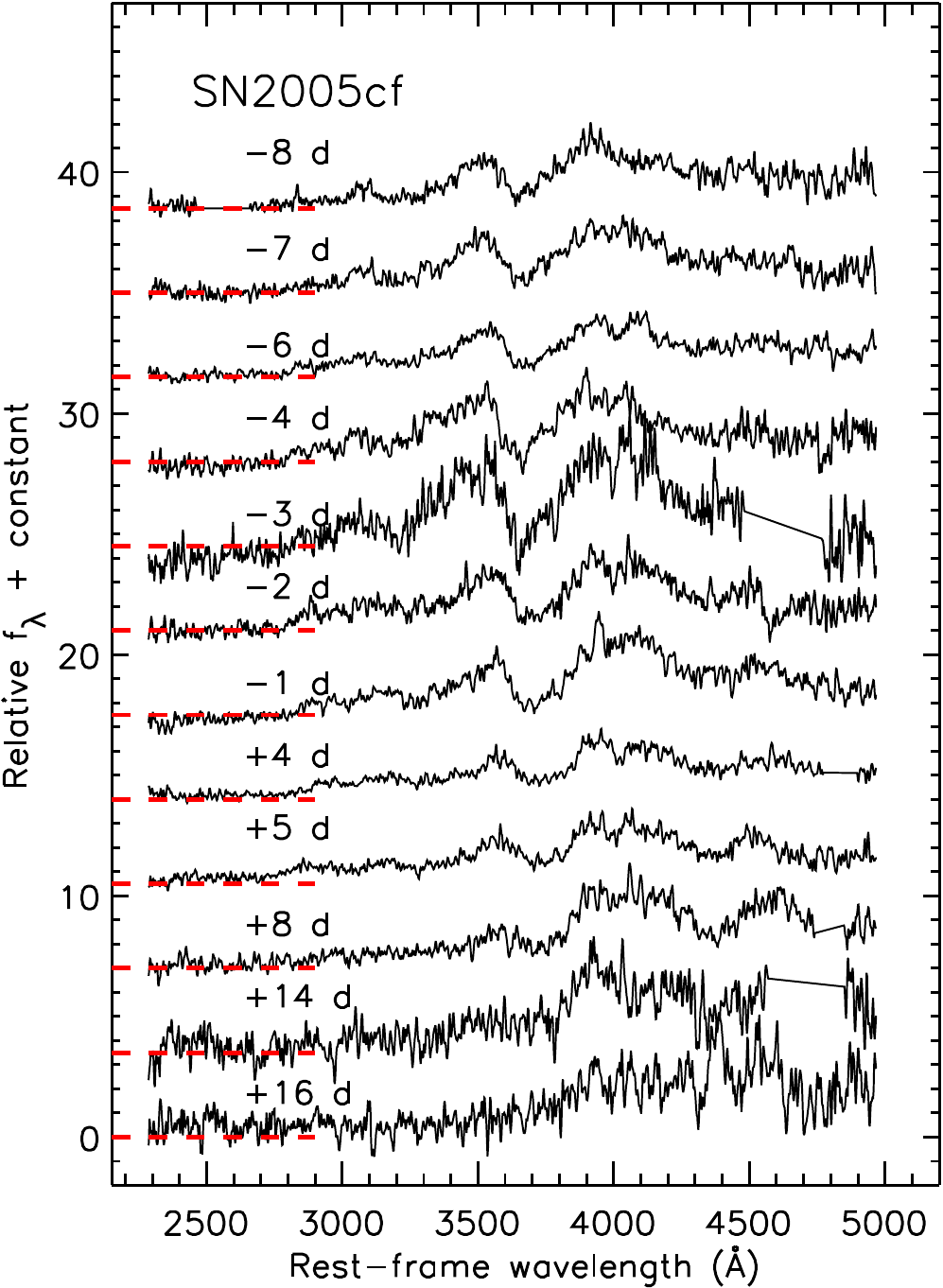}
		\hspace{0.25cm}
		\includegraphics[scale=0.8]{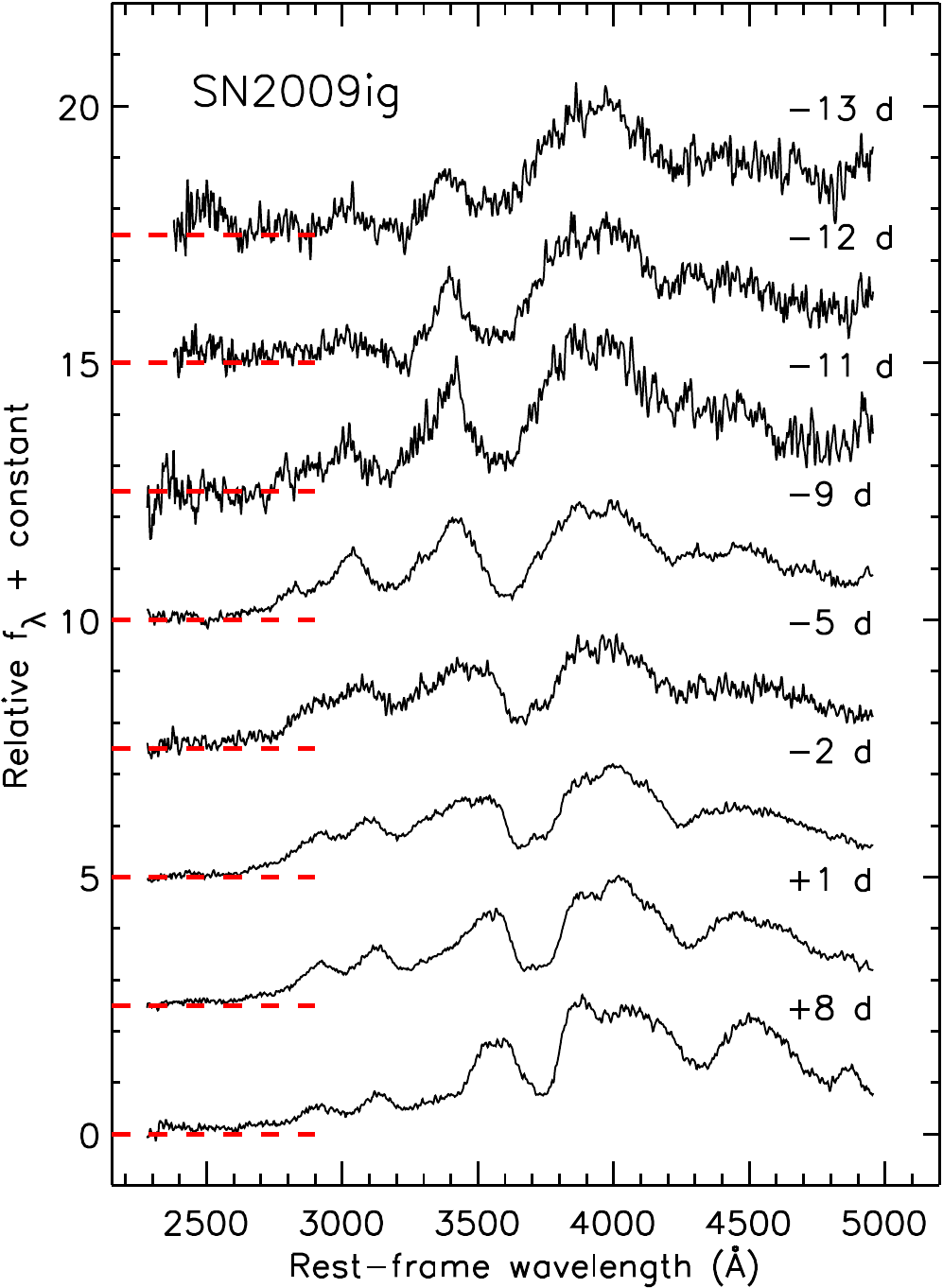}
		       \caption{{\it Swift} UVOT grism observations of SNe~Ia. The red dashed line marks the level of zero flux.
              }
      \label{uvspec1}
\end{figure*}

\begin{figure*}
	\centering
	    \includegraphics[scale=0.53]{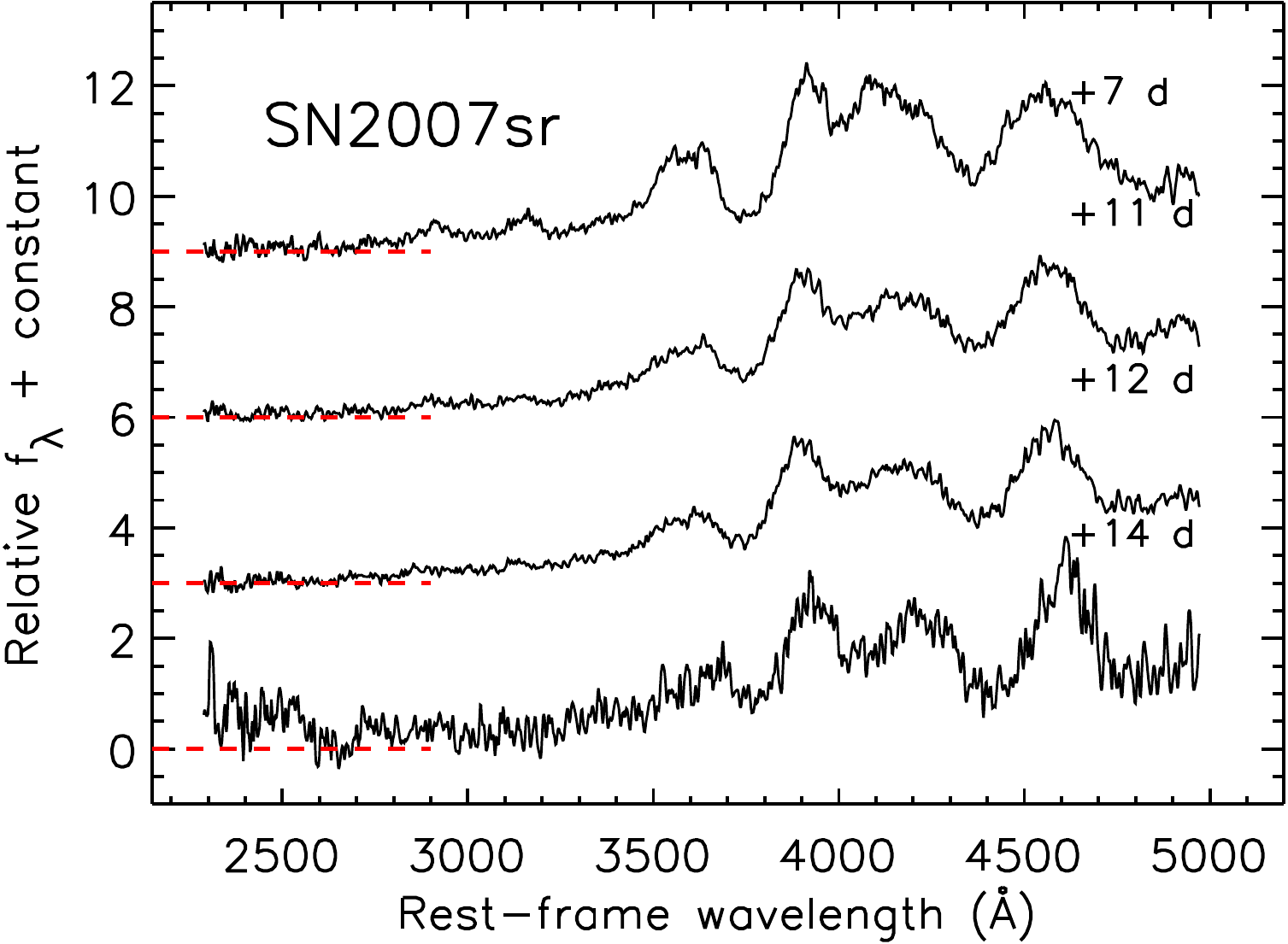}
		\hspace{0.25cm}
		\includegraphics[scale=0.53]{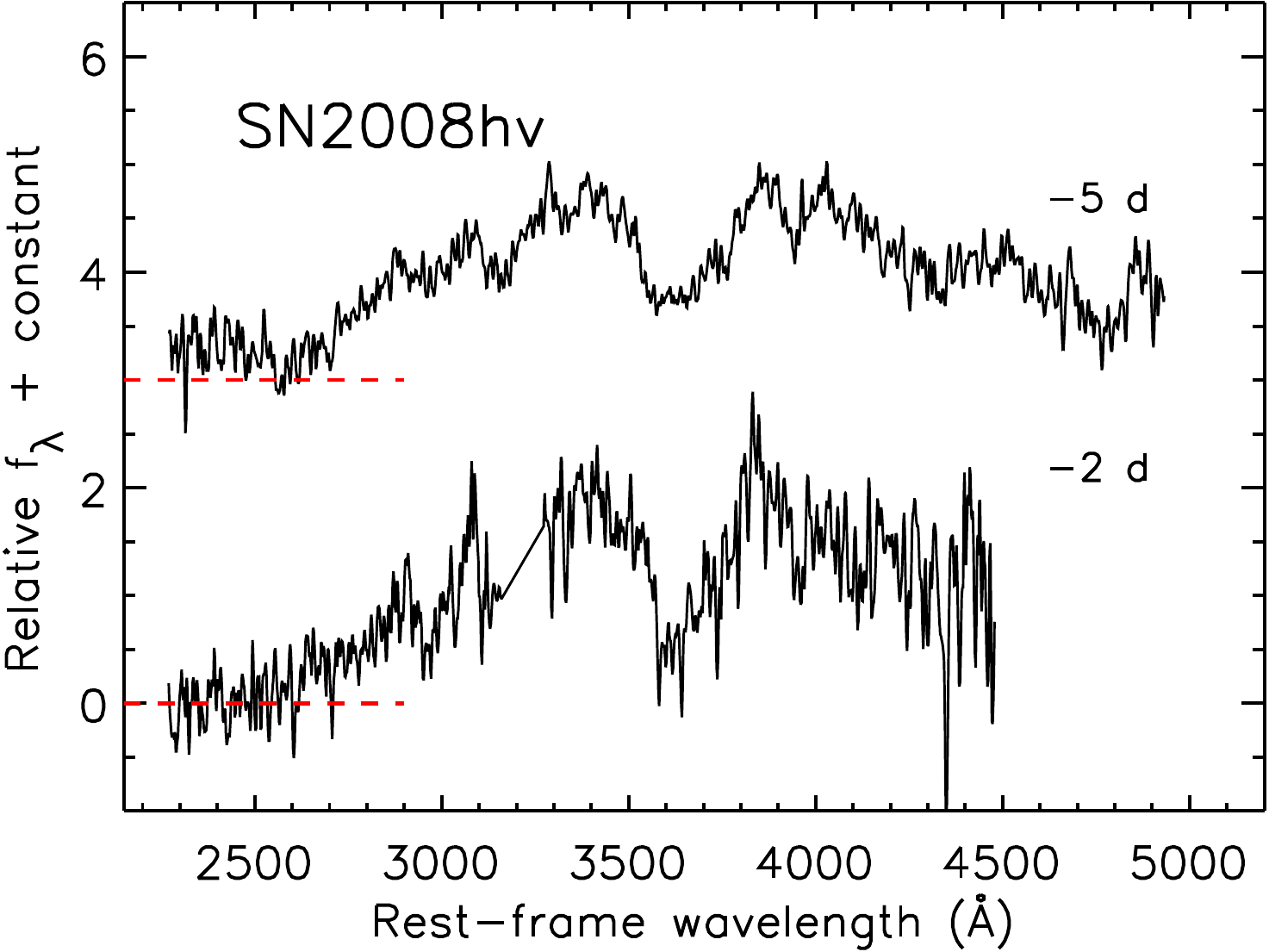}\\
		\vspace{0.25cm}
		\includegraphics[scale=0.53]{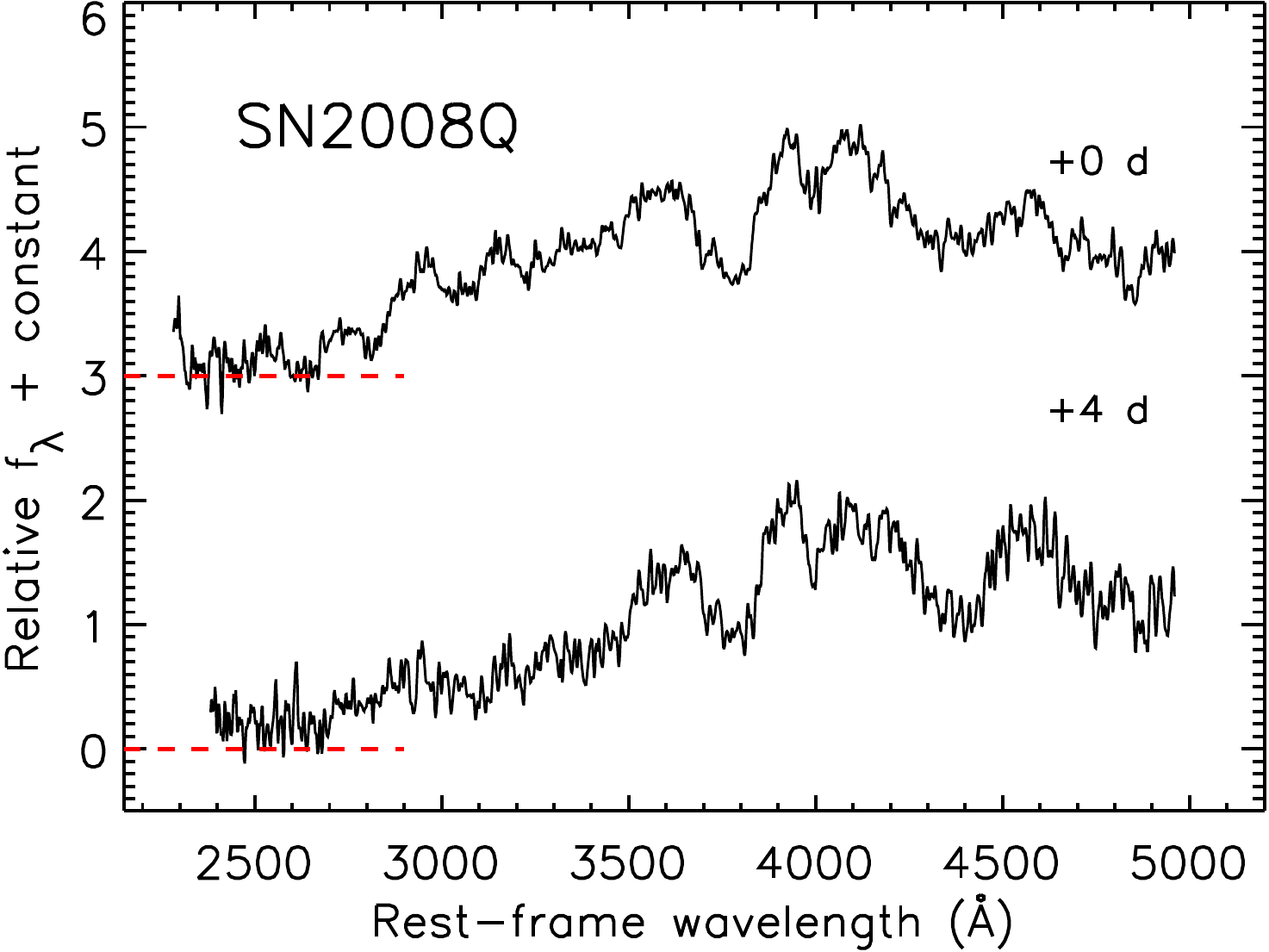}
		\hspace{0.25cm}
		\includegraphics[scale=0.53]{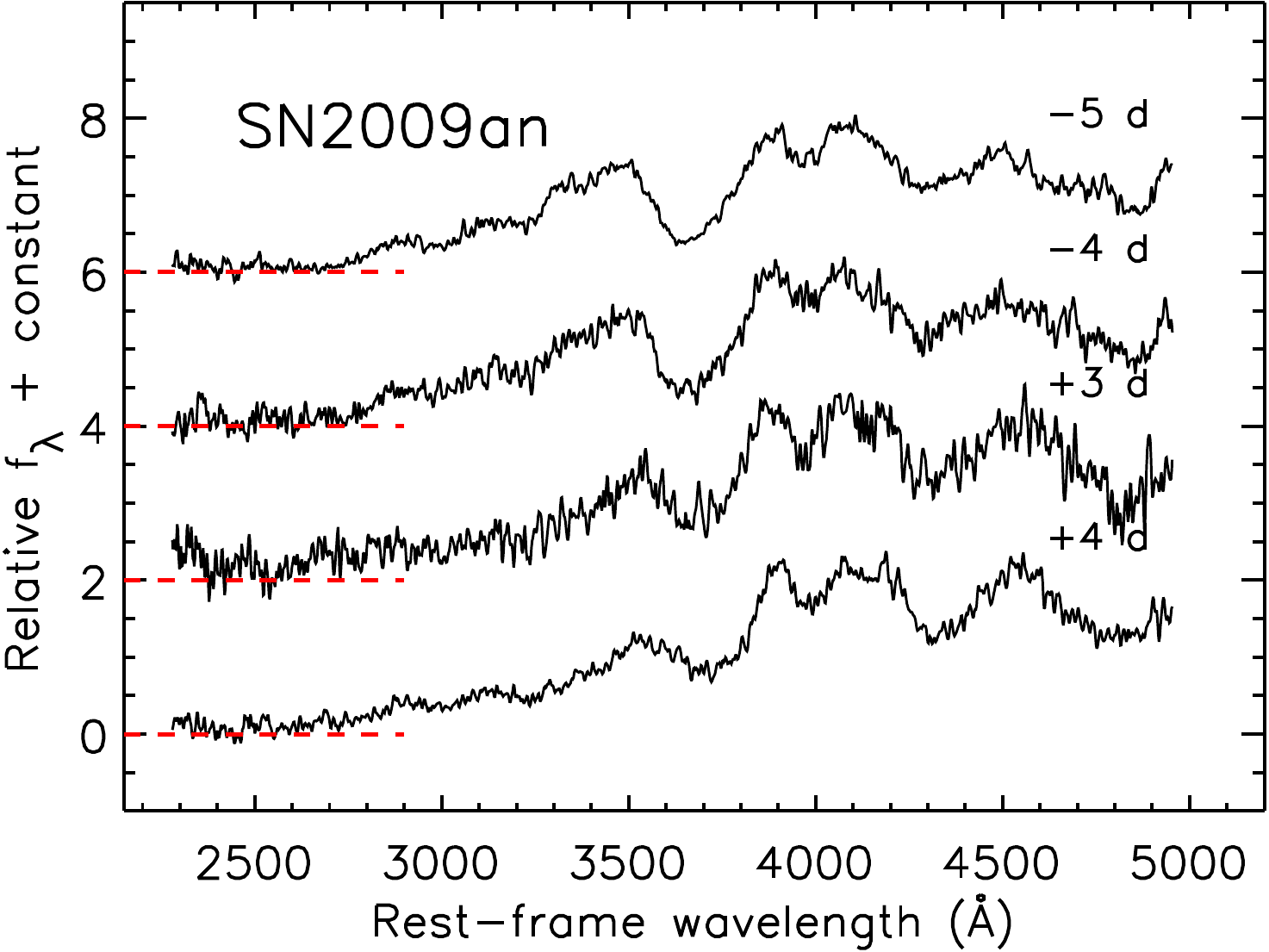}\\
        \vspace{0.25cm}
	    \includegraphics[scale=0.8]{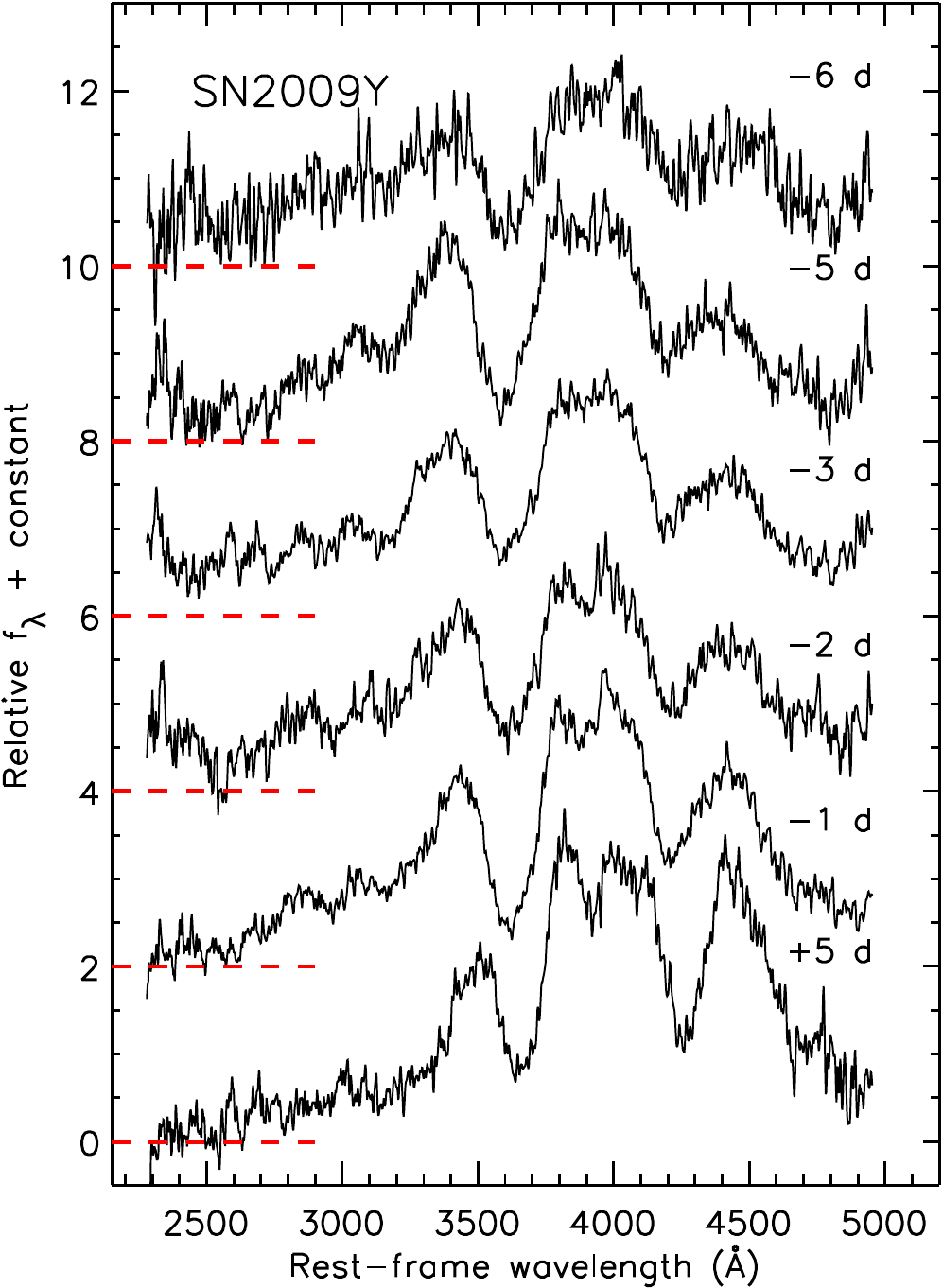}
	    \hspace{0.25cm}
	    \includegraphics[scale=0.8]{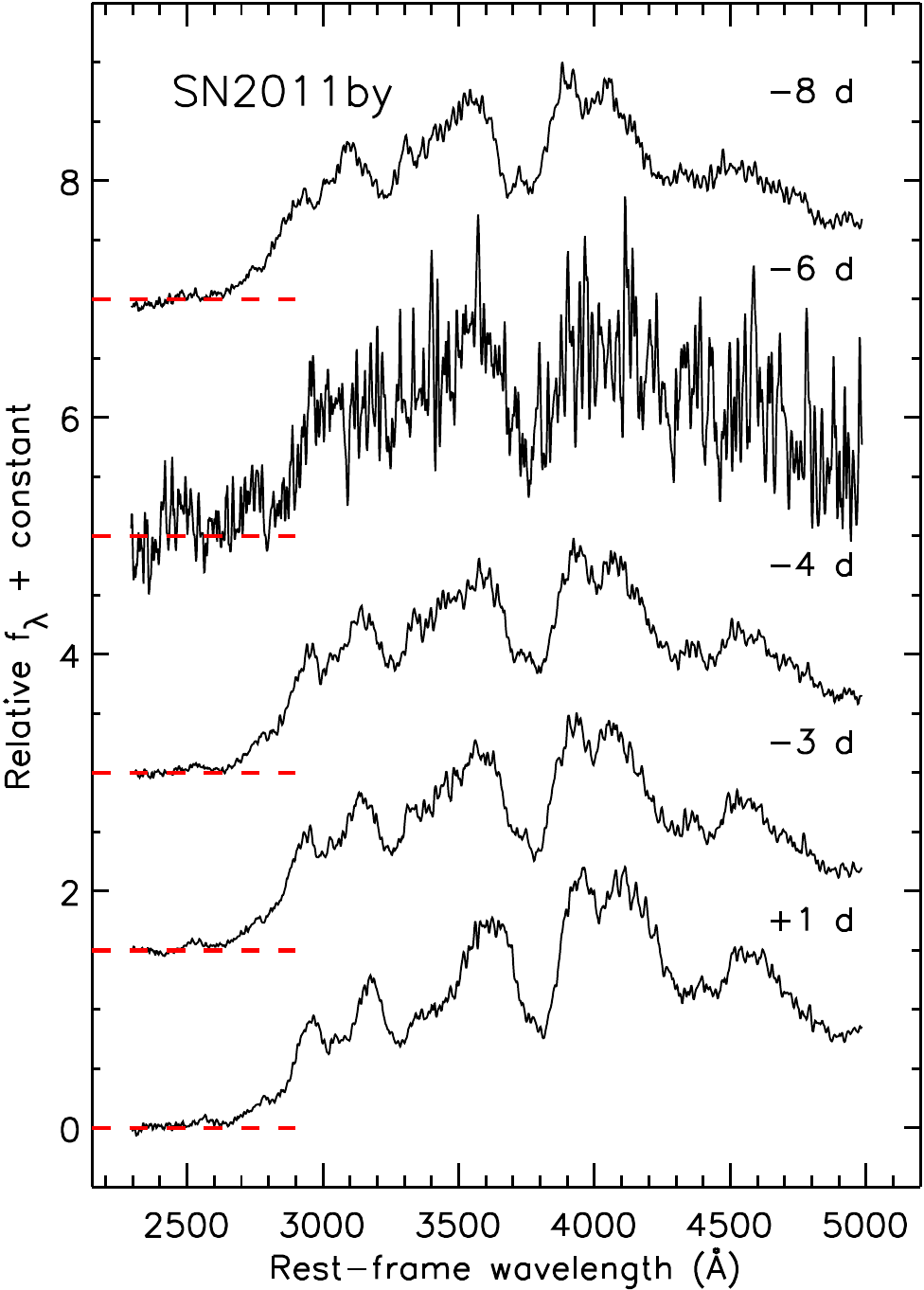}
       \caption{{\it Swift} UVOT grism observations of SNe~Ia (continued).
               }
        \label{uvspec2}
\end{figure*}

\begin{figure*}
	\centering
	    \includegraphics[scale=0.53]{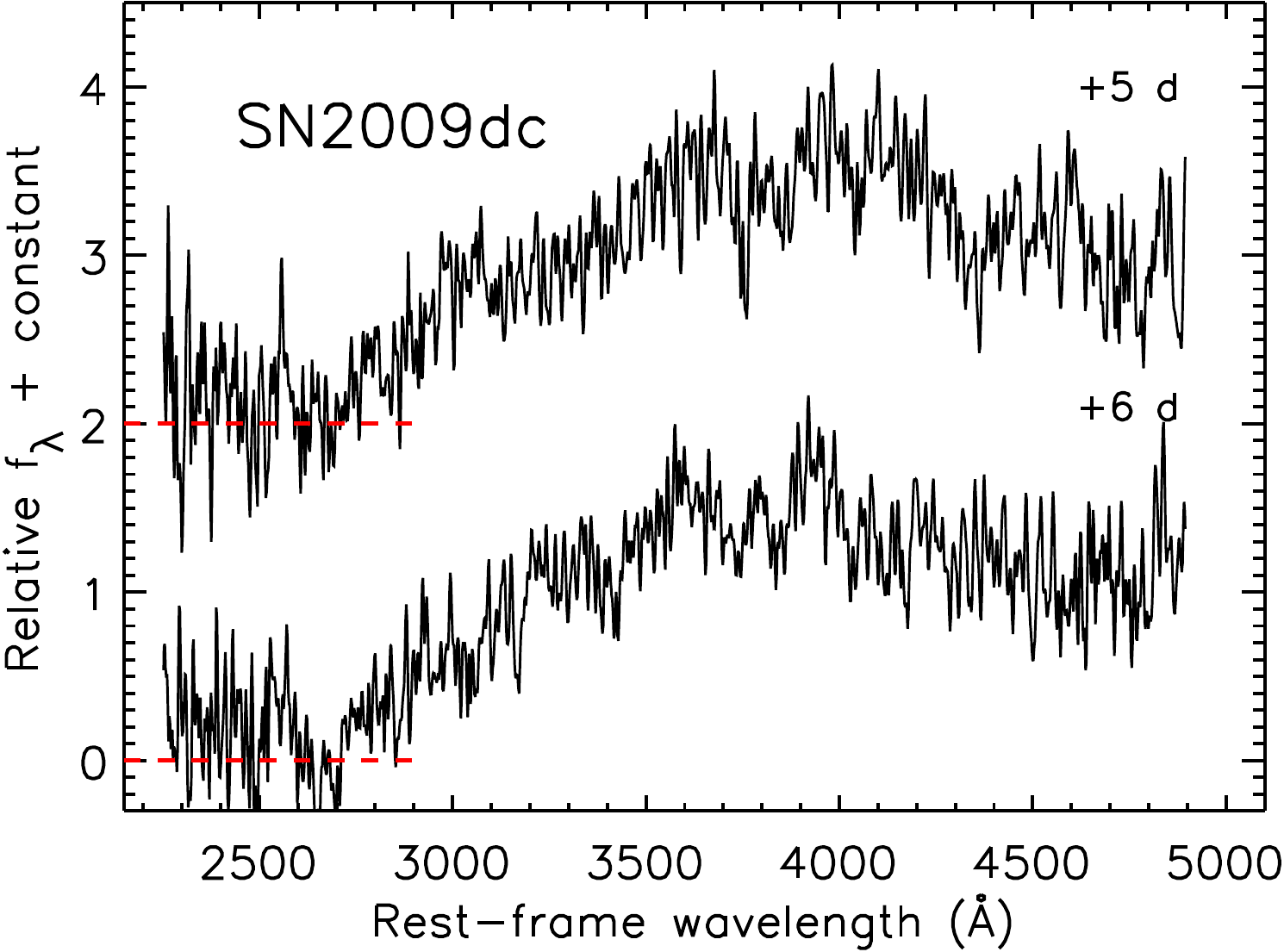}
	    \hspace{0.25cm}
	    \includegraphics[scale=0.53]{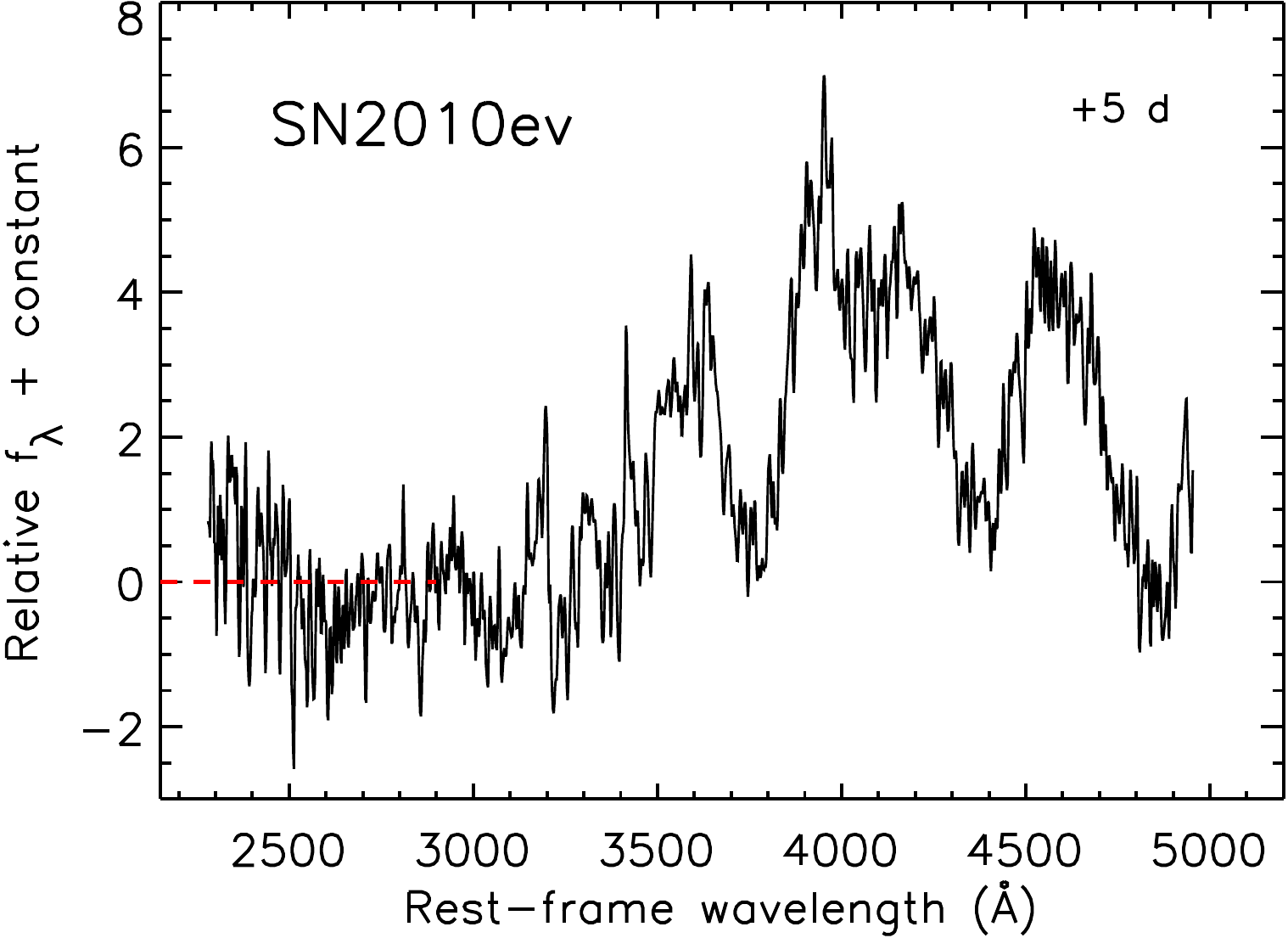}\\
	    \vspace{0.25cm}
	    \includegraphics[scale=0.53]{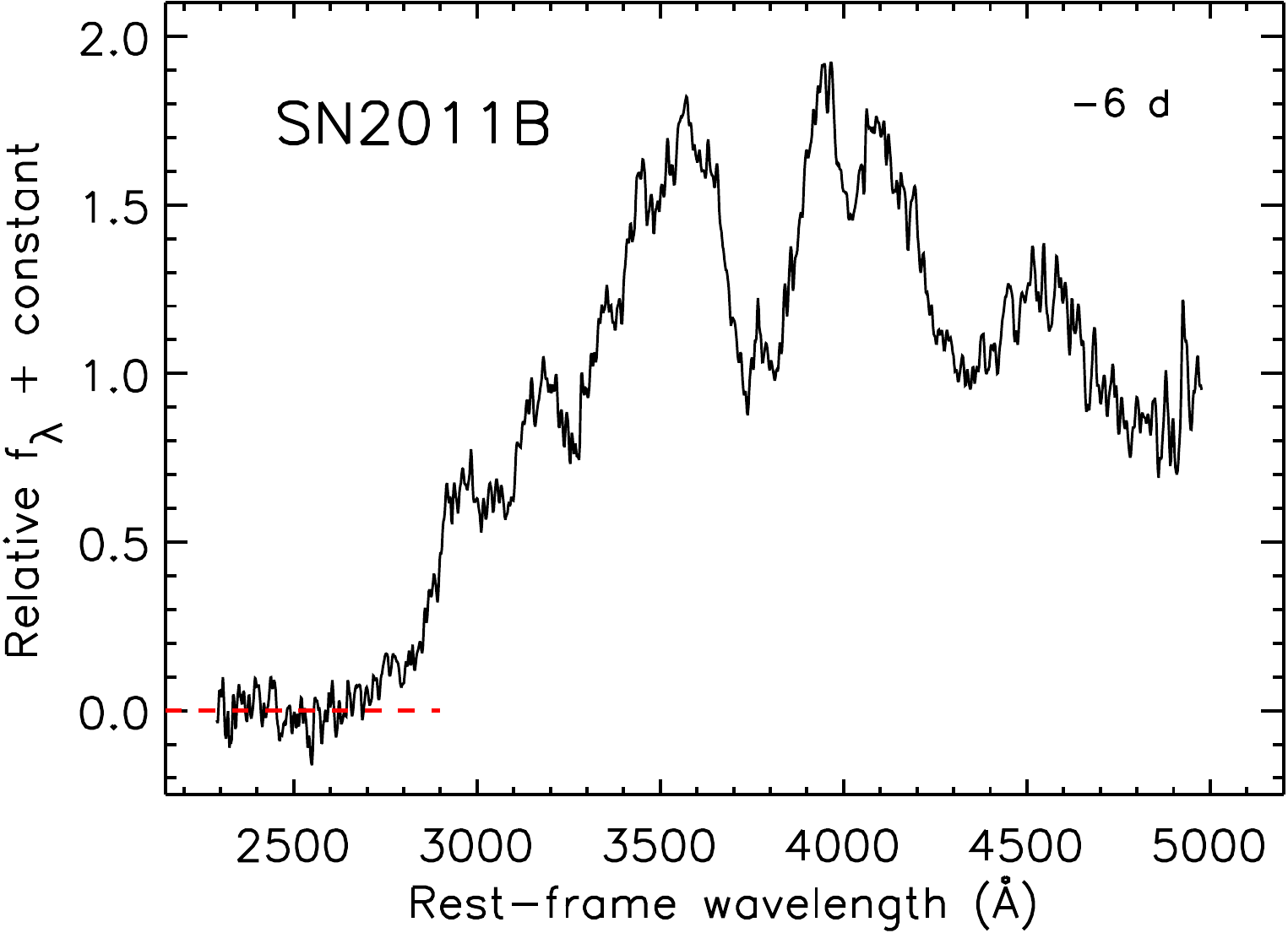}
		\hspace{0.25cm}
		\includegraphics[scale=0.53]{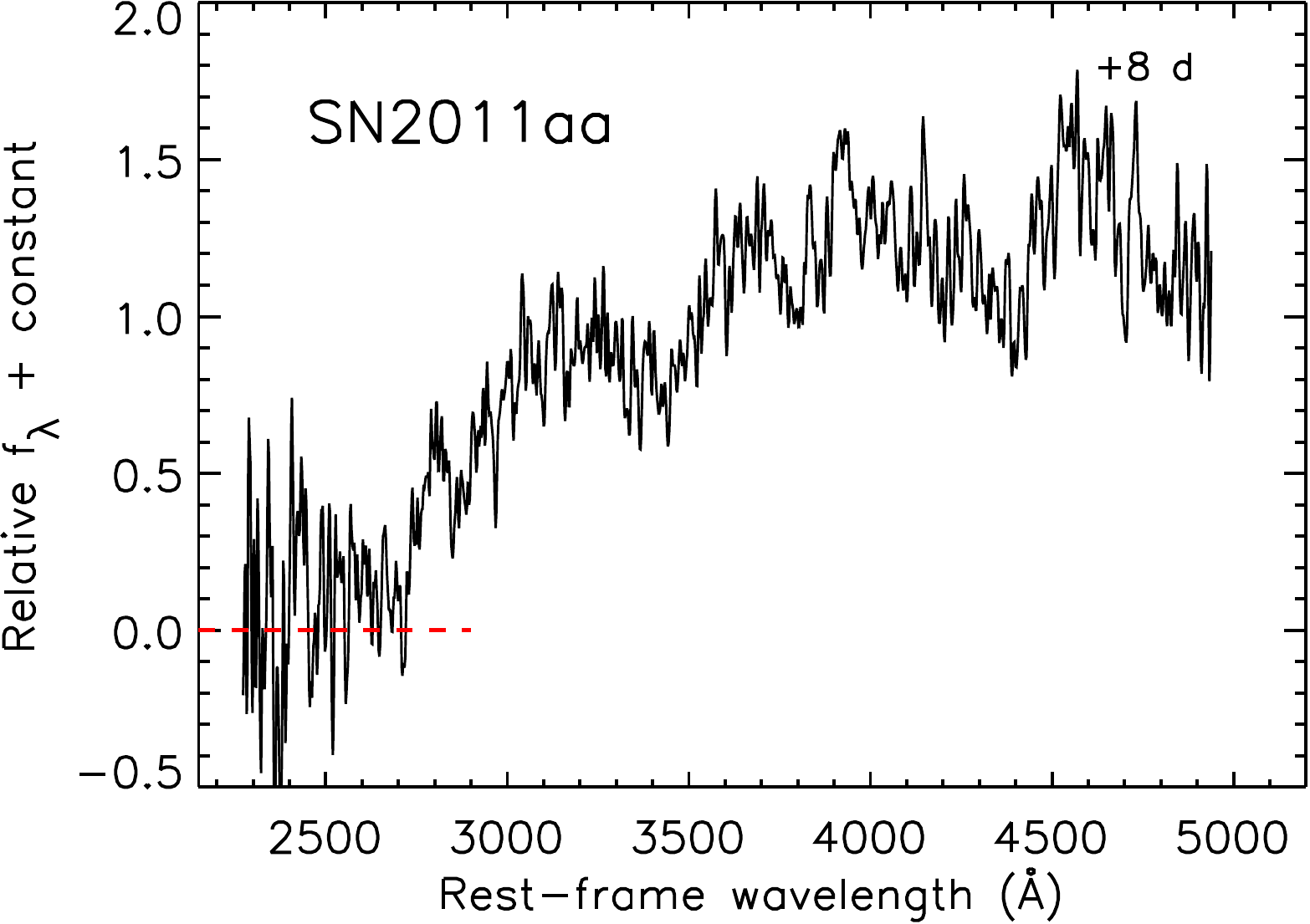}\\
		\vspace{0.25cm}
	    \includegraphics[scale=0.8]{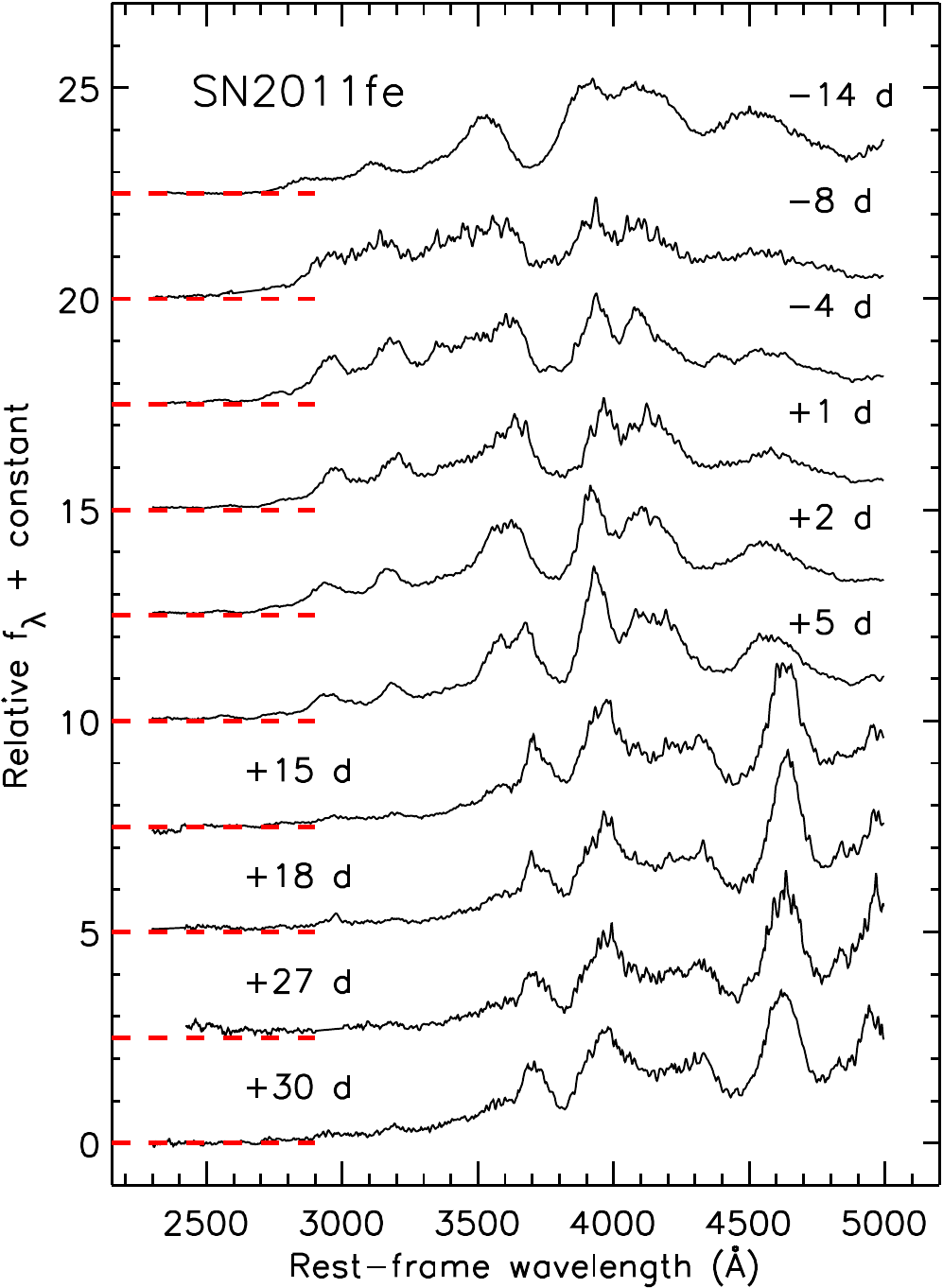}
		\hspace{0.25cm}
		\includegraphics[scale=0.8]{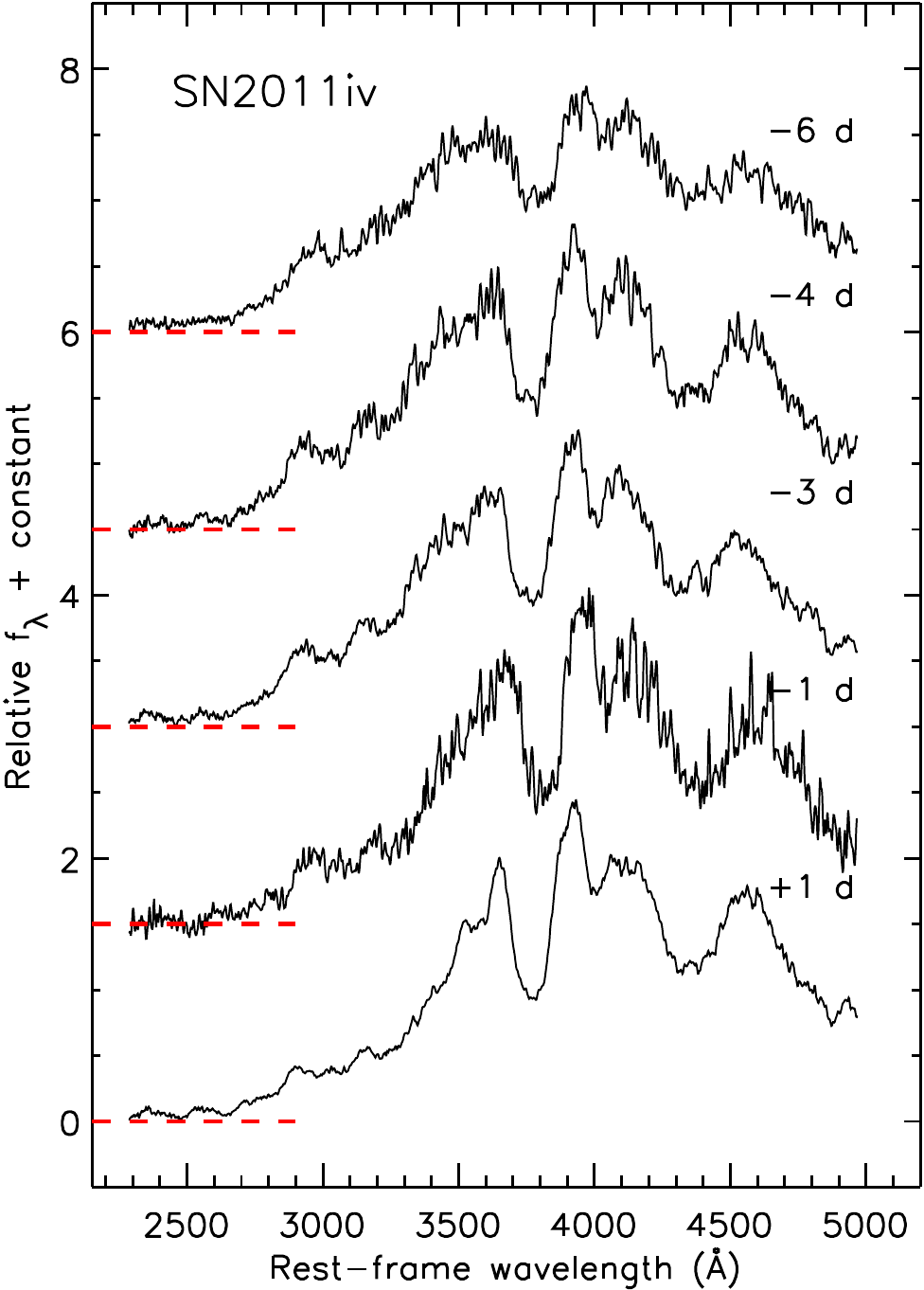}
       \label{uvspec3}
        \caption{{\it Swift} UVOT grism observations of SNe~Ia (continued).
               }
\end{figure*}

\begin{figure*}
	\centering
		\includegraphics[scale=0.53]{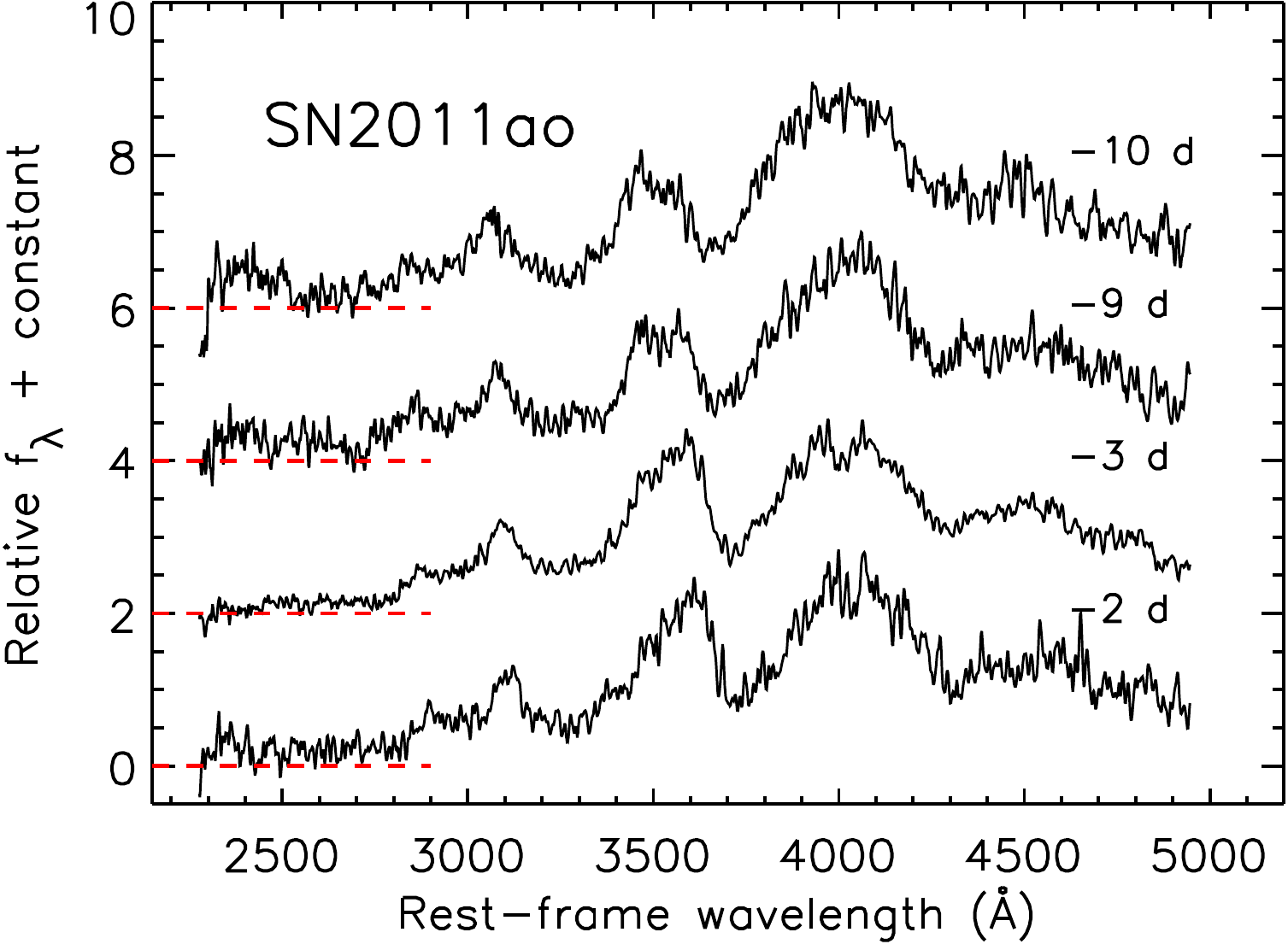}
		\hspace{0.25cm}
		\includegraphics[scale=0.53]{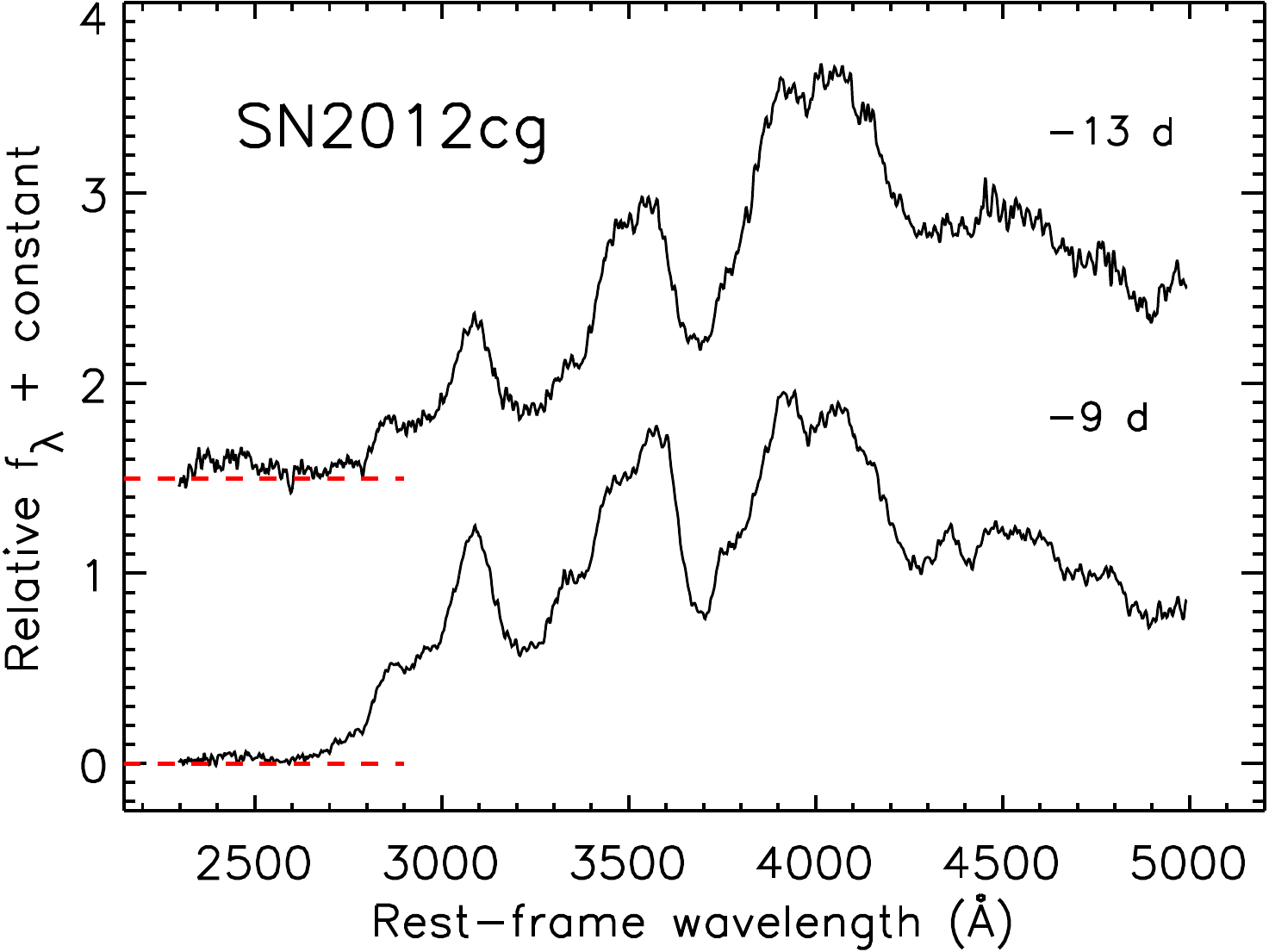}\\
		\vspace{0.25cm}
		\includegraphics[scale=0.53]{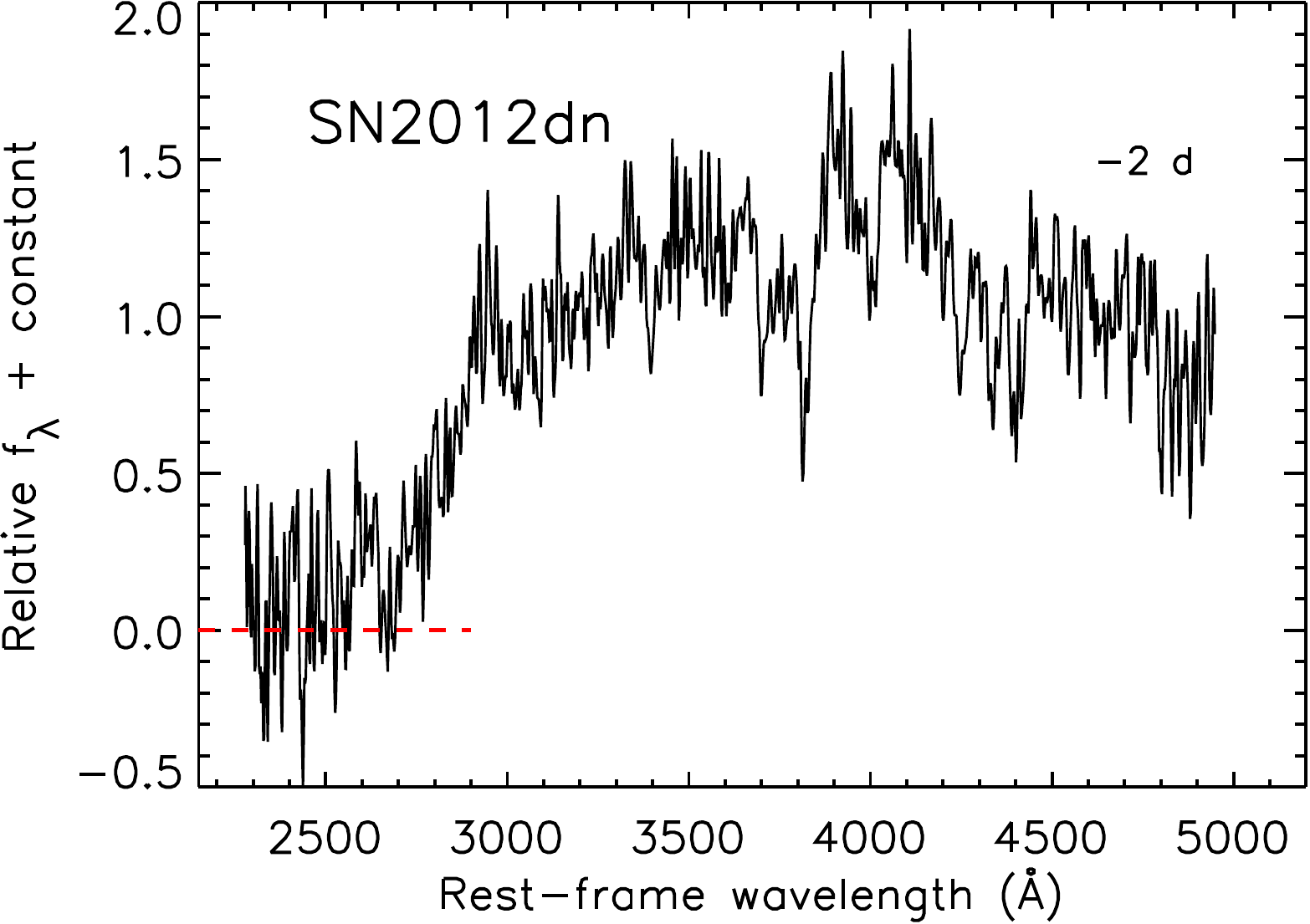}
		\hspace{0.25cm}
		\includegraphics[scale=0.53]{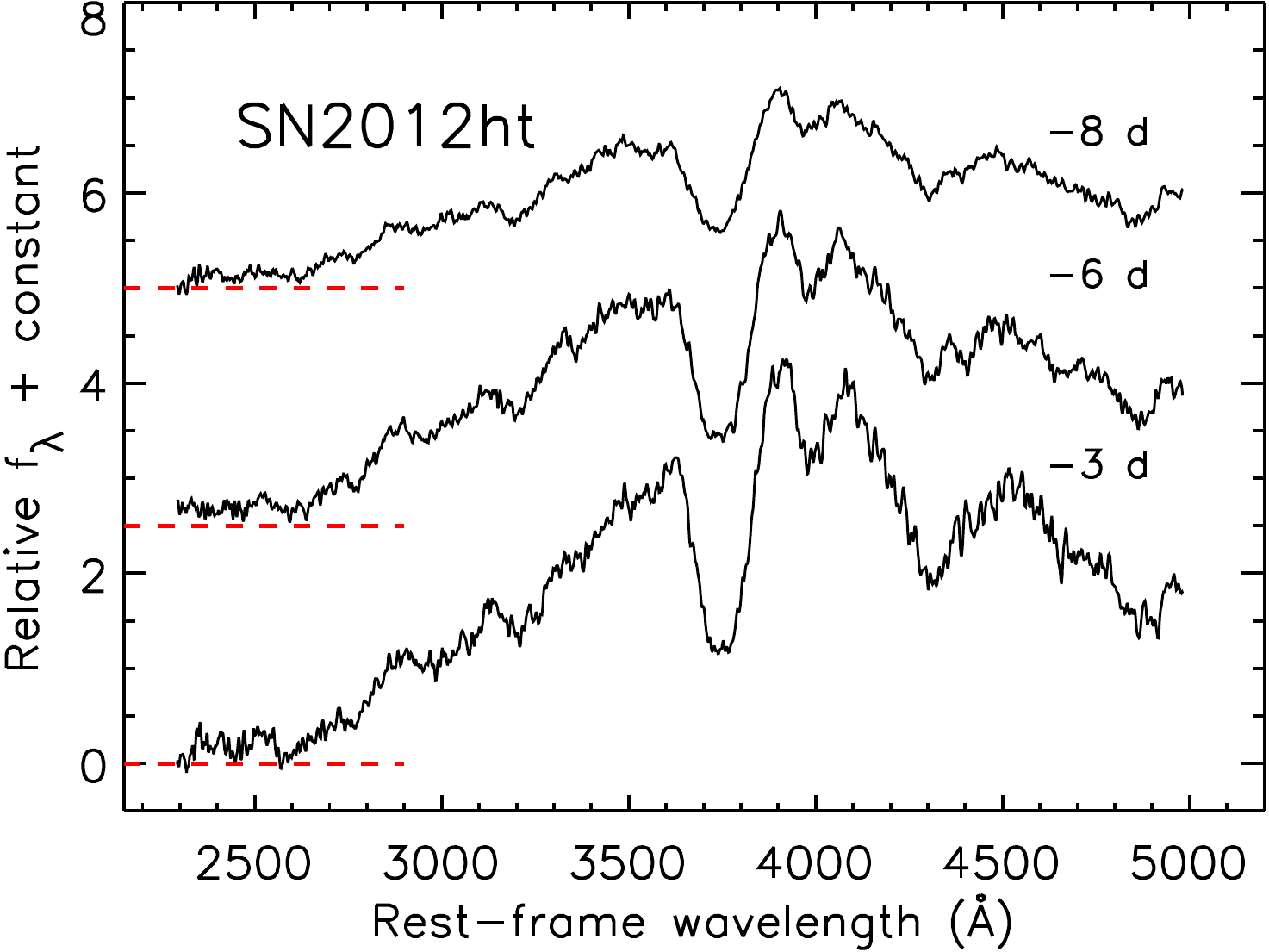}\\
		\vspace{0.25cm}
	    \includegraphics[scale=0.8]{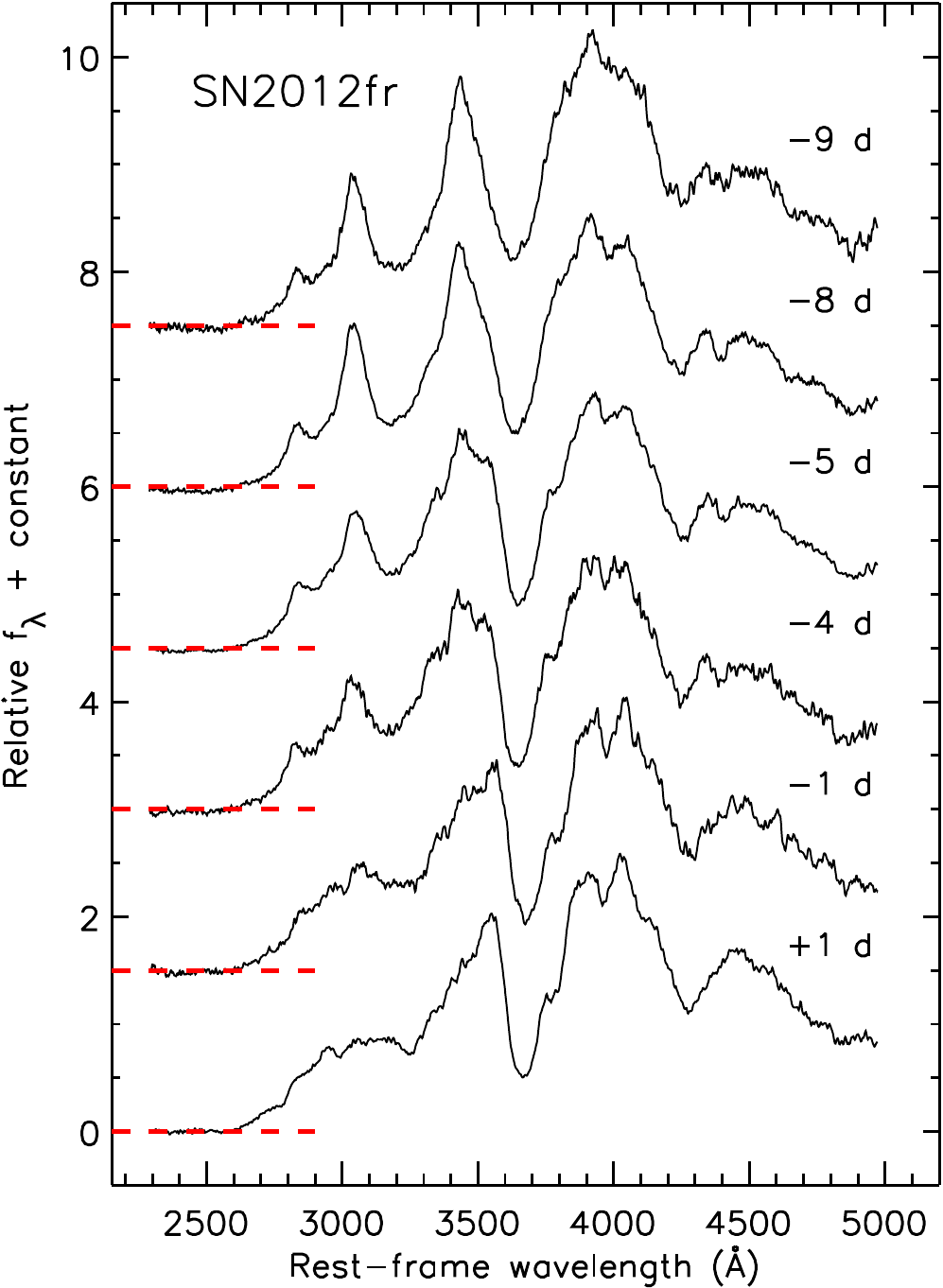}
	    \hspace{0.25cm}
	    \includegraphics[scale=0.8]{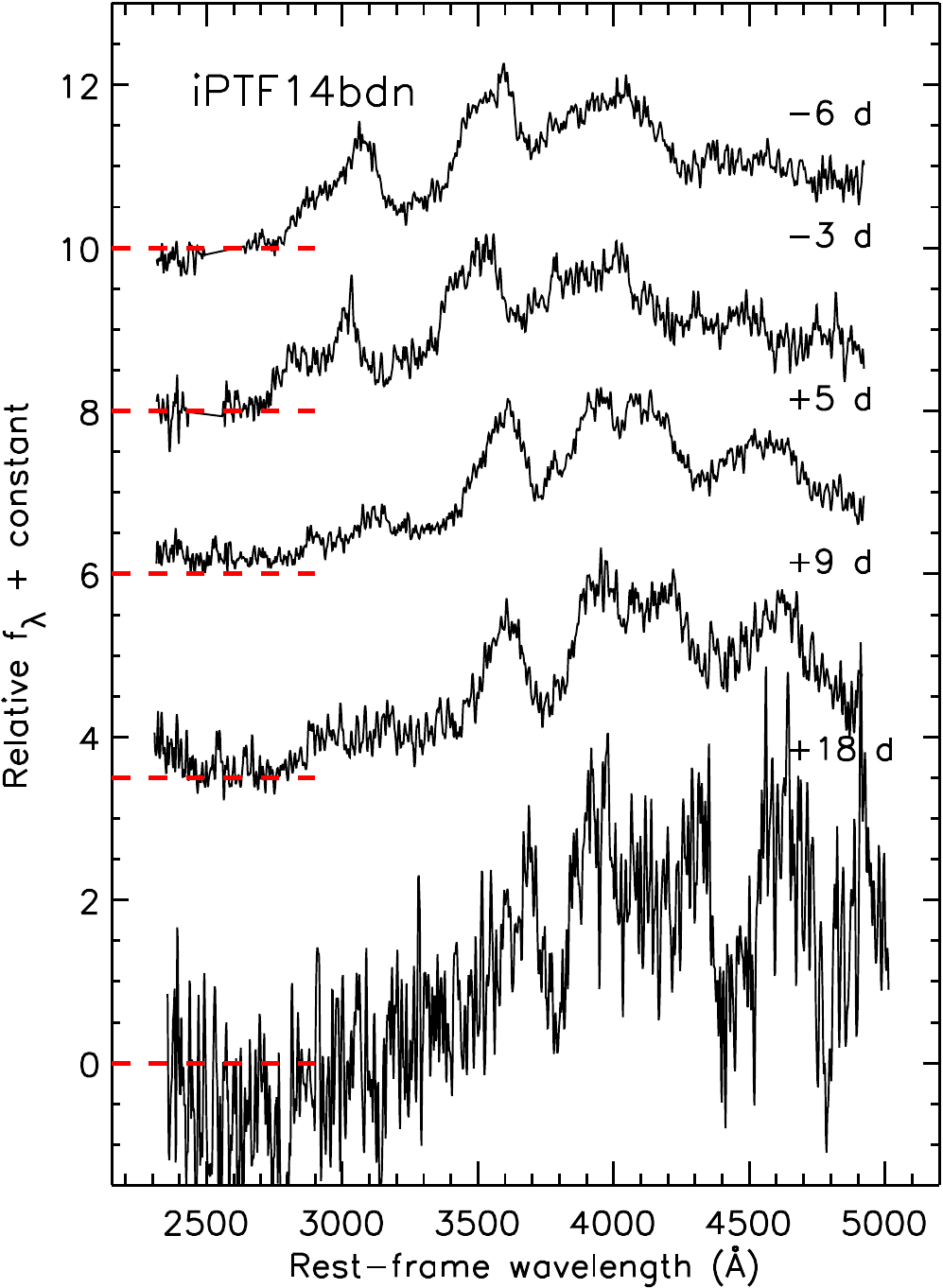}
        \caption{{\it Swift} UVOT grism observations of SNe~Ia (continued).
               }
        \label{uvspec4}
\end{figure*}

\begin{figure*}
	\centering
	    \includegraphics[scale=0.53]{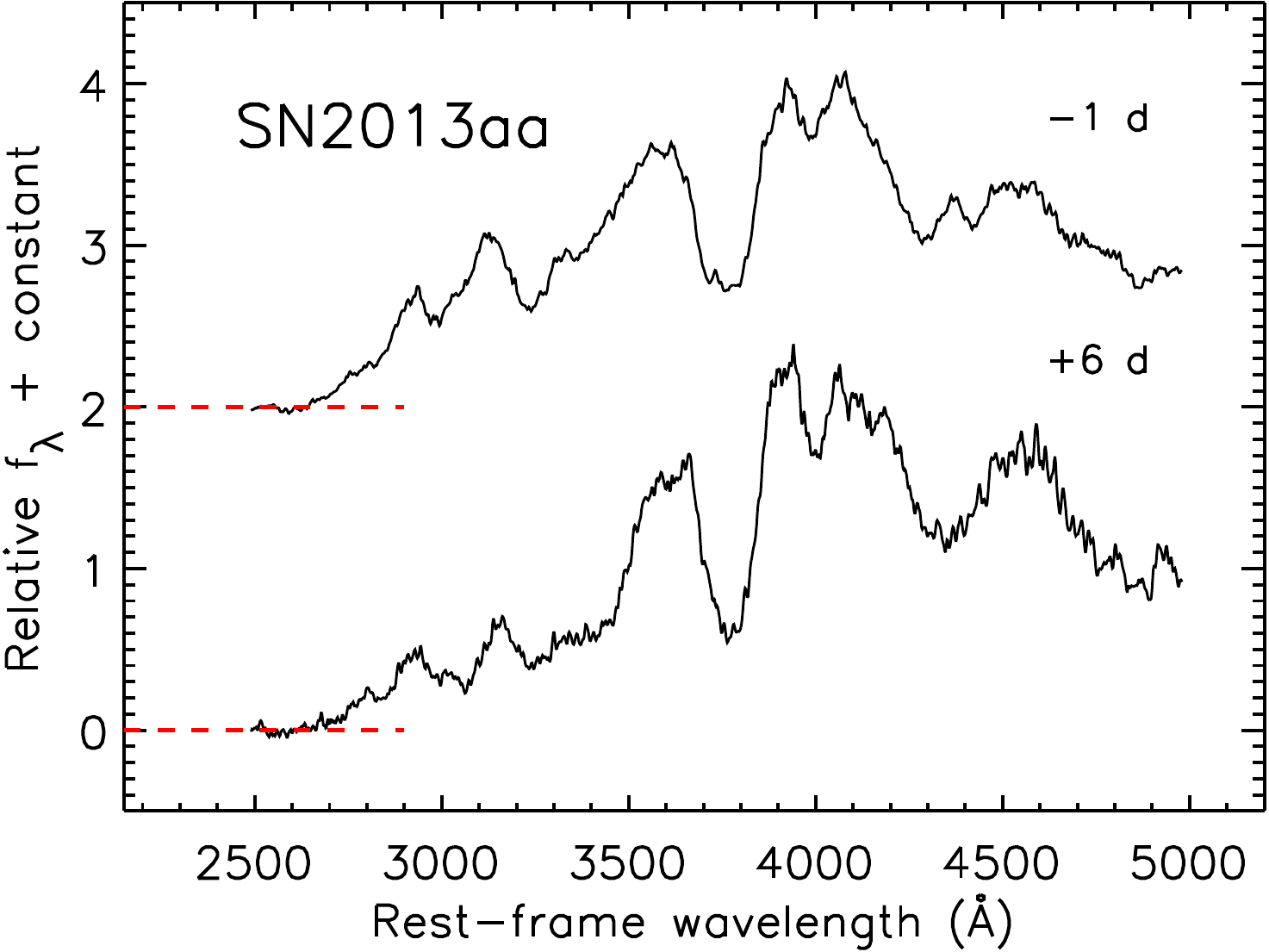}
		\hspace{0.25cm}
		\includegraphics[scale=0.53]{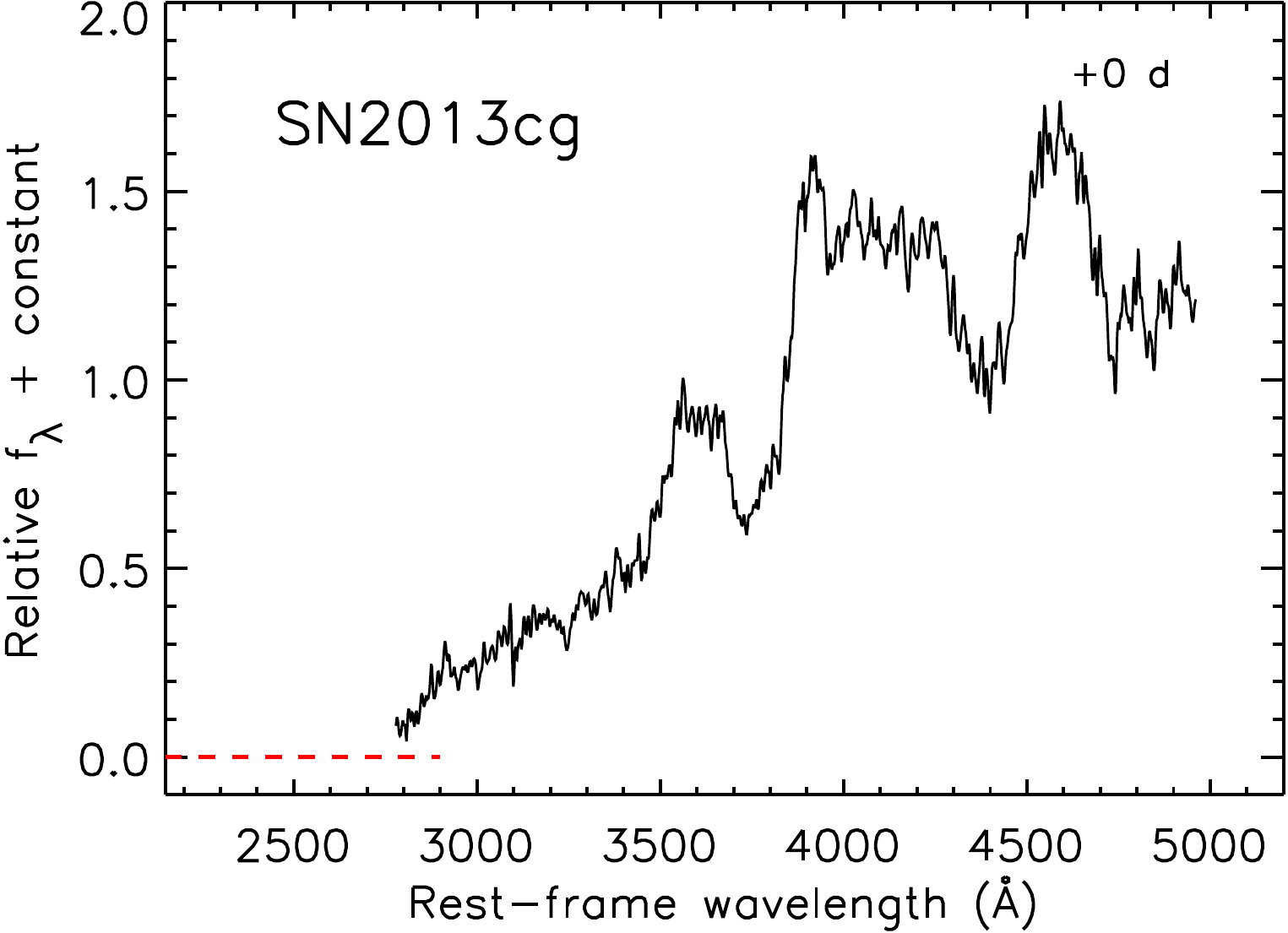}\\
	    \vspace{0.25cm}
	    \includegraphics[scale=0.53]{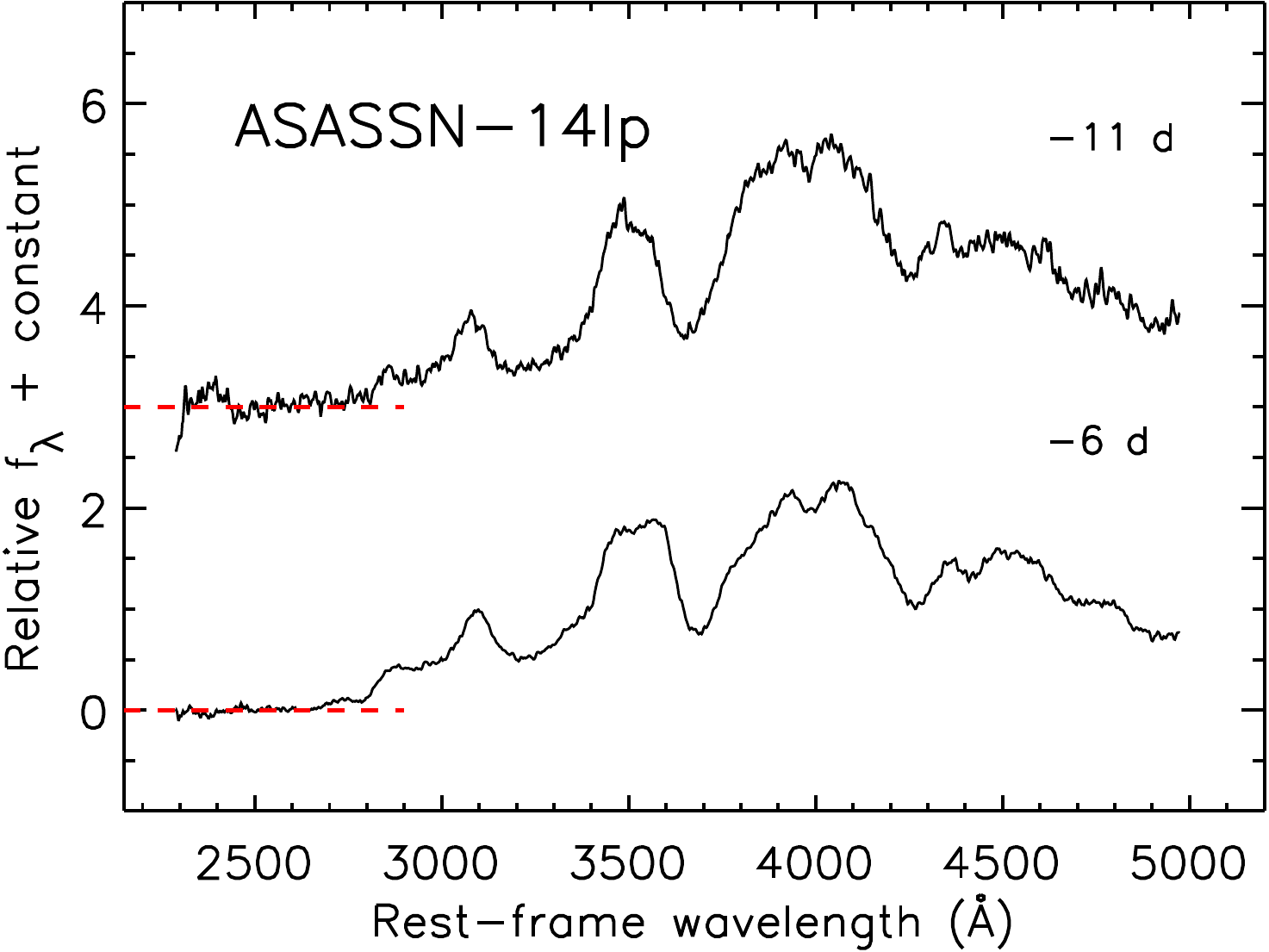}
	    \hspace{0.25cm}
	    \includegraphics[scale=0.53]{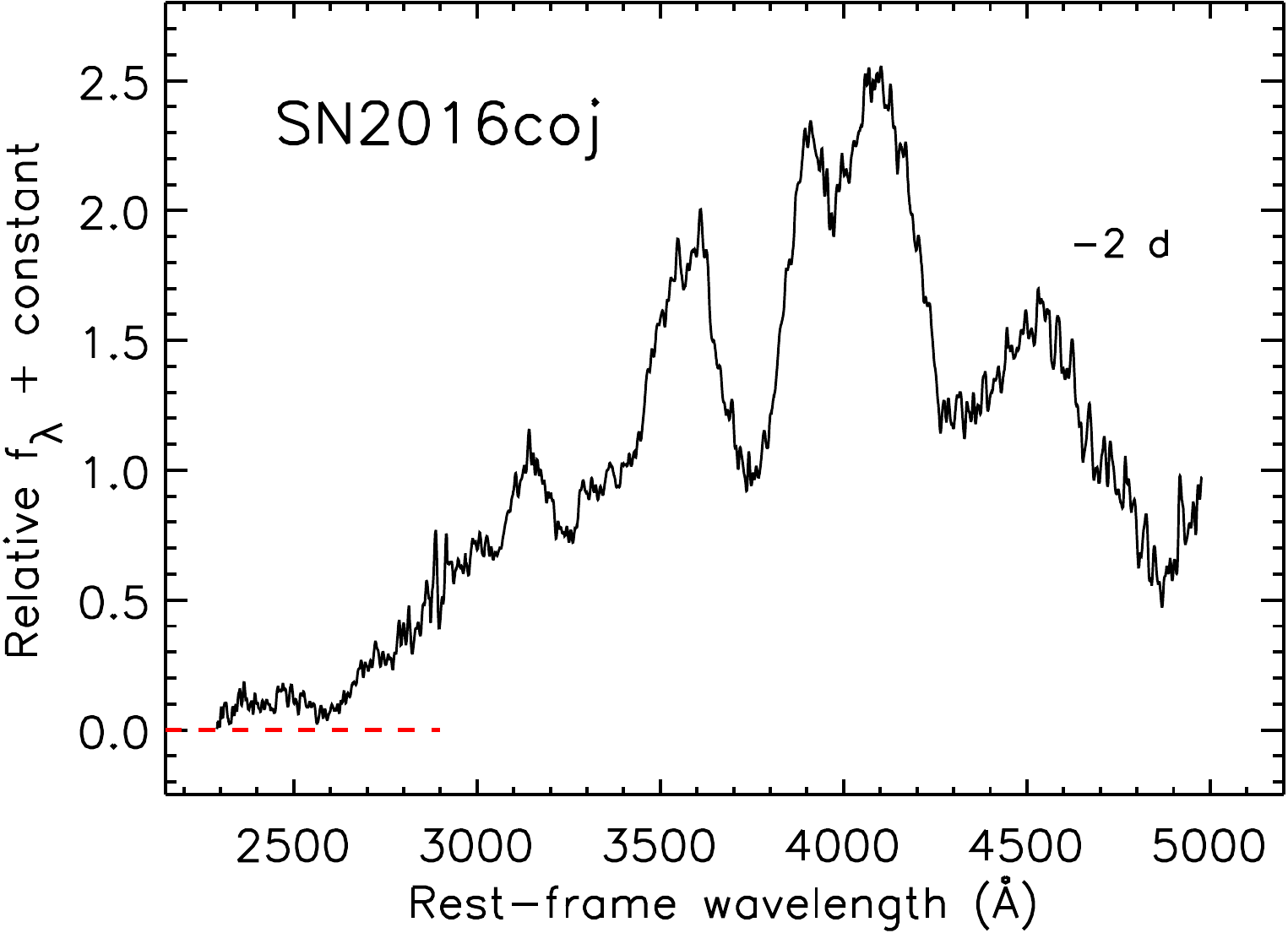}\\
	    \vspace{0.25cm}
	    \includegraphics[scale=0.8]{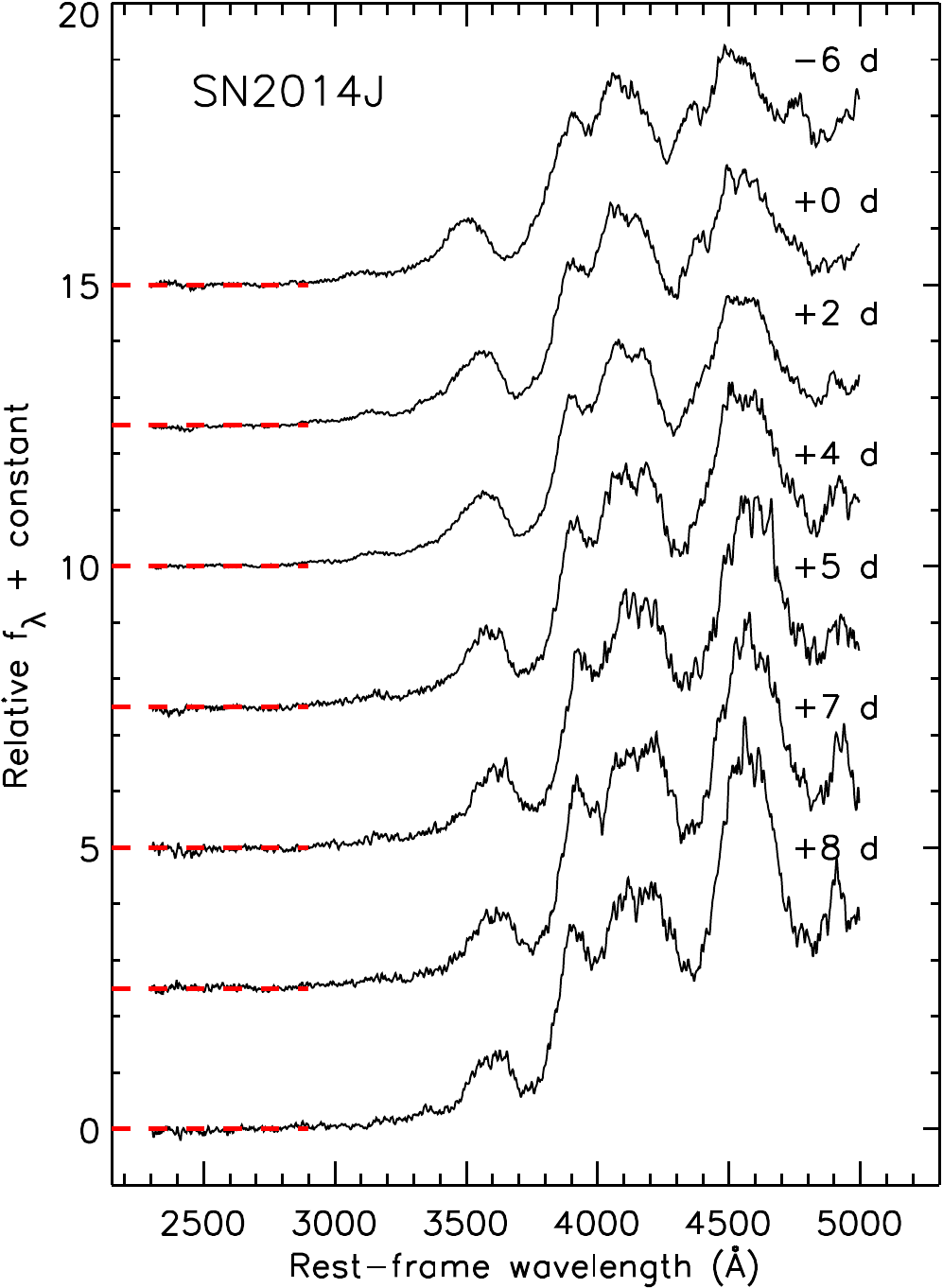}
        \caption{{\it Swift} UVOT grism observations of SNe~Ia (continued).
              }
        \label{uvspec5}
\end{figure*}

\begin{figure*}
	\centering
	    \includegraphics[scale=0.53]{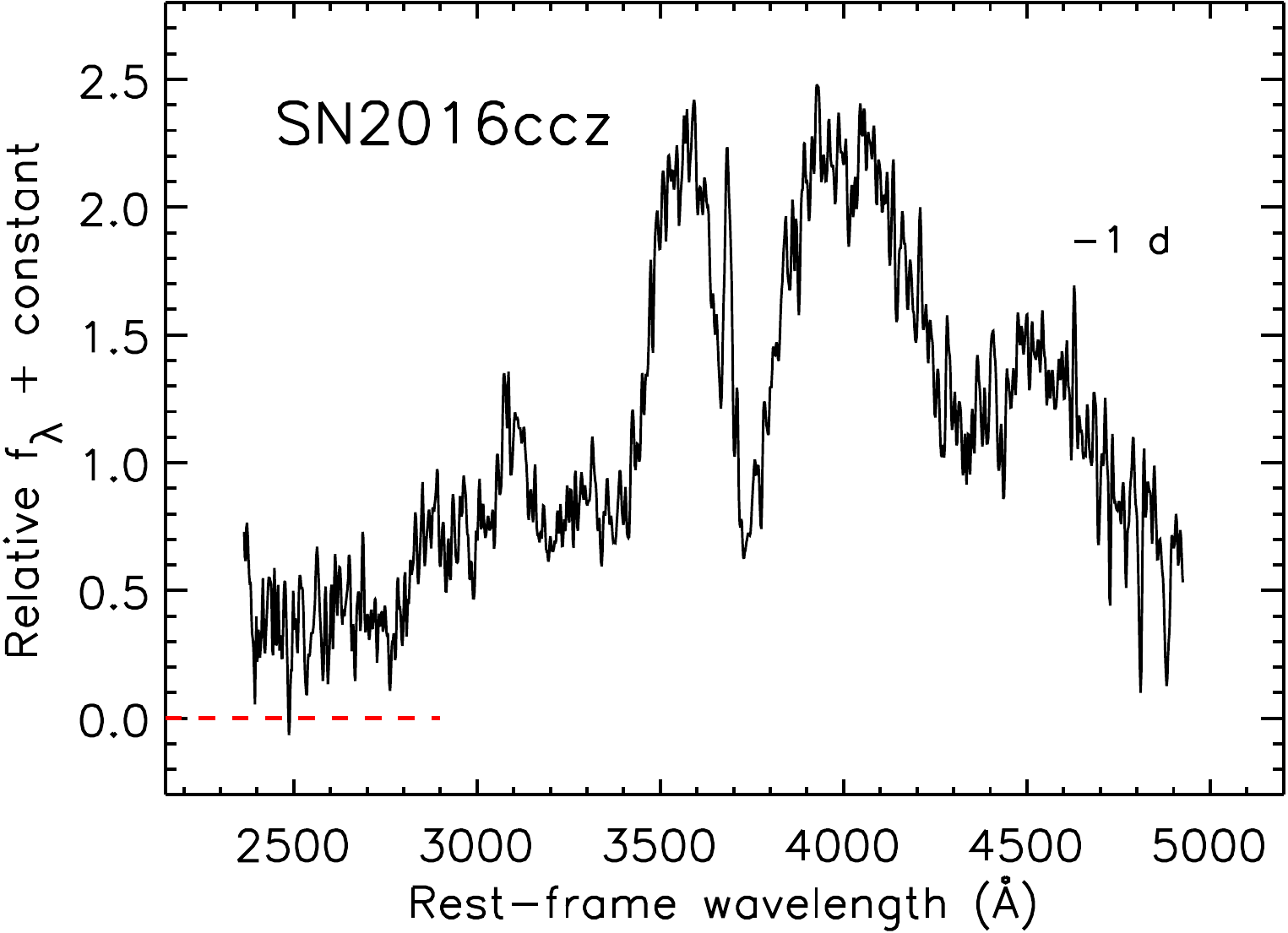}
	    \hspace{0.25cm}
	    \includegraphics[scale=0.53]{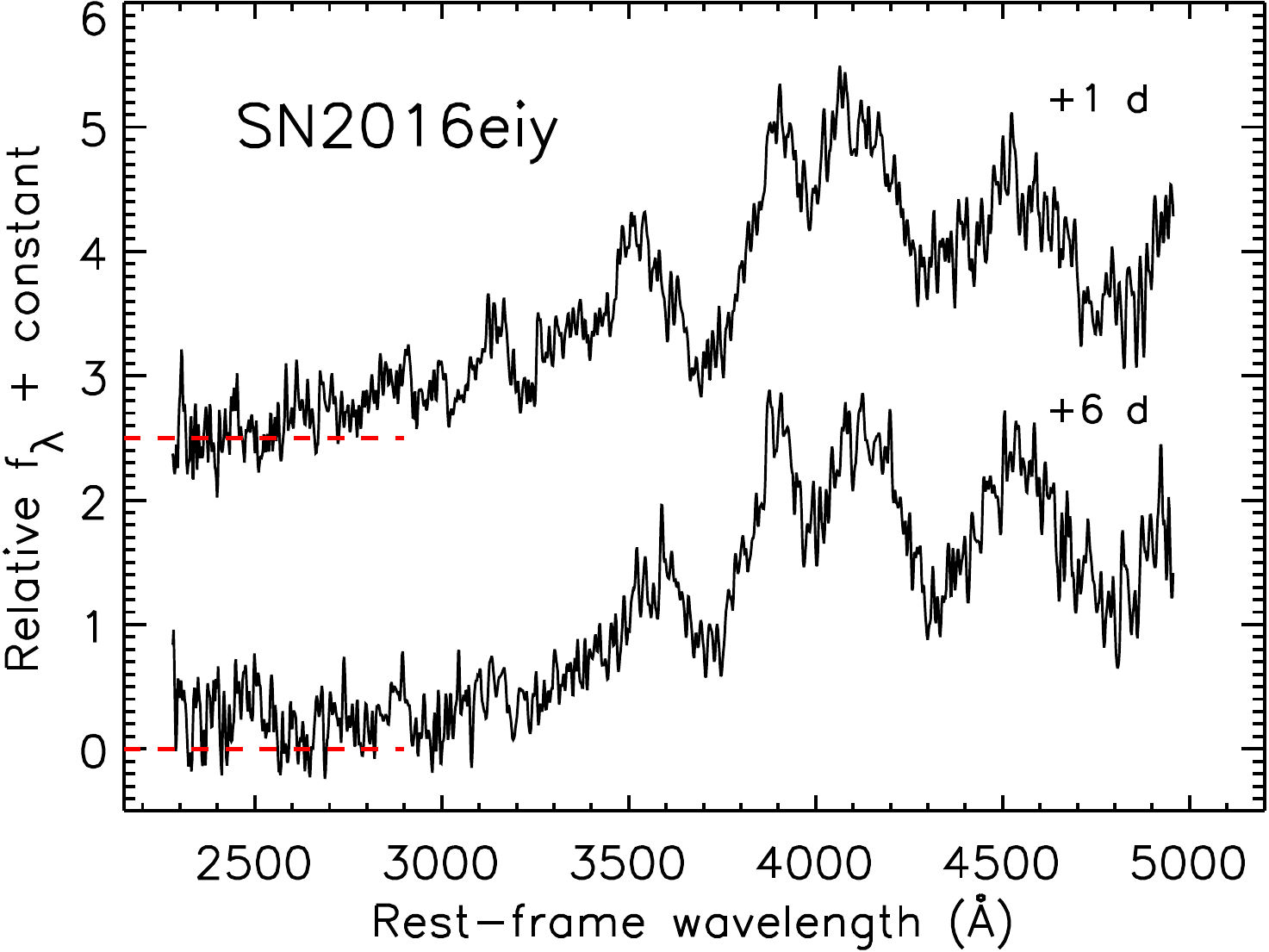}\\
	    \vspace{0.25cm}
	    \includegraphics[scale=0.53]{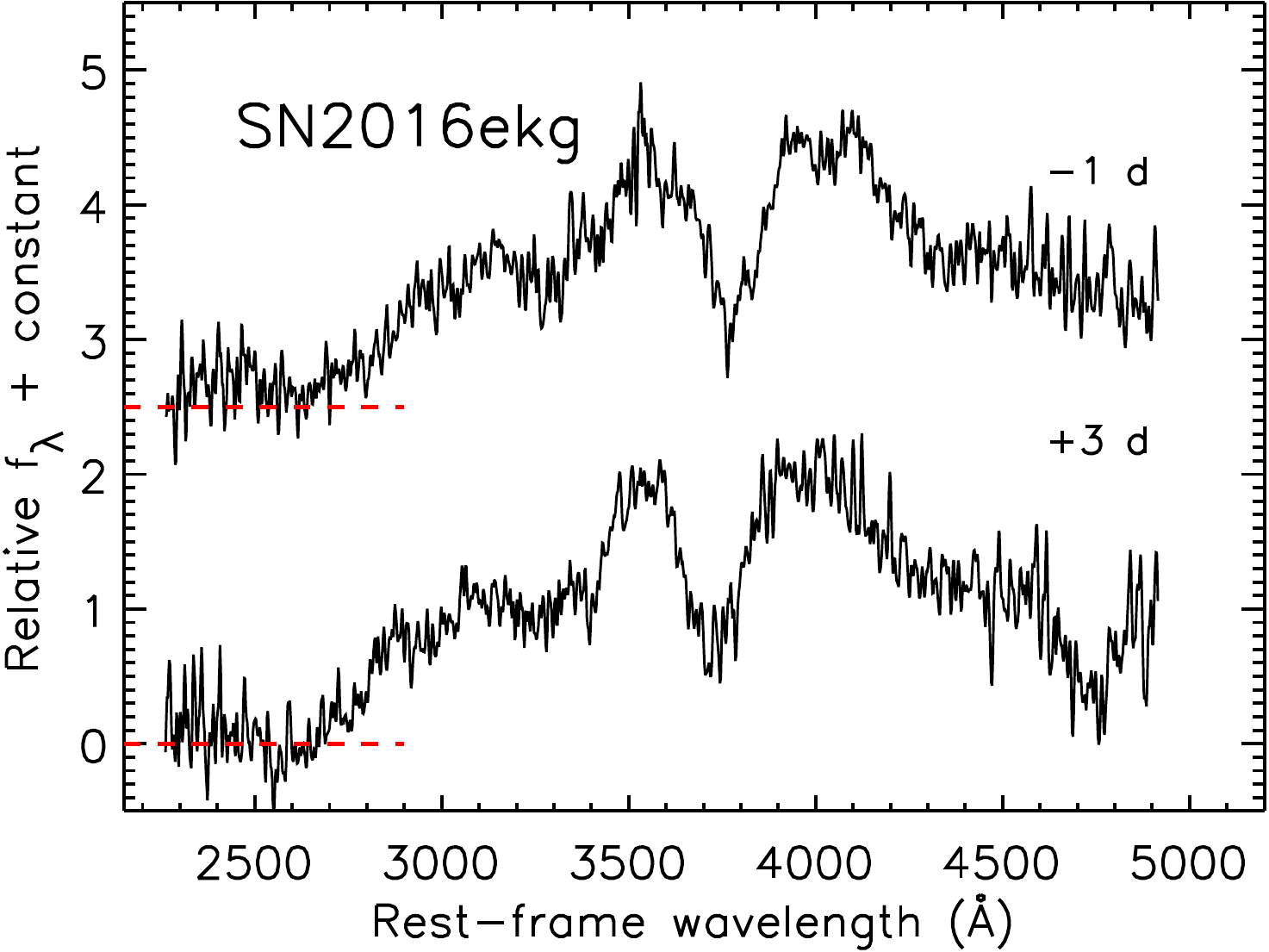}
	    \hspace{0.25cm}
	    \includegraphics[scale=0.53]{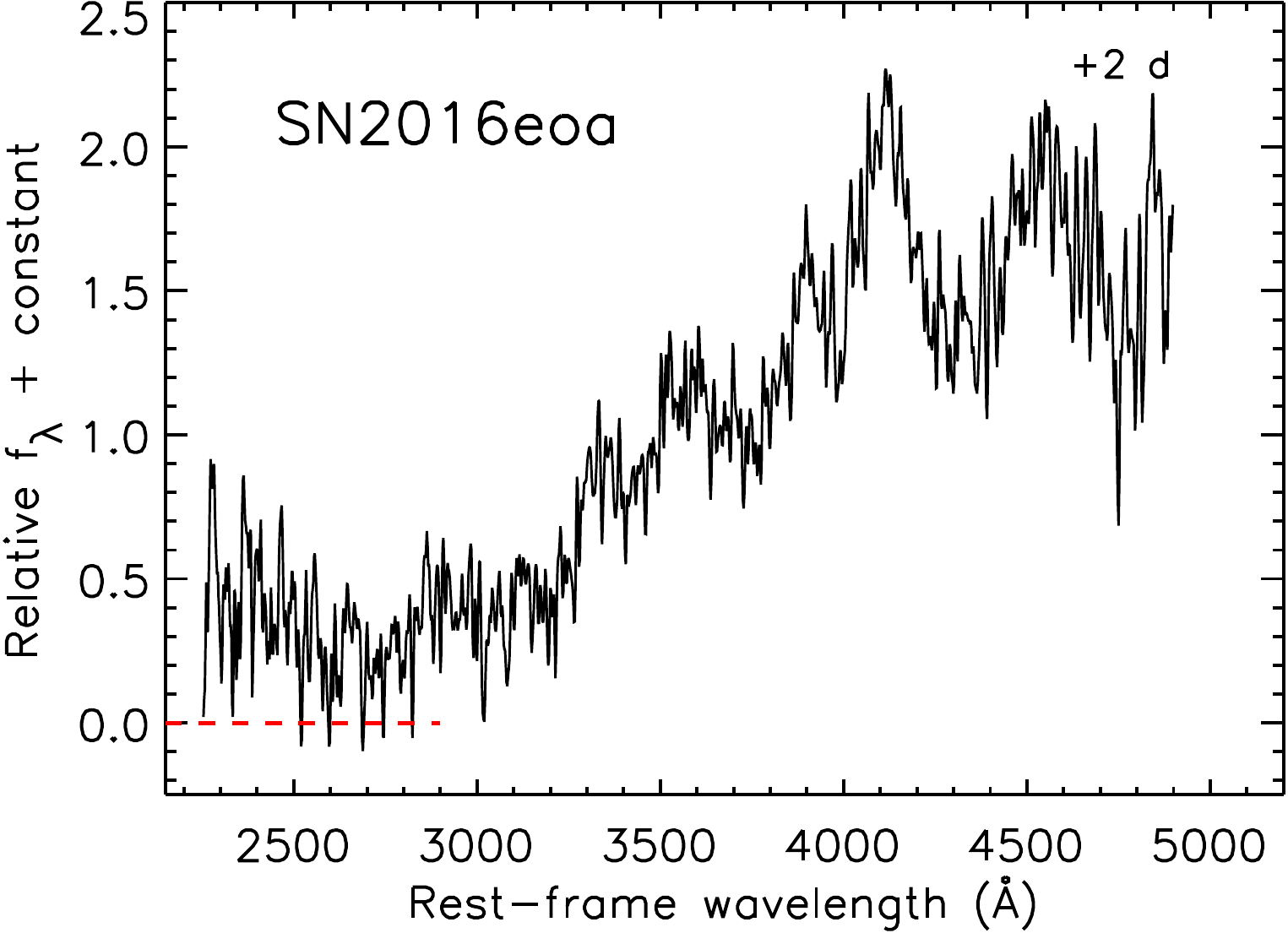}\\
	    \vspace{0.25cm}
        \includegraphics[scale=0.53]{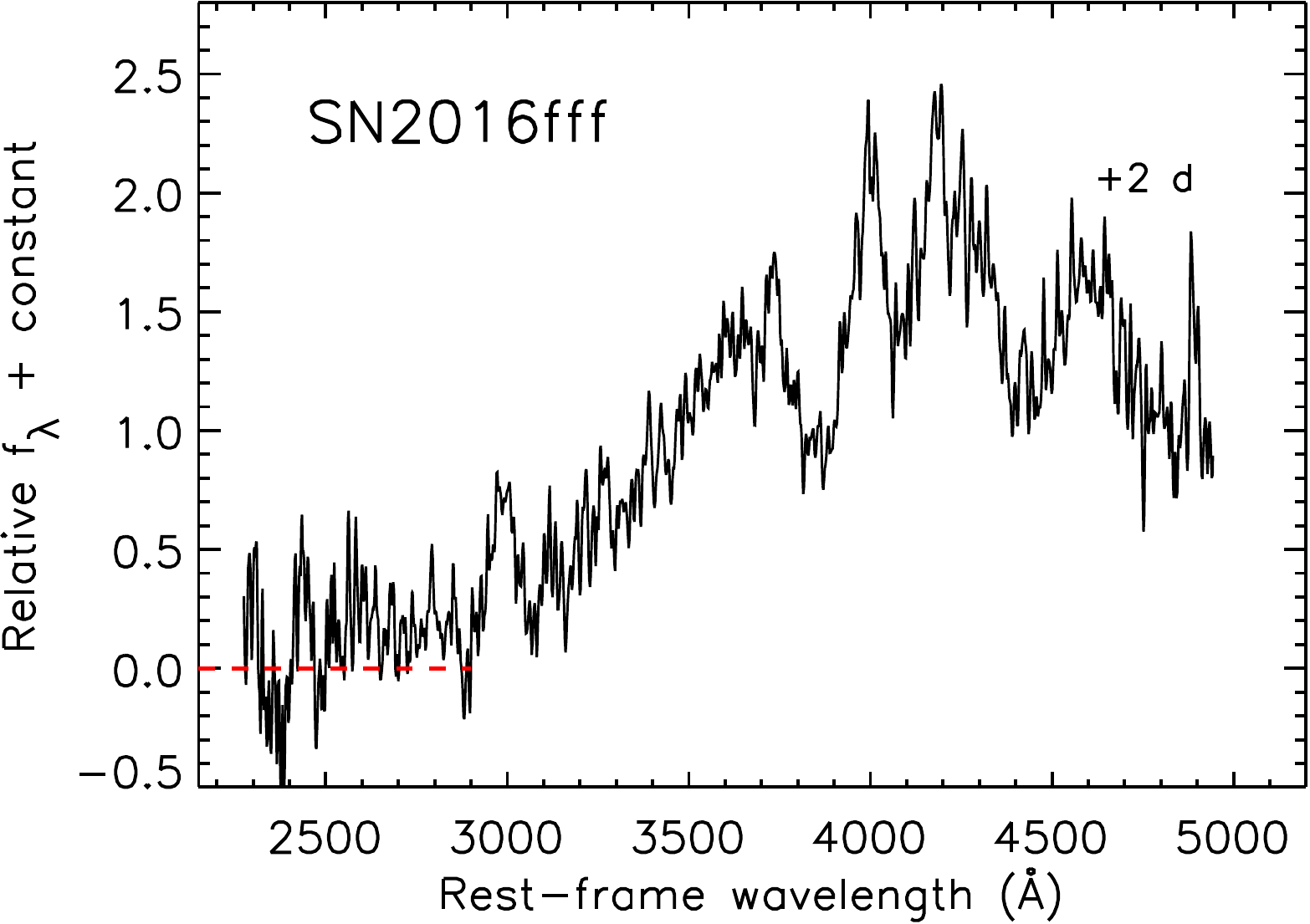}
	    \hspace{0.25cm}
	    \includegraphics[scale=0.53]{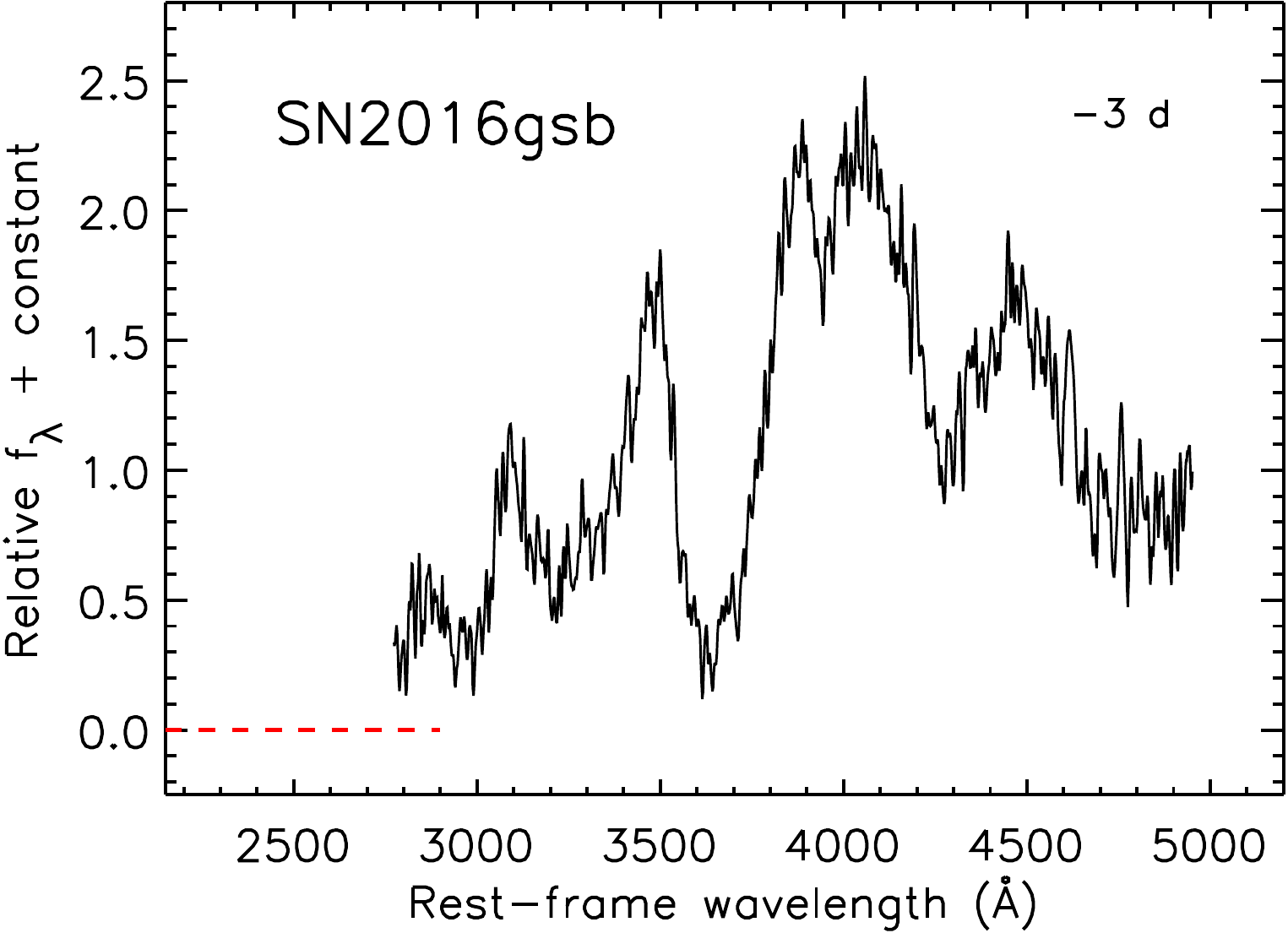}
	    \caption{{\it Swift} UVOT grism observations of SNe~Ia (continued).
              }
        \label{uvspec6}
\end{figure*}

\begin{figure*}
	\centering
	    \includegraphics[scale=0.53]{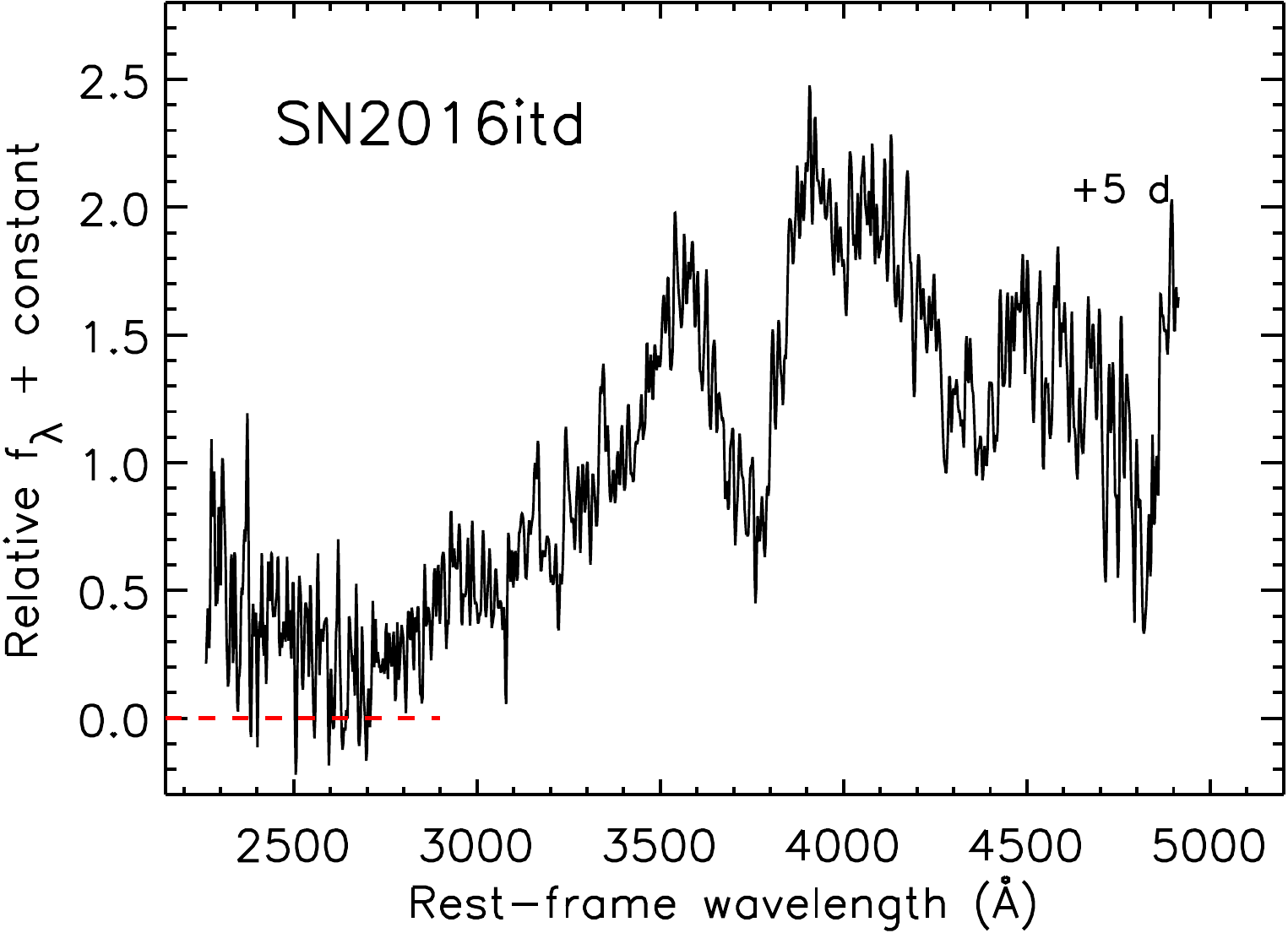}
	    \hspace{0.25cm}
        \includegraphics[scale=0.53]{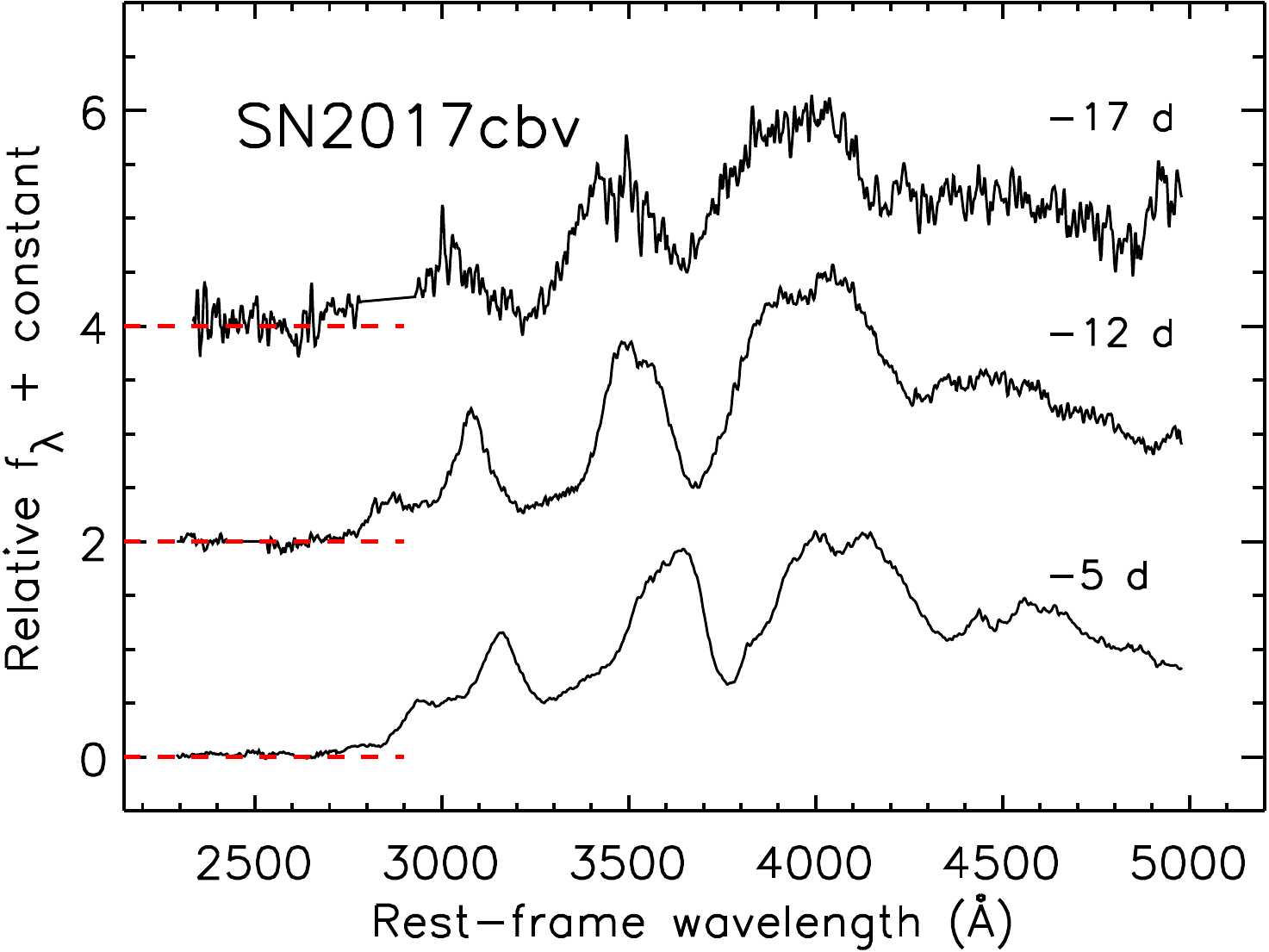}\\
	    \vspace{0.25cm}
	    \includegraphics[scale=0.53]{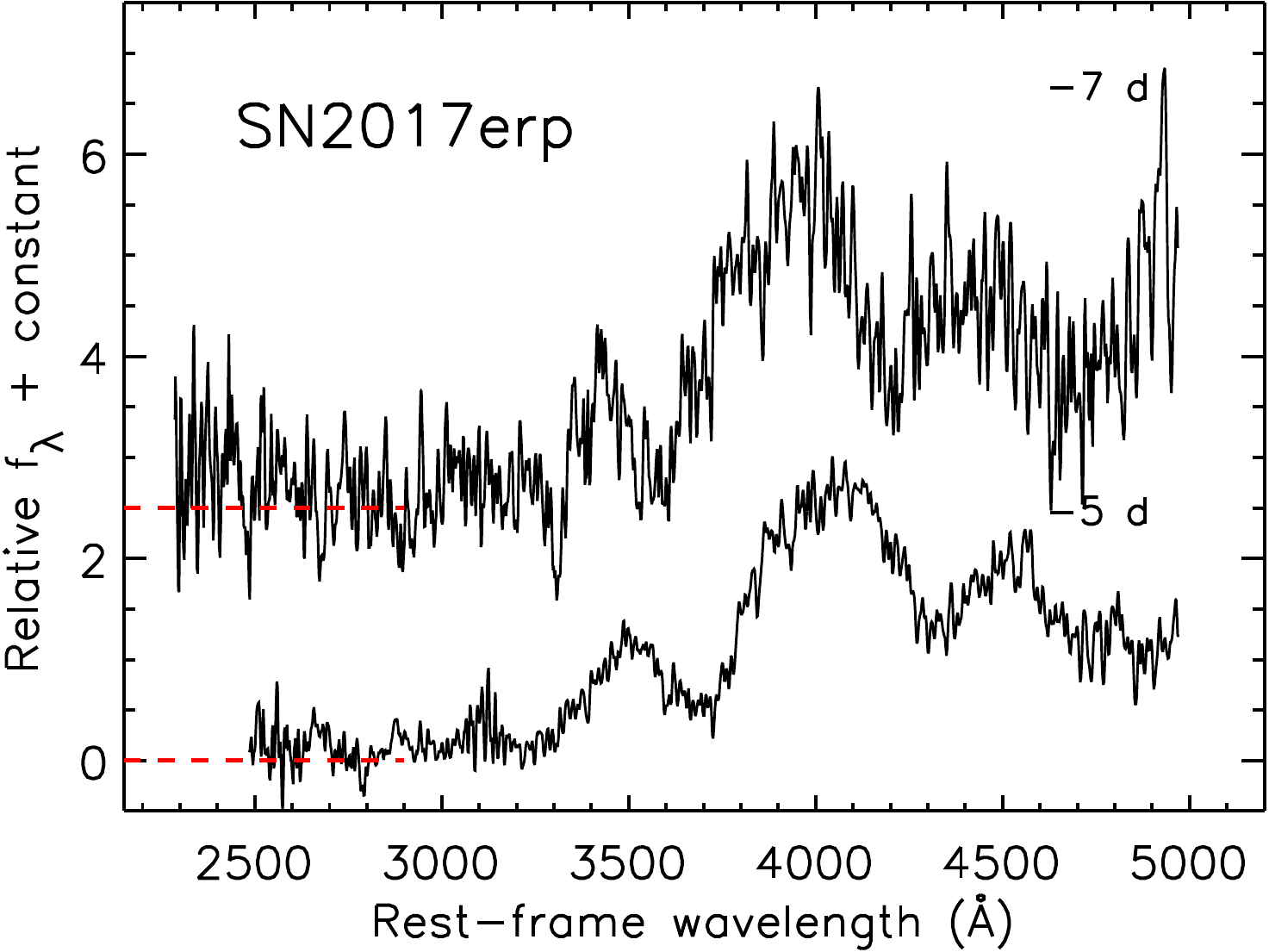}
	    \hspace{0.25cm}
	    \includegraphics[scale=0.53]{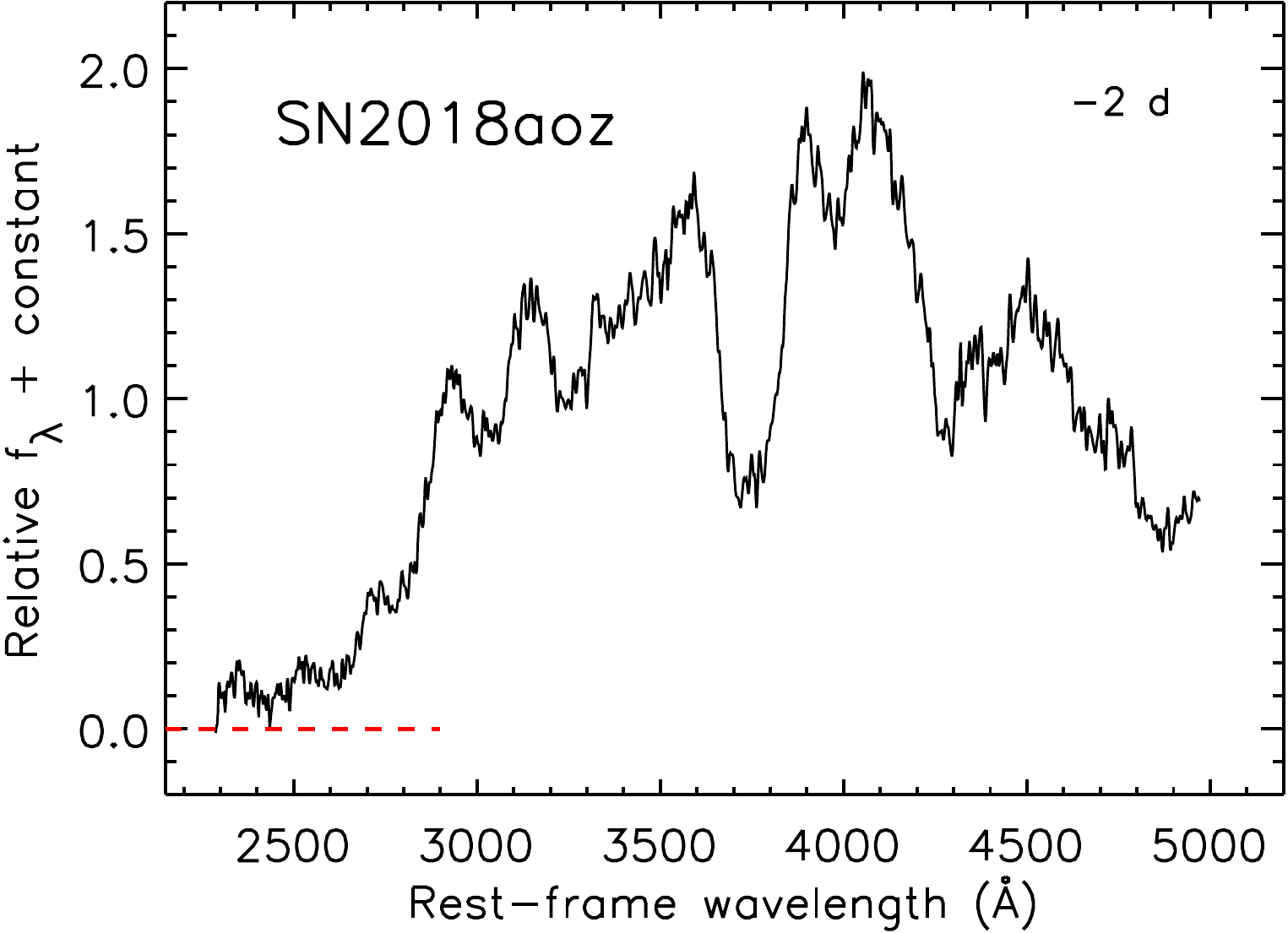}
	    \caption{{\it Swift} UVOT grism observations of SNe~Ia (continued).
              }
        \label{uvspec7}
\end{figure*}

We display the reductions of all 120 spectra in Figures~\ref{uvspec1}
through \ref{uvspec7} using the improved method. Here we trimmed each
spectrum to show only the UV flux redward of 2300\,\AA, as the SNR of
{\it Swift} spectra generally deteriorates dramatically below
$\sim$2300\,\AA.

\subsection{Data quality}
\label{sec:quality}
The SNR of each spectrum depends on many factors such as the exposure
time ($T_{\rm exp}$), distance to the SN, the amount of host-galaxy
extinction, and phase.  The sensitivity of the {\it Swift} UV grism peaks at $\sim2800$\,\AA.
However, because of a SN~Ia SED peaks near 4000\,\AA\ at maximum brightness,
the SNR also increases from the UV to the optical (in terms of the effective wavelength).  
Here we report the SNR of each spectrum as a function of redshift, luminosity, rest-frame phase,
and wavelength.  The large range of $T_{\rm exp}$ (a factor of
$\sim$10) for our sample complicates the comparison of other
properties. To compensate for these differences, we also calculate the 
SNR scaled to that expected with a $T_{\rm exp}=15$\,ks exposure. This is achieved
by multiplying the SNR by a factor of $\sqrt{15 {\rm ~ks}/T_{\rm
    exp}}$ (assuming the noise is dominated by Poisson noise).  This
simple approximation ignores other factors such as the varying
background for different SNe and detector noise, but it is sufficient
for our purposes.

To investigate the effect of distance on the SNR, we select only the
spectrum nearest to the peak brightness (within 5 days from the peak) for each SN. The result is
shown in the upper panels of Figure~\ref{snr}. Not surprisingly, the
SNR of the spectrum shows strong correlation with the redshift, in a
sense that SNe at higher redshift (thus more distant) tend to have
lower SNR than those of more nearby SNe. We find that the trend generally
follows the same direction as an inverse-square law (although with large dispersion), 
where the SN brightness is expected to be inversely proportional to the square of its luminosity distance 
(or the SNR is inversely proportional to the luminosity distance).

As noted above, the SNR of the spectrum also depends on the
wavelength. Here we calculate the SNR in two separate regions:
2300--2800\,\AA\ (mid-UV) and 2900--4000\,\AA\ (near-UV). We find that the SNR decreases
dramatically at shorter wavelengths. The SNR in the near-UV
is on average $\sim10$ times higher than that in the mid-UV region.

We show the SNR of the (near-peak) spectrum as a function of SN $V$-band peak brightness
in the middle panels of Figure~\ref{snr}. Most of the SNe in our sample are brighter than $V = 15$\,mag at peak,
with the brightest object having $V \approx 10$\,mag (SN~2011fe) and the faintest object having $V \approx 15.6$\,mag (SN~2016eoa).
The relation also appears to be tighter than that with redshift, and thus will be useful for 
scheduling future observations, in estimating the SNR of the 
spectrum given the exposure time and SN magnitude.
Note that SN~2014J has $V \approx 10.6$\,mag at peak, which is the second brightest object in our sample 
(in terms of $V$-band peak brightness). However, its spectrum has a low SNR ($\sim1$ in mid-UV) due to 
large extinction \citep[$A_{V}\approx2$\,mag; e.g.,][]{2014ApJ...784L..12G,2014ApJ...788L..21A,2014MNRAS.443.2887F} from the dust.

In the bottom panels of Figure~\ref{snr}, we show the dependence of SNR on the SN
phase. By examining the SNe which have multi-epoch observations in our
sample, we find the trend that the SNR of pre-peak spectra generally
rises with phase and peaks at the maximum light. It then decreases
after the peak when the SN gets fainter at later phases. This trend is
consistent with the SN~Ia light curve, where the peak luminosity
is generally 3--4\,mag brighter than that right after the SN explosion
\citep[e.g.,][]{2013ApJ...778L..15Z}. The resulting variation on 
SNR of the spectrum can be as large as a factor of $\gtrsim 10$
(assuming a fixed $T_{\rm exp}$).

\subsection{Comparison to HST UV sample}
\label{sec:hst-compare}
\begin{figure}
	\centering
	    \includegraphics[scale=0.51]{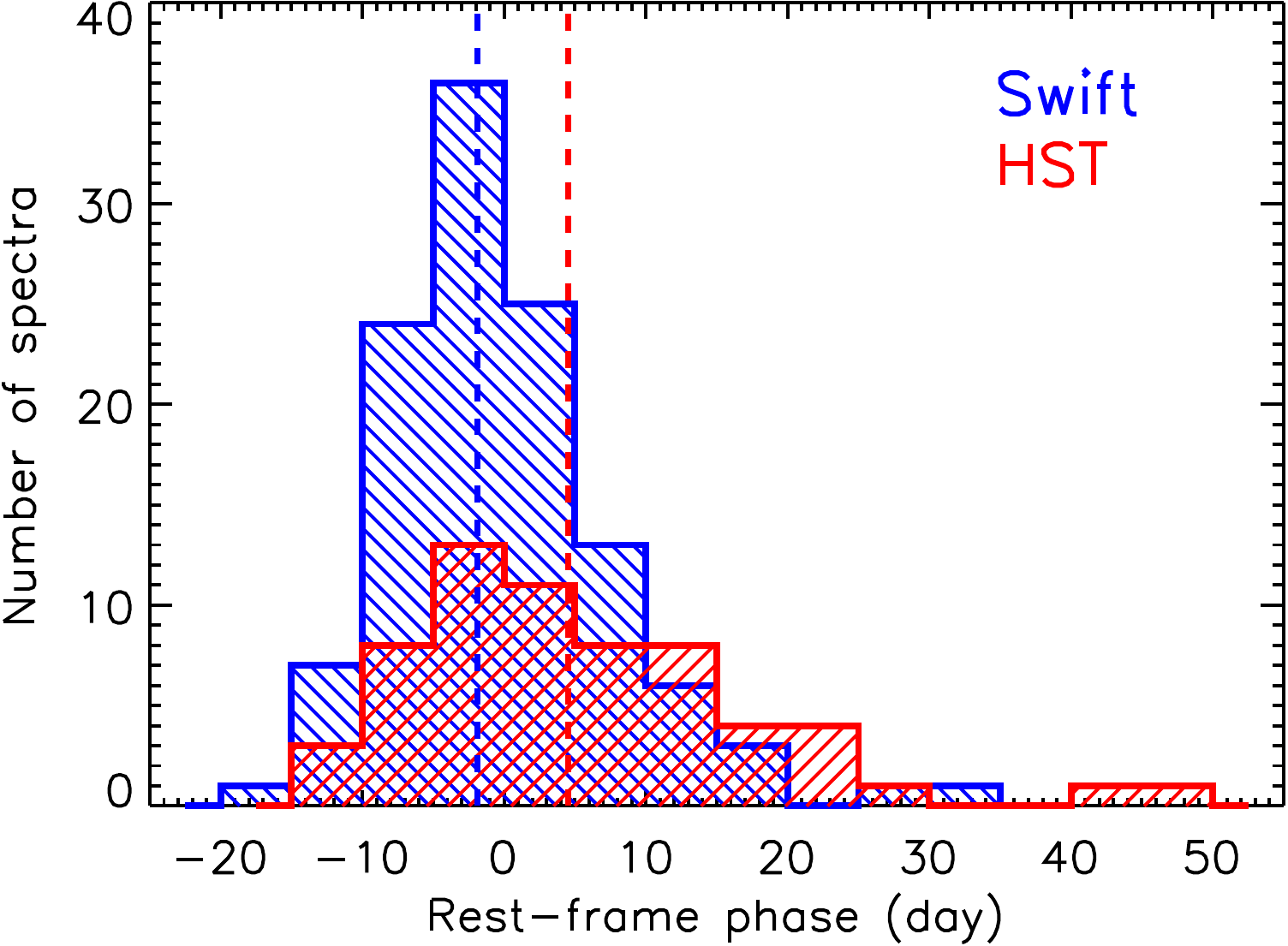}\\
	    \vspace{0.25cm}
		\includegraphics[scale=0.52]{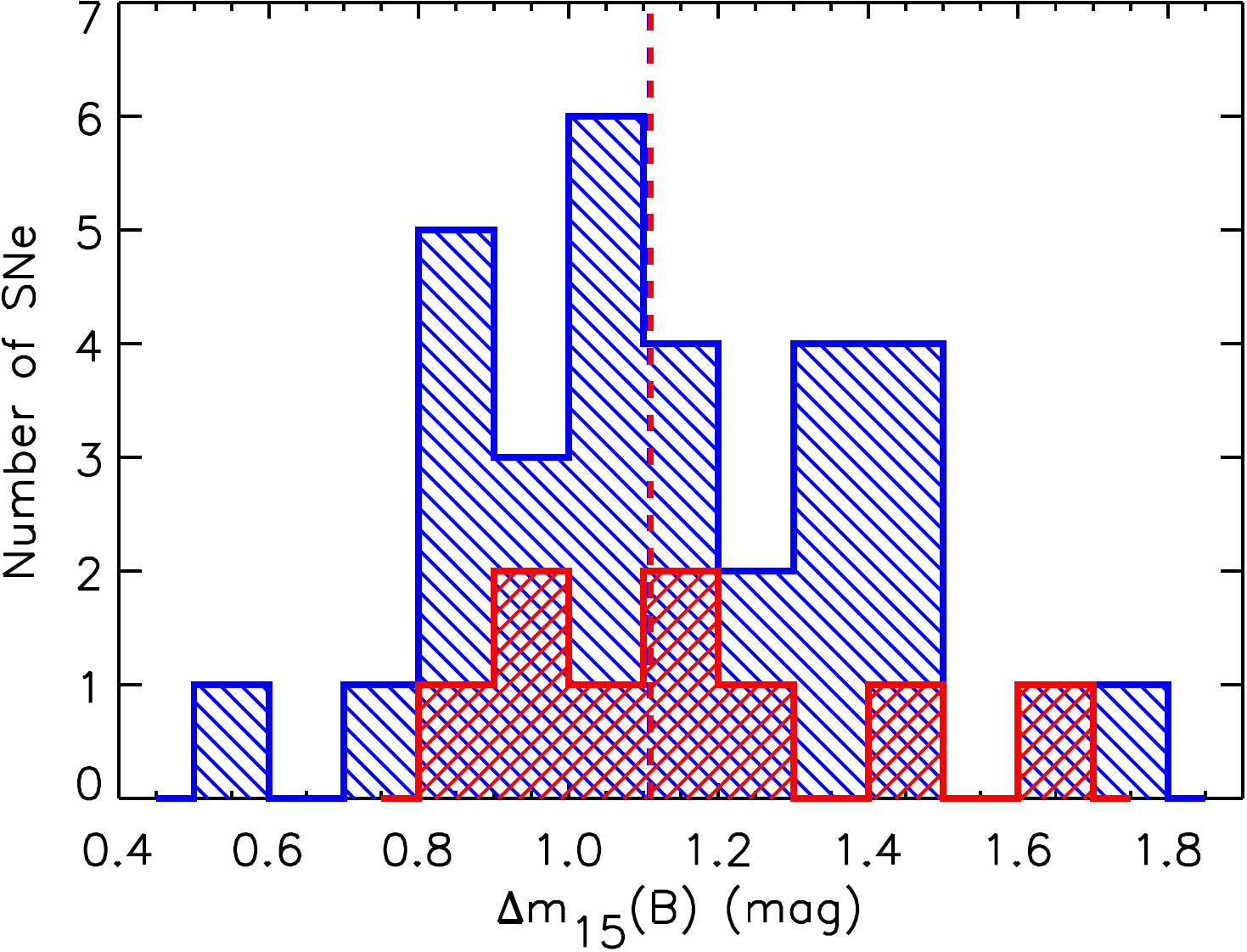}
                \caption{Phase (upper panel) and \deltam($B$) (lower
                  panel) distributions for the {\it Swift} sample
                  (blue histogram) and the current {\it HST} UV sample
                  (red histogram). The vertical dashed line in each
                  panel represents the median of the distribution.  }
        \label{swift-hst-1}
\end{figure}

Compared to {\it Swift}, the current {\it HST} SN~Ia UV sample that
probes blueward of $2900$\,\AA\ is relatively small \citep[e.g., 9 SNe
in][]{2016MNRAS.461.1308F}. The addition of 39 SNe~Ia observed by {\it
  Swift} greatly increases the number of SNe with UV data.  Here we
compare these samples.

The top panel of Figure~\ref{swift-hst-1} compares the SN phase
distributions of the {\it Swift} and \citet{2016MNRAS.461.1308F} {\it
  HST} samples.  {\it Swift} tends to observe SNe at earlier phases
than {\it HST}, with medians phases of $-1.9$ and $+5.5$~days,
respectively.  There are 6 {\it Swift}-observed SNe~Ia with their
first UV spectrum $\gtrsim$10~days before peak brightness.  In
contrast, only 2 {\it HST}-observed SNe~Ia have UV observations
$\gtrsim$10~days before peak brightness.  As earlier observations
provide critical progenitor information
\citep[e.g.,][]{2015MNRAS.452.4307P}, {\it Swift} can be a powerful
resource for studying SN~Ia physics.

The bottom panel of Figure~\ref{swift-hst-1} compares the
\deltam($B$) distributions of the {\it Swift} and
\citet{2016MNRAS.461.1308F} {\it HST} samples.  The {\it Swift} sample
has a larger range of \deltam($B$) and several more examples for
specific values of \deltam($B$).  In particular, our {\it Swift} sample
contains several slowly declining SNe~Ia (e.g., \deltam($B$)
$<0.8$\,mag).  This will greatly increase the resolution and precision
of parameter space when constructing data-driven models of SN~Ia UV
spectra \citep[e.g.,][]{2016MNRAS.461.1308F}.

\begin{figure*}
	\centering
		\includegraphics[scale=0.8]{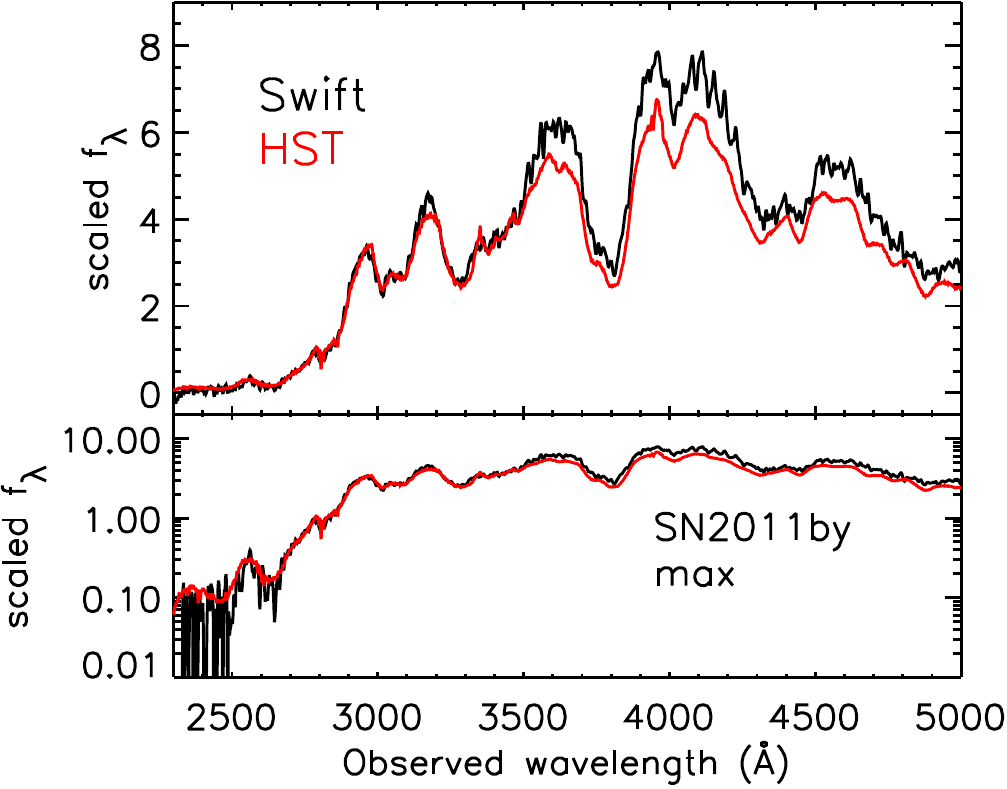}
		\hspace{0.25cm}
		\includegraphics[scale=0.8]{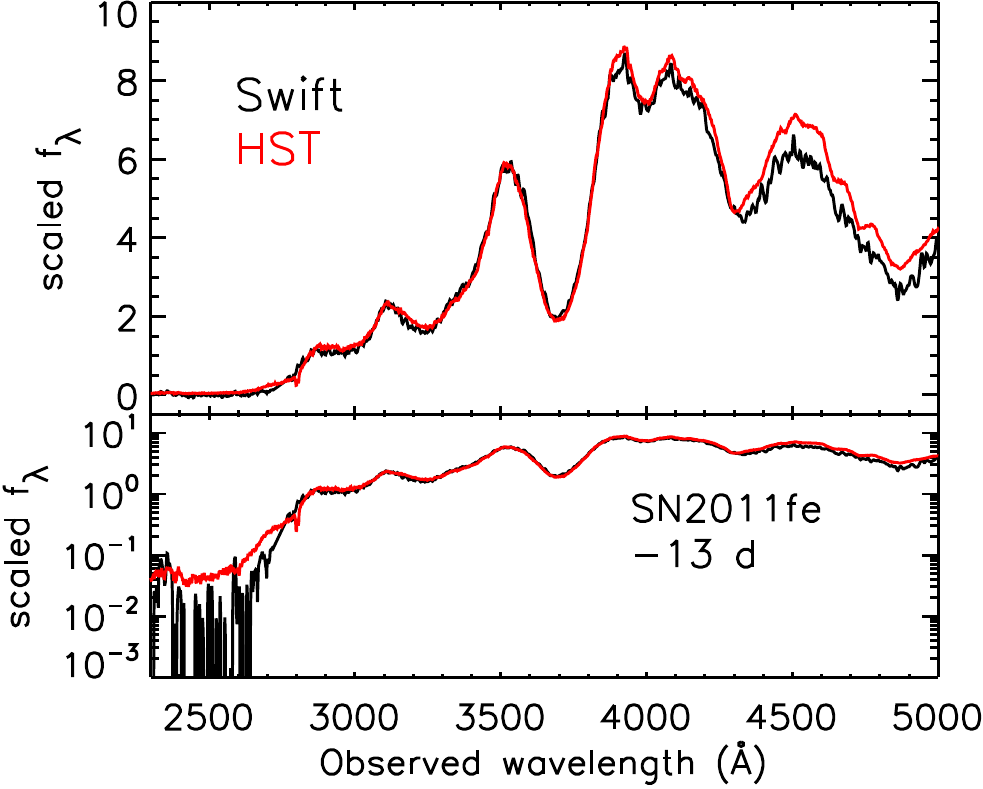}\\
		\vspace{0.25cm}
		\includegraphics[scale=0.8]{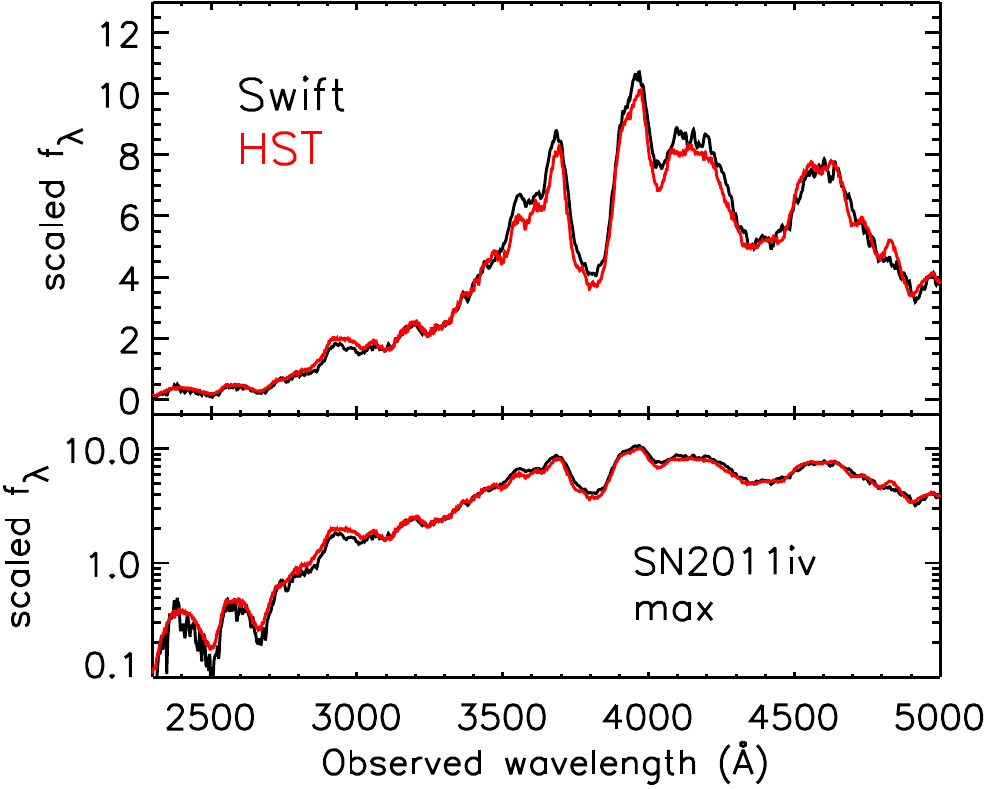}
		\hspace{0.25cm}
		\includegraphics[scale=0.8]{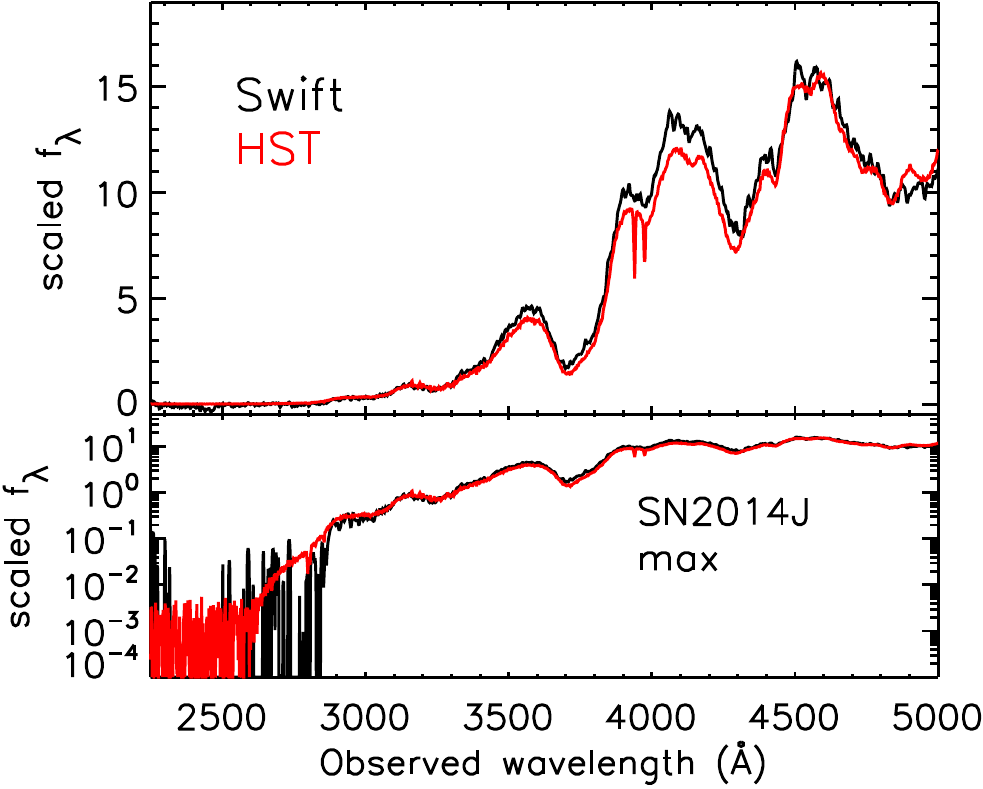}\\
		\vspace{0.25cm}
		\includegraphics[scale=0.8]{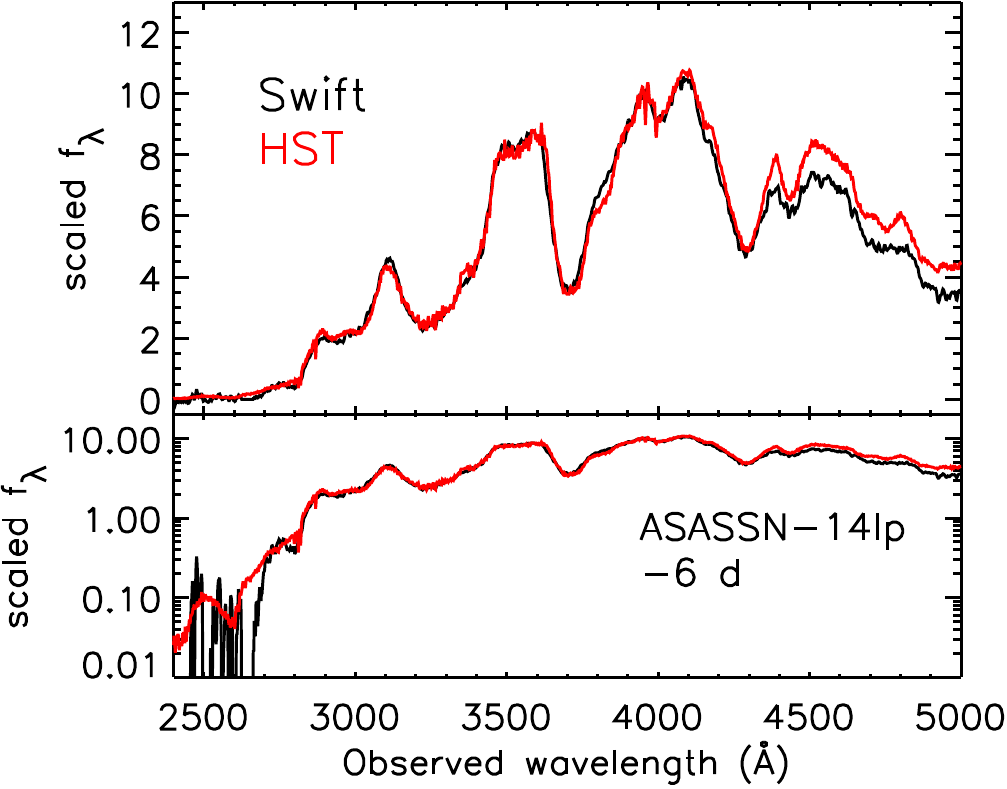}
		\hspace{0.55cm}
		\includegraphics[scale=0.8]{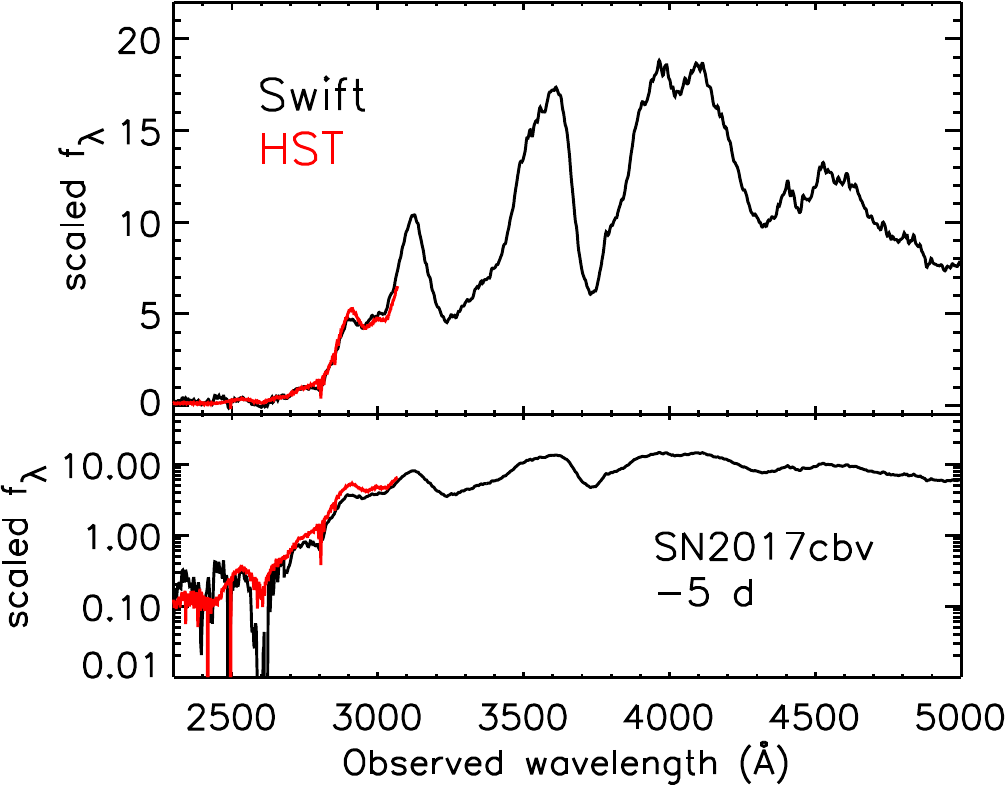}
                \caption{UV through optical spectra (2300--5000\,\AA)
                  of 6 SNe~Ia observed with both {\it Swift}/UVOT
                  (black curves) and {\it HST}/STIS (red curves). For
                  each SN, we display the flux with both linear (upper
                  subpanel) and logarithmic (lower subpanel) scales.
                  Note that for SN~2017cbv, the {\it HST} spectrum
                  only covers $\lambda \lesssim 3100$\,\AA.  }
        \label{swift-hst-2}
\end{figure*}

For the subset of SNe that were observed at similar phases with both
{\it Swift} and {\it HST}, we can directly compare the {\it Swift}
data reduction to the well-calibrated {\it HST} data.
Figure~\ref{swift-hst-2} displays the 6 SNe~Ia which have high-SNR UV
spectra and similar phases for both {\it Swift} and $\it HST$
observations.  The {\it Swift} spectra have been normalised to match
the {\it HST} spectra in the region 3000--3500\,\AA\ (for SN~2017cbv,
whose {\it HST} spectrum only covers $\lambda < 3100$\,\AA, we use
2500--3000\,\AA). We present the spectra in both linear and
logarithmic scales in the figure.  The spectra are generally well
matched, further indicating that the {\it Swift} spectral reductions
are accurate.

Compared to {\it Swift}, {\it HST} covers a similar wavelength range in
the UV, but the {\it HST} spectra have consistently higher SNR.  

\section{Summary}
\label{sec:summary}
In this work, we present {\it Swift}/UVOT observations and reductions
of 120 spectra of 39 nearby SNe~Ia. This is the largest existing sample of
SN~Ia UV spectra that probe blueward of 2900\,\AA. The new sample
doubles the number of SN~Ia UV spectra and triples the number of
SNe~Ia with UV spectra.

We outline an improved method to reduce the {\it Swift} 
spectroscopic data and perform the reductions.  This method achieves a more
precise background subtraction than the original reduction pipeline.
Our new method can effectively reduce the contamination from
background sources, which is critical for the slitless observations of
{\it Swift} UVOT.

Compared to the {\it HST} sample, the {\it Swift} sample is larger in
both number of SNe and number of spectra.  The {\it Swift} sample has
a broad \deltam($B$) distribution, spanning the entire range for SNe~Ia.
The {\it Swift} sample also has significantly more SNe at the earliest
phases, with the median phase of the first observation of a SN and all
spectra being $-4.5$ and $-1.9$~days, respectively.
 
SN~Ia UV spectra are critical to understanding the progenitor systems
and explosion mechanisms of SNe~Ia. With the addition of {\it Swift}
UV spectra, we are building a sample which has the statistical power
to investigate the UV properties of the SN~Ia sample.  A detailed
analysis of these data will be presented in the second paper of this
series (Pan et al., in preparation).

\section{Acknowledgements}
\label{sec:acknowledgements}
{\it Swift} spectroscopic observations were performed under programs
GI--04047, GI--5080130, GI--6090689, GI--8110089, GI--1013136, and
GI--1215205; we are very grateful to N.\ Gehrels, S. B.\ Cenko, and the
{\it Swift} team for executing the observations quickly.
Based in part on observations made with the NASA/ESA {\it Hubble Space
  Telescope}, obtained at the Space Telescope Science Institute
(STScI), which is operated by the Association of Universities for
Research in Astronomy, Inc., under National Aeronautics and Space
Administration (NASA) contract NAS 5--26555. These observations are
associated with Programmes GO--12298, GO--12592, GO--13286, GO--13646,
and DD--14925.  This manuscript is based upon work supported by NASA
under Contract No.\ NNG16PJ34C issued through the {\it WFIRST} Science
Investigation Teams Programme.

We thank Steven Downing for useful discussions and comments. 
The referee, Peter Brown, provided suggestions that helped improve this paper.
N.P.M.K. has been supported by the UK Space Agency.  
The UCSC group is supported in part by NASA grant 14-WPS14-0048, NSF grant
AST-1518052, the Gordon \& Betty Moore Foundation, and by fellowships
from the Alfred P.\ Sloan Foundation and the David and Lucile Packard
Foundation to R.J.F. Additional financial assistance to A.V.F. was
provided by the TABASGO Foundation, the Christopher R. Redlich Fund,
and the Miller Institute for Basic Research in Science (UC Berkeley).

\bibliographystyle{mn2e}
\bibliography{swift_ia_data}

\clearpage
\begin{table*}
\caption{Summary of {\it Swift} UVOT grism observations of the SN~Ia sample in this work.}
\begin{tabular}{lccccrrrc}
\hline\hline
SN Name & Host & Redshift  & \deltam($B$)$^a$ & UT Obs. Date  & $T_{\rm exp}$ & $T_{\rm exp}$ used$^b$  & Phase & LC Ref.$^c$\\
 &  &  & (mag) &  & (s) & (s) & (day) & \\
\hline
SN~2005am & NGC~2811 & 0.0079 & 1.45(03) & 2005-03-08.02 & 2781.73 & 1097.31 & 0.0 & \citet{2005ApJ...635.1192B}\\
          \hline
SN~2005cf & MCG-01-39-03 & 0.0065 & 1.05(03) & 2005-06-04.71 & 1897.12 & 1897.12 & $-7.9$ & \citet{2009ApJ...697..380W}\\
          &  &        & & 2005-06-05.71 & 1958.95 & 1958.95 & $-6.9$ & \\
          &  &        & & 2005-06-06.72 & 1572.61 & 1572.61 & $-5.9$ & \\
          &  &        & & 2005-06-08.80 & 1726.76 & 1726.76 & $-3.8$ & \\
          &  &        & & 2005-06-09.52 & 626.97 & 626.97 & $-3.1$ & \\
          &  &        & & 2005-06-10.74 & 713.38 & 713.38 & $-1.9$ & \\
          &  &        & & 2005-06-11.61 & 2022.68 & 2022.68 & $-1.0$ & \\
          &  &        & & 2005-06-16.36 & 1812.21 & 1812.21 & 3.7 & \\
          &  &        & & 2005-06-17.30 & 1670.84 & 1670.84 & 4.6 & \\
          &  &        & & 2005-06-20.71 & 2002.30 & 2002.30 & 8.0 & \\
          &  &        & & 2005-06-26.34 & 2004.37 & 2004.37 & 13.6 & \\
          &  &        & & 2005-06-29.11 & 2122.21 & 2122.21 & 16.3 & \\
          \hline
SN~2005df & NGC~1559 & 0.0044 & 1.12(00) & 2005-08-11.06 & 1634.74 & 1634.74 & $-6.1$ & \citet{2017RNAAS...1a..36K}\\
          &  &        & & 2005-08-14.13 & 983.46 & 983.46 & $-3.1$ & \\
          &  &        & & 2005-08-17.94 & 2018.12 & 2018.12 & 0.7 & \\
          &  &        & & 2005-08-21.57 & 422.10 & 422.10 & 4.4 & \\
          \hline
SN~2005ke & NGC~1371 & 0.0049 & 1.76(01) & 2005-11-20.44 & 1952.35 & 318.56 &$-4.7$ & \citet{2010AJ....139..120F}\\
          &  &        & & 2005-11-22.42 & 3657.86 & 3657.86 & $-2.7$ & \\
          \hline
SN~2006dd & NGC~1316 & 0.0059 & 1.08(01) & 2006-07-14.51 & 3956.35 & 3956.35 & 11.0 & \citet{2010AJ....140.2036S}\\
          \hline
SN~2007sr & NGC~4038 & 0.0055 & 0.92(04) & 2007-12-20.69 & 1992.51 & 1992.51 & 6.7 & \citet{2013ApJ...773...53F}\\
          &  &        & & 2007-12-25.12 & 3729.30 & 3729.30 & 11.1 & \\
          &  &        & & 2007-12-26.15 & 3391.09 & 3391.09 & 12.1 & \\
          &  &        & & 2007-12-28.06 & 5988.71 & 5988.71 & 14.0 & \\
          \hline
SN~2008Q  & NGC~524 & 0.0079 & 1.41(05) & 2008-02-09.23 & 5755.24 & 4916.04 & $-0.2$ & \citet{2015MNRAS.454.3816C}\\
          &  &        & & 2008-02-09.79 & 5097.28 & 3192.17 & 4.3 & \\
          \hline
SN~2008hv & NGC~2765 & 0.0136 & 1.25(01) & 2008-12-11.66 & 8522.28 & 8522.28 & $-5.1$ & \citet{2013ApJ...773...53F}\\
          &  &        & & 2008-12-14.74 & 8584.80 & 8584.80 & $-2.0$ & \\
          \hline
SN~2009Y  & NGC~5728  & 0.0093 & 1.21(00) & 2009-02-07.64 & 7602.16 & 1352.78 & $-6.2$ & \citet{2015ApJS..220....9F}\\
          &  &        & & 2009-02-08.41 & 8627.93 & 5933.63 & $-5.5$ & \\
          &  &        & & 2009-02-10.56 & 13392.66 & 10862.94 & $-3.3$ & \\
          &  &        & & 2009-02-11.86 & 6269.84 & 6269.84 & $-2.0$ & \\
          &  &        & & 2009-02-13.16 & 18123.06 & 14394.74 & $-0.7$ & \\
          &  &        & & 2009-02-18.49 & 13208.81 & 11059.49 & 4.6 & \\
          \hline
SN~2009an & NGC~4332 & 0.0092 & 1.44(00) & 2009-03-03.63 & 14473.74 & 12650.73 & $-4.5$ & \citet{2015ApJS..220....9F}\\
          &  &        & & 2009-03-04.17 & 4265.17 & 4265.17 & $-4.0$ & \\
          &  &        & & 2009-03-10.96 & 8431.56 & 1624.79 & 2.7 & \\
          &  &        & & 2009-03-12.24 & 10282.27 & 8659.64 & 4.0 & \\
          \hline
SN~2009dc & UGC~10064 & 0.0214 & 0.72(03) & 2009-04-30.85 & 3707.84 & 3283.08 & 5.3 & \citet{2011MNRAS.410..585S}\\
          &  &        & & 2009-05-01.22 & 5110.87 & 3978.95 & 5.8 & \\
          \hline
SN~2009ig & NGC~1015 & 0.0088 & 0.89(00) & 2009-08-24.12 & 7023.63 & 5943.86 & $-12.8$ & \citet{2012ApJ...744...38F}\\
          &  &        & & 2009-08-24.76 & 5516.39 & 5516.39 & $-12.1$ & \\
          &  &        & & 2009-08-25.69 & 4547.29 & 3031.54 & $-11.2$ & \\
          &  &        & & 2009-08-27.57 & 14024.93 & 12737.63 & $-9.3$ & \\
          &  &        & & 2009-09-01.34 & 5857.20 & 1624.77 & $-4.6$ & \\
          &  &        & & 2009-09-03.70 & 13531.05 & 13531.05 & $-2.3$ & \\
          &  &        & & 2009-09-07.34 & 18060.624 & 18060.62 & 1.3 & \\
          &  &        & & 2009-09-14.26 & 18009.77 & 16657.00 & 8.2 & \\
          \hline
SN~2010ev & NGC~3244 & 0.0092 & 1.12(02) & 2010-07-12.13 & 17820.81 & 6325.86 & 5.0 & Guti{\'e}rrez et al. (2016)\\ 
          \hline
SN~2011B  & NGC~2655 & 0.0047 & 1.38(16) & 2011-01-15.20 & 10263.21 & 9654.62 & $-6.0$ & \citet{2017ApJ...836..232B}\\
          \hline
SN~2011aa & UGC~3906 & 0.0124 & 0.59(07) & 2011-02-28.15 & 7519.65 & 6271.86 & 8.1 & \citet{2014ApJ...787...29B}\\
          \hline
SN~2011ao & IC~2973 & 0.0107 & 1.00(00) & 2011-03-08.69 & 10654.46 & 10654.46 & $-9.5$ & \citet{2015ApJS..220....9F}\\
          &  &        & & 2011-03-09.48 & 6633.80 & 6633.80 & $-8.7$ & \\
          &  &        & & 2011-03-15.74 & 7878.21 & 7878.21 & $-2.5$ & \\
          &  &        & & 2011-03-16.27 & 1694.51 & 1694.51 & $-2.0$ & \\
          \hline
SN~2011by & NGC~3972 & 0.0028 & 1.14(03) & 2011-05-01.58 & 9456.18 & 9456.18 & $-8.3$ & \citet{2013MNRAS.430.1030S}\\
          &  &        & & 2011-05-04.01 & 37.16 & 37.16 & $-5.9$ & \\
          &  &        & & 2011-05-05.50 & 7788.52 & 7144.74 & $-4.4$ & \\
          &  &        & & 2011-05-07.26 & 9237.24 & 9237.24 & $-2.6$ & \\
          &  &        & & 2011-05-10.54 & 9560.79 & 9560.79 & 0.6 & \\
\hline
\label{obs-log}
\end{tabular}
\end{table*}

\begin{table*}
\caption{Summary of {\it Swift} UVOT grism observations of the SN~Ia sample in this work. (continued)}
\begin{tabular}{lccccrrrc}
\hline\hline
SN Name & Host & Redshift  & \deltam($B$)$^a$ & UT Obs. Date  & $T_{\rm exp}$ & $T_{\rm exp}$ used$^b$ & Phase & LC ref.$^c$\\
 &  &  &  & (mag) & (s) & (day) & \\
\hline
SN~2011fe & NGC~5457  & 0.0008 & 1.11(00) & 2011-08-28.52 & 9437.18 & 9437.18 & $-13.8$ & \citet{2013NewA...20...30M}\\
          &  &        & & 2011-09-03.63 & 8591.08 & 8591.08 & $-7.7$ & \\
          &  &        & & 2011-09-07.28 & 7218.58 & 5878.83 & $-4.0$ & \\
          &  &        & & 2011-09-10.65	& 7679.62 & 7679.62 & 0.6 & \\   
          &  &        & & 2011-09-13.56	& 6156.47 & 6156.47 & 2.3 & \\
          &  &        & & 2011-09-16.24	& 6127.14 & 5195.36 & 5.0 & \\
          &  &        & & 2011-09-26.19	& 5507.78 & 5507.78 & 14.9 & \\
          &  &        & & 2011-09-29.26	& 3838.31 & 3838.31 & 18.0 & \\
          &  &        & & 2011-10-08.32	& 2596.54 & 2596.54 & 27.0 & \\
          &  &        & & 2011-10-11.73	& 6490.82 &	6490.82 & 30.4 & \\
          \hline
SN~2011iv & NGC~1404 & 0.0065 & 1.69(05) & 2011-12-05.26 & 8015.13 & 3048.35 & $-5.7$ & \citet{2012ApJ...753L...5F}\\
          &  &        & & 2011-12-06.85 & 2858.51 & 2858.51 & $-4.1$ & \\
          &  &        & & 2011-12-07.48 & 5627.53 & 5387.54 & $-3.5$ & \\
          &  &        & & 2011-12-09.83 & 644.11 & 622.89 & $-1.2$ & \\
          &  &        & & 2011-12-11.56 & 11926.23 & 11926.23 & 0.5 & \\
          \hline
SN~2012cg & NGC~4424 & 0.0014 & 1.04(00) & 2012-05-23.38 & 17685.15 & 15126.62 & $-12.6$ & \citet{2013NewA...20...30M}\\
          &  &        & & 2012-05-26.86 & 16944.41 & 12267.43 & $-9.1$ & \\
          \hline
SN~2012dn & ESO~462-G016 & 0.0101 & 1.08(03) & 2012-07-23.28 & 3811.96 & 1188.78 & $-1.5$ & \citet{2014ApJ...787...29B}\\
          \hline
SN~2012fr & NGC~1365 & 0.0055 & 0.85(05) & 2012-11-03.34 & 5927.61 & 5927.61 & $-9.1$ & \citet{2014AJ....148....1Z}\\
          &  &        & & 2012-11-04.67 & 11276.43 & 11276.43 & $-7.8$ & \\
          &  &        & & 2012-11-07.67 & 12656.91 & 12656.91 & $-4.8$ & \\
          &  &        & & 2012-11-08.14 & 3512.32 & 3512.32 & $-4.3$ & \\
          &  &        & & 2012-11-11.62 & 4547.31 & 4547.31 & $-0.9$ & \\
          &  &        & & 2012-11-13.39 & 16431.98 & 15312.46 & 0.9 & \\
          \hline
SN~2012ht & NGC~3447 & 0.0036 & 1.39(05) & 2012-12-26.35 & 13501.33 & 13501.33 & $-8.3$ & \citet{2014ApJ...782L..35Y}\\
          &  &        & & 2012-12-28.32 & 6860.62 & 6860.62 & $-6.3$ & \\
          &  &        & & 2012-12-31.46 & 5590.03 & 4057.27 & $-3.1$ & \\
          \hline
SN~2013aa & NGC~5643 & 0.0040 & 0.80(03) & 2013-02-19.73 & 5625.08 & 5625.08 & $-1.3$ & \citet{2013MNRAS.436..222M}\\
          &  &        & & 2013-02-26.74 & 5807.38 & 1503.57 & 5.7 & \\
          \hline
SN~2013cg & NGC~2891 & 0.0080 & \nodata & 2013-05-12.95 & 11846.95 & 11846.95 & 0.0 & Spectrum$^d$\\
          \hline
SN~2014J  & NGC~3034 & 0.0007 & 0.95(01) & 2014-01-25.67 & 11245.94 & 8649.39 & $-6.3$ & \citet{2014MNRAS.443.2887F}\\
          &  &        & & 2014-02-01.37 & 9043.35 & 7999.93 & 0.4 & \\
          &  &        & & 2014-02-03.08 & 9347.16 & 9347.16 & 2.1 & \\
          &  &        & & 2014-02-04.88 & 4401.08 & 4401.08 & 3.9 & \\
          &  &        & & 2014-02-05.87 & 4807.05 & 3639.29 & 4.9 & \\
          &  &        & & 2014-02-07.61 & 4175.04 & 3402.30 & 6.6 & \\
          &  &        & & 2014-02-08.77 & 4099.32 & 3239.30 & 7.8 & \\
          \hline
iPTF14bdn & UGC~8503 & 0.0156 & 0.84(05) & 2014-06-08.70 & 7272.02 & 7272.02 & $-5.7$ & \citet{2015ApJ...813...30S}\\
          & &         & & 2014-06-11.52 & 4589.50 & 4589.50 & $-2.9$ & \\
          & &         & & 2014-06-19.60 & 9142.26 & 9142.26 & 5.0 & \\
          & &         & & 2014-06-23.25 & 8621.41 & 8621.41 & 8.6 & \\
          & &         & & 2014-07-02.32 & 5659.88 & 5659.88 & 17.5 & \\
\hline
ASASSN-14lp & NGC~4666 & 0.0051 & 0.80(05) & 2014-12-13.54 & 16848.62 & 16848.62 & $-11.2$ & \citet{2016ApJ...826..144S}\\
          &  &        & & 2014-12-18.47 & 17369.47 & 10806.60 & $-6.3$ & \\
\hline
SN~2016ccz & MRK~685 & 0.0150 & 1.00(02) & 2016-05-28.87 & 11505.85 & 11505.85 & $-1.2$ & \citet{2017arXiv171102474F}\\
\hline
SN~2016coj & NGC~4125 & 0.0045 & 1.33(03) & 2016-06-07.02 & 12877.91 & 7549.49 & $-2.1$ & \citet{2017arXiv171102474F}\\
\hline
SN~2016eiy & ESO~509-IG064 & 0.0087 & \nodata & 2016-08-05.58 & 6697.64 & 4556.10 & 0.6 & Spectrum$^e$\\
           &  &        & & 2016-08-10.77 & 4031.31 & 4031.31 & 5.7 & \\
           \hline
SN~2016ekg & PGC~67803 & 0.0171 & \nodata & 2016-08-06.31 & 7185.58 & 6105.81 & $-0.7$ & Spectrum$^f$\\
           &  &        & & 2016-08-09.56 & 5138.10 & 5138.10 & 2.5 & \\
           \hline
SN~2016eoa & NGC~0083 & 0.0208 & 1.35(03) & 2016-08-17.51 & 13133.66 & 13133.66 & 1.8 & \citet{2017arXiv171102474F}\\
\hline
SN~2016fff & UGCA~430 & 0.0114 & 1.49(05) & 2016-08-28.87 & 6315.70 & 5408.91 & 2.4 & Swope$^g$\\
\hline
SN~2016gsb & ESO~555-G029 & 0.0097 & \nodata & 2016-10-04.44 & 11189.93 & 7405.62 & $-2.5$ & Spectrum$^h$\\
\hline
SN~2016itd & UGC~9165 & 0.0175 & \nodata & 2016-12-13.42 & 11628.96 & 11628.96 & 5.3 & Spectrum$^i$\\
\hline
\label{obs-log2}
\end{tabular}
\end{table*}

\begin{table*}
\caption{Summary of {\it Swift} UVOT grism observations of the SN~Ia sample in this work. (continued)}
\begin{tabular}{lccccrrrc}
\hline\hline
SN Name & Host & Redshift  & \deltam($B$)$^a$ & UT Obs. Date  & $T_{\rm exp}$ & $T_{\rm exp}$ used$^b$ & Phase & LC ref.$^c$\\
 &  &  &  & (mag) & (s) & (day) & \\
\hline
SN~2017cbv & NGC~5643 & 0.0040 & 1.06(00) & 2017-03-12.53 & 11914.14 & 5005.92 & $-16.5$ & \citet{2017ApJ...845L..11H}\\
           &  &        & & 2017-03-17.58 & 11081.64 & 9239.51 & $-11.5$ & \\
           &  &        & & 2017-03-24.48 & 12373.66 & 12373.66 & $-4.6$ & \\
\hline
SN~2017erp & NGC~5861 & 0.0062 & \nodata & 2017-06-17.35 & 8495.45 & 4743.82 & $-7.1$ & Swope$^g$\\
           &  &        & & 2017-06-19.18 & 10173.46 & 8581.91 & $-5.3$ & \\
\hline
SN~2018aoz & NGC~3923 & 0.0058 & \nodata & 2018-04-12.82 & 2323.53 & 2323.53 & $-2.2$ & Spectrum$^j$\\
\hline
\label{obs-log3}
\end{tabular}
\begin{flushleft}
$^a${The $B$-band decline 15~days after the peak brightness.}\\
$^b${The total exposure time of the exposures actually used for data reduction and SNR calculation in this work.}\\
$^c${The reference of SN photometric properties adopted in this work.}\\
$^d${The epoch of peak brightness is estimated from the optical spectrum observed on May 26, 2013 UT under the ANU WiFeS SuperNovA Programme \citep[AWSNAP;][]{2016PASA...33...55C}}.\\
$^e${The epoch of peak brightness is estimated from the optical spectrum observed on July 26, 2016 UT nder the Public ESO Spectroscopic Survey for Transient Objects \citep[PESSTO;][]{2015A&A...579A..40S}} program.\\
$^f${The epoch of peak brightness is estimated from the optical spectrum observed on Aug. 01, 2016 UT under the PESSTO program.}\\
$^g${Using the light curves obtained from Swope telescope at Cerro Las Campanas.}\\
$^h${The epoch of peak brightness is estimated from the optical spectrum observed on Oct. 01, 2016 UT under the PESSTO program.}\\
$^i${The epoch of peak brightness is estimated from the optical spectrum observed on Dec. 06, 2016 UT by KANATA 1.5-m telescope.}
$^j${The epoch of peak brightness is estimated from the optical spectrum observed on April 02, 2018 UT by FLOYDS-S telescope.}
\end{flushleft}
\end{table*}

\begin{table*}
\caption{Summary of {\it Swift} UVOT grism observations of the SNe~Ia {\bf NOT} included in this work owing to either extremely low SNR or serious background contamination in the UV.}
\begin{tabular}{lcc}
\hline\hline
SN Name & UT Obs. Date & Swift Obsid\\
\hline
SN~2005am & 2005-03-09 & 30010007\\
          & 2005-03-18 & 30010036\\
          & 2005-03-22 & 30010051\\
          & 2005-03-24 & 30010057\\
          \hline
SN~2005hk & 2005-11-08 & 30338004\\
          \hline
SN~2007sr & 2007-12-21 & 31073004\\
          & 2007-12-22 & 31073007\\
          & 2007-12-23 & 31073010\\
          & 2007-12-24 & 31073013\\
          \hline
SN~2010ev & 2010-07-05 & 31751001\\
          \hline
SN~2014J  & 2014-01-23 & 33124001\\
          \hline
SN~2016gfr & 2016-09-21 & 34732002\\
          \hline
OGLE16dha & 2016-09-30 & 34742002\\
          & 2016-10-05 & 34742004\\
          \hline  
SN~2018gv & 2018-01-23 & 10521003\\
          & 2018-01-23 & 10521005\\
          \hline
SN~2018xx & 2013-02-23 & 10572004\\
\hline
\label{bad-log}
\end{tabular}
\end{table*}

\clearpage

\label{lastpage}

\end{document}